%% file: Draft_Mesons.tex
\documentclass[prd,superscriptaddress,nofootinbib,eqsecnum]{revtex4}

\pdfoutput=1

\usepackage{tabularx}  
\usepackage{graphicx}  
\usepackage{color}  
\usepackage{bm}  
\usepackage{amssymb}  
\usepackage{times}  

\newcommand{\mc}[3]{\multicolumn{#1}{#2}{#3}}
 
\newcommand{\bi}{\begin{itemize}}
\newcommand{\ei}{\end{itemize}}

\newcommand{\bt}[2]{\begin{table}[!b] \caption{#1}\begin{tabular}{#2} \hline\hline  \\[-0.5em]}
\newcommand{\et}[1]{\hline\hline \end{tabular} \label{#1} \end{table}}

\newcommand{\btd}[1]{\begin{table}[!b] \begin{tabular}{#1} \hline\hline  \\[-0.5em]}
\newcommand{\etd}[2]{\hline\hline \end{tabular} \caption{#1}\label{#2} \end{table}}

\newcommand{\be}{\begin{equation}}
\newcommand{\ee}{\end{equation}}

\newcommand{\bea}{\begin{eqnarray}}
\newcommand{\eea}{\end{eqnarray}}

\newcommand{\err}[3]{\ensuremath{{#1}^{+#2}_{#3}}}

\newcommand{\kc}{\ensuremath{\kappa_{\rm crit}}}
\newcommand{\kp}{\ensuremath{ \kappa }}
\newcommand{\kpch}{\ensuremath{\kappa_{\rm c} }}
\newcommand{\kpbot}{\ensuremath{\kappa_{\rm b} }}

\newcommand{\dk}{\ensuremath{ \delta\kappa }}
\newcommand{\dkk}{\ensuremath{ \frac{\delta \kappa}{\kappa} }}

\newcommand{\amq}{\ensuremath{am_q'}}

\newcommand{\ams}{\ensuremath{ am_s}}

\newcommand{\aMps}{\ensuremath{aM}}
\newcommand{\aMvm}{\ensuremath{aM^*}}

\newcommand{\aMr}{\ensuremath{M_1}}
\newcommand{\aMrps}{\ensuremath{aM_1}}
\newcommand{\aMrvm}{\ensuremath{aM^*_1}}

\newcommand{\ahf}{\ensuremath{a\Delta_1}}
\newcommand{\ronehf}{\ensuremath{r_1 \Delta_1}}
\newcommand{\hf}{\ensuremath{\Delta_1}}

\newcommand{\Mkbar}{\ensuremath{\overline{M}_2}}
\newcommand{\aMkps}{\ensuremath{aM_2}}
\newcommand{\aMkvm}{\ensuremath{aM^*_2}}
\newcommand{\aMkbar}{\ensuremath{a\overline{M}_2}}

\newcommand{\Ep}{\ensuremath{ E(\p) }}
\newcommand{\aEp}{\ensuremath{ aE(\p) }}

\newcommand{\p}{ {\text {\boldmath $p$}} }

\newcommand{\n}{\ensuremath{\bm n}}

\newcommand{\chisq}{\ensuremath{\chi^2/\text{dof}}}
\newcommand{\chiaug}{\ensuremath{\chi^2_{\rm aug}}}
\newcommand{\chiaugdof}{\ensuremath{\chi^2_{\rm aug}/\text{dof}}}

\newcommand{\Dslash}{\ensuremath{{D\kern -0.65em /}}}
\newcommand{\half}{\ensuremath{{\textstyle\frac{1}{2}}}}
\newcommand{\ihalf}{\ensuremath{{\textstyle\frac{i}{2}}}}

\newcommand{\sixth}{\ensuremath{{\textstyle\frac{1}{6}}}}
\newcommand{\third}{\ensuremath{{\textstyle\frac{1}{3}}}}

\begin{document}

\title{Tuning Fermilab Heavy Quarks in 2+1 Flavor Lattice QCD with Application to Hyperfine Splittings }

\date{\today}

\begin{abstract}
We report the non-perturbative tuning  of  parameters---$\kpch$, $\kpbot$, and \kc---that
are related to the bare heavy-quark mass in the Fermilab action.
This requires the computation of the masses of $D_s^{(*)}$ and $B_s^{(*)}$ mesons comprised of a Fermilab heavy quark and a staggered light quark.
Additionally, we report the hyperfine splittings for $D_s^{(*)}$ and $B_s^{(*)}$ 
mesons as a cross-check of our simulation and analysis methods.
We find a splitting of $145 \pm 15$~MeV for the 
$D_s$ system and $40 \pm 9$~MeV for the $B_s$ system.  
These are in good agreement with the Particle Data Group average values of  $143.9\pm0.4$~MeV and $46.1\pm1.5$~MeV, respectively.   
The calculations are carried out with 
the MILC 2+1 flavor gauge configurations at three lattice spacings $a \approx 0.15, 0.12$ and $0.09$~fm.
\end{abstract}

\author{C.~Bernard}
\affiliation{Department of Physics, Washington University, St.~Louis, Missouri 63130, USA}

\author{C.~DeTar}
\affiliation{Physics Department, University of Utah, Salt Lake City, Utah 84112, USA}

\author{M.~Di Pierro}
\affiliation{School of Computing, DePaul University, Chicago, Illinois 60604, USA}

\author{A.X.~El-Khadra}
\affiliation{Physics Department, University of Illinois, Urbana, Illinois 61801, USA}

\author{R.T.~Evans}
\affiliation{Physics Department, University of Illinois, Urbana, Illinois 61801, USA}
\affiliation{Institut f\"ur Theoretische Physik, Universit\"at Regensburg, 93040 Regensburg, Germany}

\author{E.D.~Freeland}
\email[]{eliz@fnal.gov}
\affiliation{Department of Physics, Washington University, St.~Louis, Missouri 63130, USA}

\author{E.~G\'amiz}
\affiliation{Physics Department, University of Illinois, Urbana, Illinois 61801, USA}
\affiliation{Fermi National Accelerator Laboratory, Batavia, Illinois 60510, USA}

\author{Steven~Gottlieb}
\affiliation{Department of Physics, Indiana University, Bloomington, Indiana 47405, USA}
\affiliation{National Center for Supercomputing Applications, University of Illinois, Urbana 61801, Illinois, USA}

\author{U.M.~Heller}
\affiliation{American Physical Society, One Research Road, Ridge, New York 11961, USA}

\author{J.E.~Hetrick}
\affiliation{Physics Department, University of the Pacific, Stockton, California 95211, USA}

\author{A.S.~Kronfeld}
\affiliation{Fermi National Accelerator Laboratory, Batavia, Illinois 60510, USA}

\author{J.~Laiho}
\affiliation{Department of Physics, Washington University, St.~Louis, Missouri 63130, USA}
\affiliation{Department of Physics and Astronomy, University of Glasgow, Glasgow, Scotland, UK}

\author{L.~Levkova}
\affiliation{Physics Department, University of Utah, Salt Lake City, Utah 84112, USA}

\author{P.B.~Mackenzie}
\affiliation{Fermi National Accelerator Laboratory, Batavia, Illinois 60510, USA}

\author{J.N.~Simone}
\affiliation{Fermi National Accelerator Laboratory, Batavia, Illinois 60510, USA}

\author{R.~Sugar}
\affiliation{Department of Physics, University of California, Santa Barbara, California 93106, USA}

\author{D.~Toussaint}
\affiliation{Department of Physics, University of Arizona, Tucson, Arizona 85721, USA}

\author{R.S.~Van~de~Water}
\affiliation{Physics Department, Brookhaven National Laboratory, Upton, New York 11973, USA}

\collaboration{Fermilab Lattice and MILC Collaborations}
\noaffiliation

\maketitle


\section{Introduction}
	\input{Introduction}

\section{Theoretical Background}   \label{Sec:theorybackground}

\input{TheoreticalBackground}

\section{Simulations}   \label{Sec:simulations}
	\input{Simulations}

\section{Analysis overview}  \label{Sec:analysis}
	\input{Analysis}

\section{Fitting Details for $E(\bm{p}), M_1, M_2$}   \label{Sec:FitDetails}
	\input{Fit_Details_Ep_M1_M2}


\section{Results}   \label{Sec:results}
	\input{Results}

\section{Summary and Outlook}   \label{Sec:conclusion}
	\input{Future_Conclusions}

\begin{acknowledgments}
Computations for this work were carried out on facilities of the USQCD Collaboration, which are funded by the
Office of Science of the U.S. Department of Energy.
This work was supported in part by the U.S. Department of Energy under Grants No.~DE-FC02-06ER41446 (C.D., L.L.),
No.~DE-FG02-91ER40661 (S.G.), No.~DE-FG02-91ER40677 (A.X.K., E.G., R.T.E.),
No.~DE-FG02-91ER40628 (C.B, E.D.F.), No.~DE-FG02-04ER-41298 (D.T.); the National Science Foundation
under Grants No.~PHY-0555243, No.~PHY-0757333,
No.~PHY-0703296 (C.D., L.L.), No.~PHY-0757035 (R.S.), No.~PHY-0704171 (J.E.H.) and No.~PHY-0555235 (E.D.F.);
URA visiting scholars awards (E.G., R.T.E.),
and the M. Hildred Blewett Scholarship of the American Physical Society (E.D.F.).
This manuscript has been co-authored by an employee of Brookhaven Science Associates, LLC, 
under Contract No. DE-AC02-98CH10886 with the U.S. Department of Energy.
R.S.V. acknowledges support from BNL via the Goldhaber Distinguished Fellowship.
Fermilab is operated by Fermi Research Alliance, LLC, under Contract
No.~DE-AC02-07CH11359 with the U.S. Department of Energy.
\end{acknowledgments}

\begin{appendix}


\input{App_M2DiscErr}

	\clearpage
	 \input{App_M2}

	 \clearpage
	\input{App_hyperfine}

 	\clearpage
	\input{App_hfsPQChiPT}

\end{appendix}

%


\end{document}

%% file: Introduction.tex


Lattice QCD calculations play a critical role in the study of standard
model physics and the search for new physics.
For a set of lattice QCD calculations to be viable, several basic tasks
are necessary.  
The bare gauge coupling must be eliminated in favor of an observable
allowing the conversion from lattice to physical units;
the bare masses in the lattice action must 
be tuned to correspond to physical quarks; and 
experimentally established quantities must be 
calculated in order to
substantiate the method's accuracy and reliability.
Once these tasks are complete, a variety of quantities inaccessible to
or not yet determined by experiment may be calculated, such as decay
constants, form factors, and mass spectra.

The Fermilab Lattice and MILC Collaborations have reported several
calculations~\cite{Davies:2003ik,Aubin:2004ck,Aubin:2005ar,Aubin:2004ej,
Evans:2008zz,Freeland:2007wk,Allison:2004be,Burch:2009az} based on 
ensembles of lattice gauge fields with 2+1 flavors of sea quarks, 
generated by the MILC Collaboration~\cite{Bernard:2001av,Aubin:2004wf}.
Details of the scale setting can be found in Refs.~\cite{r1, Bazavov:2009bb}, 
and details of the light-quark mass tuning in Ref.~\cite{Bazavov:2009bb}.
In this paper, we report on the necessary tuning of the heavy-quark 
action for charmed and bottom quarks.
In particular, we describe calculations of the heavy-light pseudoscalar 
and vector meson masses using, for light quarks, the asqtad staggered action ~\cite{stag_fermion} and, for 
heavy quarks, the Fermilab interpretation~\cite{El-Khadra:1996mp} of the 
Sheikholeslami-Wohlert (``clover'') action~\cite{Sheikholeslami:1985ij}
for Wilson fermions~\cite{wilson:1977}. 
We use the spin-average of these meson masses to nonperturbatively tune
the hopping parameter $\kappa$, which is equivalent to the bare heavy-quark mass.
We also describe the determination of $\kc$, the value of $\kappa$ for which a 
degenerate Wilson pseudoscalar's mass vanishes.
The value of $\kc$ plays a minor role in the calculation of heavy-light matrix elements~\cite{Aubin:2005ar,Aubin:2004ej,Evans:2008zz}, 
and a more important role when determining a renormalized quark mass~\cite{Freeland:2007wk}. 
Finally, as a by-product of these calculations, we report the 
spin-dependent hyperfine splittings for $B_s$ and $D_s$ mesons, which 
test how well we have improved the chromomagnetic interaction.

Two aspects of the Fermilab method are important here.
First, the Fermilab interpretation makes no assumptions about the size
of the quark mass.
Therefore, we are able to treat both charm and bottom quarks within the same
framework.
Second, since the Sheikoleslami-Wohlert action maintains the spin and
flavor symmetries of heavy quarks, 
heavy-quark effective theory (HQET) can be used to interpret and improve
lattice discretization effects~\cite{Kronfeld:2000ck,HQETradiative}.
HQET techniques can be used to show how the improvement works for 
observables, such as meson masses, in a way
simpler than, though equivalent to, the Symanzik
improvement program~\cite{Symanzik:1979ph}.

This paper is organized as follows.
Section~\ref{Sec:theorybackground}  reviews the theoretical framework
upon which these calculations are based.
Section~\ref{Sec:simulations} contains specific descriptions of the gauge
configurations, actions, and operators used for the meson masses.  
Section~\ref{Sec:analysis} covers the components of the numerical
analysis.
Section~\ref{Sec:FitDetails} details the fitting procedures.
Section~\ref{Sec:results} presents the results for the non-perturbative
tuning of the heavy-quark hopping parameters $\kpch$ and $\kpbot$, the hyperfine
splitting, and the critical hopping parameter~\kc.
Section~\ref{Sec:conclusion} summarizes with a discussion of
improvements to these calculations that are currently underway.
Details of the meson-mass discretization error estimation are given in
Appendix~\ref{App_M2DiscErr}.
Appendices~\ref{App_M2}  and~\ref{App_hyperfine} tabulate intermediate numerical results.
The partially quenched chiral perturbation theory expression for the hyperfine splitting is derived in Appendix~\ref{App_hfsPQChiPT}.


%% file: TheoreticalBackground.tex
 
The hopping-parameter form of the heavy-quark action
is~\cite{El-Khadra:1996mp}
\be  \label{eq:actionS}
	S = S_0 + S_B + S_E ,
\ee
where
\bea
	S_0 & = & \sum_n \overline{\psi}_n \psi_n 
	           -\kappa \sum_{n,\mu} \left[ \, \overline{\psi}_n ( 1 -
\gamma_\mu )U_{n,\mu} \psi_{n+\hat{\mu}}  +
\overline{\psi}_{n+\hat{\mu}} (1 + \gamma_\mu) U^{\dag}_{n,\mu} \psi_n
\right]  ,  \\[1.0em]
 	S_B & = & \frac{i}{2} c_{B} \kappa \sum_{n; i,j,k}   \epsilon_{ijk}
\overline{\psi}_n  \sigma_{ij} B_{n;k} \psi_n   ,  \label{eq:SB} \\[1.0em]
       S_E & = & i c_{E} \kappa \sum_{n; i}  \overline{\psi}_n
\sigma_{0i} E_{n;i} \psi_n ,    \label{eq:SE}
 \eea
where $\sigma_{\mu \nu} = \ihalf[\gamma_\mu, \gamma_\nu]$.
The chromomagnetic and chromoelectric fields $B_{n;i}$ and $E_{n;i}$  are
standard and given in Ref.~\cite{El-Khadra:1996mp}. 
The term $S_0$ includes dimension-five terms to alleviate the fermion doubling problem~\cite{wilson:1977}.
The couplings $c_E$ and $c_B$ of the dimension-five operators in $S_B$
and $S_E$ are chosen to reduce discretization
effects~\cite{El-Khadra:1996mp,Sheikholeslami:1985ij}.

The hopping parameter $\kappa$ is related to the tadpole-improved bare
quark mass by
\be  \label{eq:baremass}
	am_0 = \frac{1}{u_0} \left(  \frac{1}{2 \kappa} - \frac{1}{2\kc}  \right) ,
\ee
where $a$ is the lattice spacing, 
$u_0$ is the tadpole-improvement factor~\cite{Lepage:1992xa}, and 
$\kc$ is the value of $\kappa$
for which the pseudoscalar meson mass (of two degenerate Wilson quarks)
vanishes.
Our nonperturbative determination of $\kc$ is discussed
in Sec.~\ref{sec:kcrit}.
%
To motivate our method of tuning $\kappa$, we first discuss the meson
dispersion relation.
We then turn to the HQET description of our Lagrangian to understand
how to best use the dispersion relation.

The meson dispersion relation can be written, for 
$|\p| \ll m_0, a^{-1}$, as~\cite{El-Khadra:1996mp}
\be
    E(\p) = M_1 + \frac{\p^2}{2 M_2} + O( \p^4) .
    \label{eq:disprel}
\ee
Here, and throughout this work, we use lower-case $m$ for quark masses
and upper-case $M$ for meson masses. 
$M_1$ and $M_2$ are known as the rest mass and kinetic mass,
respectively.
Because the lattice breaks Lorentz invariance, $M_1\neq M_2$, 
although $M_1\to M_2$ as $a\to0$ for the action
in Eq.~(\ref{eq:actionS}).
By tuning $\kappa$, one could adjust the bare, heavy-quark mass such that
\emph{either} $M_1$ or $M_2$ is equal to the physical meson mass.
(To set $M_1=M_2$ requires the introduction, and
tuning, of an additional parameter in the action.  This is possible
but, as discussed below, not necessary~\cite{El-Khadra:1996mp}.)

To clarify the role of the different masses in Eq.~(\ref{eq:disprel}),
it is useful to introduce an effective Lagrangian.
This  also sets up a language for discussing discretization errors
later.
Because the action in Eq.~(\ref{eq:actionS}) has the same heavy-quark
spin and flavor symmetries as continuum QCD, HQET is an obvious
candidate for its description~\cite{Kronfeld:2000ck,HQETradiative}.  
To employ HQET, one separates the
short-distance physics at the scale of the inverse heavy-quark mass
$1/m_Q$ from the long-distance physics at 
the characteristic scale of QCD, $\Lambda_{\rm QCD}$.
The fact that we have a 
lattice does not change the validity or utility of this separation.
It simply means that the lattice spacing $a$ must be
included in the description of the short-distance physics.  
Thus, the short-distance coefficients of HQET
applied to Eq.~(\ref{eq:actionS})  differ from those arrived at by
applying HQET to continuum QCD; these differences are the heavy-quark
discretization errors.
Parameters in the lattice action can be chosen to minimize them.

We introduce the heavy-quark effective Lagrangian for our lattice gauge theory  by
writing~\cite{Kronfeld:2000ck,HQETradiative}
\be
	\mathcal{L}_{\rm LGT} \doteq     \mathcal{L}_{\rm light}  +
\mathcal{L}_{\rm HQET},
\ee
where $\mathcal{L}_{\rm light}$ is the 
Symanzik local effective Lagrangian for the light degrees of freedom
and $\doteq$ means the Lagrangian on the right-hand 
side describes the on-shell matrix elements of the Lagrangian on the 
left-hand side.
The HQET Lagrangian has a power-counting scheme, denoted by
\be  \label{eq:LHQET}
	\mathcal{L}_{\rm HQET} =  \sum_s  \mathcal{L}_{\rm HQET}^{(s)},
\ee
where $\mathcal{L}_{\rm HQET}^{(s)}$ includes all operators of
dimension $4+s$, with coefficients of dimension $-s$ 
consisting of
powers of the short distances, $1/m_Q$ or $a$.
The first few terms in $\mathcal{L}_{\rm HQET}$
are~\cite{Kronfeld:2000ck}
\bea
	\mathcal{L}_{\rm HQET}^{(0)} &=&  - \bar{h}^{(+)} ( D_4 + m_1 )
h^{(+)} ,   \label{eq:L0} \\
	\mathcal{L}_{\rm HQET}^{(1)} &=&  \bar{h}^{(+)} \frac {\bm{D}^2}{2m_2}
h^{(+)}  +  \bar{h}^{(+)}  \frac {i\bm{\sigma}\cdot \bm{B} }{2m_B}  h^{(+)}
,  \label{eq:L1} \\
	\mathcal{L}_{\rm HQET}^{(2)} &=&  \bar{h}^{(+)}  \frac { i\bm{\sigma} \cdot
(\bm{D} \times \bm{E}) }{8m_E^2}  h^{(+)}  +  \bar{h}^{(+)} \frac
{\bm{D} \cdot \bm{E} }{8m_D^2}  h^{(+)} ,
    \label{eq:L2}
\eea
where $h^{(+)}$ is a two-component heavy-quark field, $\bm{\sigma}$ are the Pauli matrices,  
and $\bm{B}$ and $\bm{E}$ are the continuum 
gauge fields.
The masses $m_1, m_2, m_B, m_E$, and $m_D$ are functions of the bare-quark
mass $m_0$ and the gauge coupling.
For example, the masses $m_1$ and $m_2$ are defined
to all orders in perturbation theory by Eq.~(\ref{eq:disprel}), applied now to the pole energy of a 
one-quark state~\cite{Mertens:1997wx}.
The entries in Eqs.~(\ref{eq:L0})--(\ref{eq:L2}) are commonly
referred to as follows. 
$\mathcal{L}_{\rm HQET}^{(0)}$ gives the rest mass.
The first term of $\mathcal{L}_{\rm HQET}^{(1)}$ is the kinetic energy
and the second is the chromomagnetic, or hyperfine, interaction.
The first term of $\mathcal{L}_{\rm HQET}^{(2)}$ is the spin-orbit
interaction while the second is known as the Darwin term.

For the pseudoscalar and vector meson rest masses, 
the HQET formalism can be used to show that~\cite{Kronfeld:2000ck}
\begin{equation}     \label{eq:HQETM1}
  	M_1^{(*)} = m_1 + \bar{\Lambda}_{\rm lat} - \frac{\lambda_{1, \rm
lat}}{2m_2} 
		- d_J \frac{ \, \lambda_{2,\rm lat}}{2m_B} + O(1/m^2),
\end{equation}
where $J$ is the total meson angular momentum with $d_0 = 3$ and $d_1 =
-1$ for the pseudoscalar ($M_1$) and vector ($M_1^*$) mesons,
respectively.
The quantities $\bar{\Lambda}_{\rm lat}$, $\lambda_{1, \rm lat}$, and
$\lambda_{2,\rm lat}$ are HQET matrix
elements. At
non-zero lattice spacing they contain discretization effects from $\mathcal{L}_{\rm light}$, hence the subscript ``lat''. 
The continuum limit of these quantities yields their
counterparts in HQET applied to continuum 
QCD~\cite{Kronfeld:2000ck},
which provides a basis for computing the continuum-QCD 
quantities $\bar\Lambda$ and $\lambda_1$~\cite{Kronfeld:2000gk}.

Mass splittings and matrix elements such as decay constants and form
factors are not affected by the value of $m_1$~\cite{Kronfeld:2000ck}.
Thus, Eqs.~(\ref{eq:L0}) and~(\ref{eq:L1}) show that the kinetic mass $m_2$
is the first mass in the expansion that does play a role in the
dynamics.
We therefore would like to associate $m_2$, and hence $M_2$, with the 
physical mass, tolerating $m_1 \neq m_2$ (and $M_1\neq M_2$) for 
nonzero lattice spacings.  
%
The nonperturbative tuning of $\kappa$ then
entails adjusting $\kappa$ until the meson kinetic mass---%
determined by fits of Monte Carlo lattice data to the dispersion
relation, Eq.~(\ref{eq:disprel})---equals that of the physical meson
mass.
A relation similar to Eq.~(\ref{eq:HQETM1}) holds for $M_2$
\begin{equation}
  	M_2^{(*)} = m_2 + \bar{\Lambda}_{\rm lat} + O(1/m),
    \label{eq:HQETM2}
\end{equation}
with the leading discretization errors appearing in the $1/m$ 
contribution.
Final values for the nonperturbative tuning of  $\kappa$ are given in
Sec.~\ref{sec:tuning}.

To calculate the hyperfine splitting of the $D_s$ or $B_s$ meson, consider
\begin{equation}
	\Delta_1 \equiv  M_1^* - M_1.
    \label{eq:hf}
\end{equation}
From Eq.~(\ref{eq:HQETM1}),
\begin{equation}
	M_1^* - M_1 = 4 \frac{ \, \lambda_{2,\rm lat}}{2m_B} + \cdots  ,
    \label{eq:hf-hqet}
\end{equation}
which differs from the continuum splitting only by
discretization errors in the light quarks and gluons appearing in $\lambda_{2,\rm lat}$, the mismatch of $m_B$ and its continuum
counterpart (or, equivalently, the choice of $c_B$), and similar contributions from higher-dimension
operators~\cite{Kronfeld:2000ck,Oktay:2008ex}.
The splitting of kinetic masses, $\Delta_2 \equiv M_2^*-M_2$, does not depend on $m_B$; rather, it depends on other generalized masses which are not tuned in our simulations.\footnote{Tree-level expressions for these masses, and hence their mismatch, can be found in Ref.~\cite{Oktay:2008ex}.}
Thus, $\Delta_1$ formally has smaller discretization errors than $\Delta_2$. 
$\Delta_1$ is also statistically cleaner than $\Delta_2$.
In Eq.~(\ref{eq:hf-hqet}), $1/m_B$ is sensitive to the clover coupling 
$c_B$ in Eq.~(\ref{eq:SB}), so $\Delta_1$ tests how well it has been 
chosen.
The $B_s$ and $D_s$ hyperfine splittings are given in Sec.~\ref{sec:hfs}.

%% file: Simulations.tex
In this section, we describe the gauge configurations used and the
details of the actions, operators, and correlation functions that
describe the heavy-light mesons.
In Section~\ref{sec:configs}, we discuss the gauge configurations and
the parameters that describe each ensemble.
We also review how the lattice spacing is determined and the values of
the conversion factors $r_1$ and $r_1/a$.
In Section~\ref{sec:correlators}, we discuss parameter choices for the valence quarks and  the smearing of the heavy-quark 
wave function and how correlators are built from heavy and light quark
fields.

\subsection{Gauge Configurations and Related Parameters}
\label{sec:configs}

We use the MILC gauge configurations~\cite{Bernard:2001av,Aubin:2004wf} that have 2+1 flavors of asqtad-improved staggered sea
quarks~\cite{stag_fermion} and a Symanzik-improved gluon
action~\cite{Alford:1995hw, more_imp_glue}. 
Discretization errors from the sea quarks and gluons start at
$O(\alpha_s a^2, a^4)$.
The four-fold degeneracy of staggered sea quarks is removed by taking
the fourth root of the determinant.
To support the legitimacy of this procedure,
Shamir has developed a renormalization-group framework for lattice QCD
with staggered fermions, which he uses to argue that
non-local effects of the rooted staggered theory are absent in the
continuum limit~\cite{Shamir:2006nj}.  
Additional support for this procedure comes from chiral perturbation theory arguments~\cite{Bernard:2006zw, Bernard:2007ma}.
Reviews of these papers and of other evidence that this procedure reproduces the correct continuum limit appear in~\cite{Bazavov:2009bb,staggered:reviews,Kronfeld:2007ek}.

Table~\ref{Tbl:vacuum} lists the parameters of the gauge configurations
used in this work. 
All configurations have been gauge-fixed to Coulomb gauge.
Ensembles of configurations are grouped by their approximate lattice
spacing and are referred to as ``fine'' ($a \approx 0.09$~fm),
``coarse'' ($a \approx 0.12$~fm), and 
``medium-coarse'' ($a \approx0.15$~fm).
The simulation bare masses of the light and strange sea quarks are
denoted by $am'_l$ and $am'_s$, respectively, where $am'_l$ is the mass of
the two lighter sea-quarks.
The range of $am'_l$ is light enough that the
physical up- and down-quark masses can be reached by a chiral
extrapolation, while $am'_s$ is close to the physical strange-quark mass.
For convenience below, we write $(am'_l, am'_s)$ to identify ensembles,
e.g., ``the (0.0031, 0.031) fine ensemble''. 
Also in Table~\ref{Tbl:vacuum} are  the tadpole factors
$u_0$~\cite{Lepage:1992xa, MILC07}, determined from the mean plaquette
and used to improve the gauge-configuration
actions~\cite{Bernard:2001av,Aubin:2004wf}.
The value of the physical strange-quark mass is denoted by the unprimed
$m_s$~\cite{MILC07}.

To convert between lattice and physical units, the physical value of
the lattice spacing must be determined.
We define the distance $r_1$~\cite{r1} by
\be
	r_1^2 F(r_1) = 1,
\ee
where $F(r)$ is the force between static quarks, calculated on the
lattice.
For each ensemble, this yields a value of $r_1$ in lattice units,
$r_1/a$.     
The values are then ``smoothed'' by fitting $\ln(r_1/a)$, from all
ensembles, to a polynomial in $\beta$ and $2am'_l+am'_s$~\cite{MILC07}.
The physical value of $r_1$ is obtained via the lattice calculation of
an experimentally measurable quantity.  
We consider two current determinations here.   
One uses a lattice calculation of the $\Upsilon(2S)$-$\Upsilon(1S)$
splitting~\cite{Gray:2005ur} to arrive at 
$r_1 =0.318(7)$~fm~\cite{MILCr1_upsilon,Aubin:2004wf}.
A more recent determination using $r_1f_\pi$ gives 
$r_1 =0.3108(15)(^{+26}_{-79})$~fm~\cite{Bernard:2007ps}.   
These two determinations are consistent within errors.   
Because the determination of $r_1$ from $f_\pi$ uses finer lattice
spacings, we take that value, 
\be  \label{eq:r1}
	r_1 = 0.3108 (^{+30}_{-80}) {\rm ~fm} 
\ee
with no additional error.
While this work was being completed, a new determination of $r_1$ that
uses two mass splittings and one decay constant became available; $r_1
= 0.3133(^{+23}_{-3})$~\cite{Davies:2009tsa}, which
is consistent with the value used in this work.   
Quantities can now be converted from lattice to physical units by using
$r_1$ and the appropriate value of $r_1/a$ given in
Table~\ref{Tbl:vacuum}~\cite{MILC07}.

\begin{table*}
	\caption{Parameters describing the ensembles used.  
	The dimensions of the lattice are given in terms of the spatial ($N_{L}$) and temporal ($N_{T}$) size in lattice units.
	The gauge coupling is given by $\beta = 10/g^2$.
	The bare masses of the light and strange sea quarks are given by
$am'_l$ and $am'_s$, respectively. 
	$L = a N_L$ is the linear spatial dimension of the lattice in fm.
	The column labeled $N_{\rm cf}$ is the number of configurations used
in this work.
	The plaquette-determined tadpole-improvement factor is
$u_0$~ \protect\cite{MILC07}.
	The physical strange quark mass is $am_s$~\protect\cite{MILC07} with errors, statistical and systematic, of less than one percent.
	The ratio $r_1/a$ is described in the text; errors are Hessian from
the smoothing fit.
	The final column lists the value of the inverse lattice spacing $a^{-1}$ using 
    $r_1= 0.3108(^{+30}_{-80})$~fm 
    to convert from $r_1/a$; errors are from the error on $r_1$ and $r_1/a$. }
	\begin{tabular*}{\textwidth}{l*{10}{@{\extracolsep{\fill}}c}}
		\hline \hline
		    & $N_{L}^3 \times N_{T}$  &   $\beta$   &  $\;am'_l$  &  $am'_s$ & $L$ (fm) &  $N_{\rm cf}$  & $u_0$     &   $am_s$  & $r_1/a$  & $a^{-1}$~GeV   \\
        \hline 
``Fine'' $a \approx$  0.09 fm 
		   & $40^3 \times 96$  &  7.08  & 0.0031  & 0.031 & 3.5 &   435   &  0.8779   &  0.0252  & 3.692(6)  &\err{2.344}{60}{-23}  \\  
		   & $28^3 \times 96$  &  7.09  & 0.0062  &  0.031 & 2.4 &   557   & 0.8782   &  0.0252  & 3.701(5)  & \err{2.349}{61}{-23} \\  
		   & $28^3 \times 96$  &  7.11  & 0.0124  &  0.031 & 2.4 &   518   & 0.8788   &  0.0252  & 3.721(5)  & \err{2.362}{61}{-23}  \\  
``Coarse'' $a \approx$  0.12 fm 
		   & $24^3 \times 64$  &  6.76   & 0.005 &  0.05    & 2.9 &   529   &  0.8678  &  0.0344  & 2.645(3)  & \err{1.679}{43}{-16} \\  
		   & $20^3 \times 64$  &  6.76   & 0.007   &  0.05    & 2.4 &   836 &  0.8678  &  0.0344  & 2.635(3)  &  \err{1.672}{43}{-16} \\  
		   & $20^3 \times 64$  &  6.76   & 0.010   &  0.05    & 2.4 &   592 &  0.8677  &  0.0344  & 2.619(3)  &  \err{1.663}{43}{-16}  \\  
		   & $20^3 \times 64$  &  6.79   & 0.020   &  0.05    & 2.4 &   460 &  0.8688  &  0.0344  & 2.651(3)  &  \err{1.683}{43}{-16} \\  
		   & $20^3 \times 64$  &  6.81   & 0.030   &  0.05    & 2.4 &   549 &  0.8696  &  0.0344  & 2.657(4)  &  \err{1.687}{43}{-16} \\ 
``Medium-coarse'' $a \approx$  0.15 fm 
		    & $16^3 \times 48$  &  6.572 & 0.0097 & 0.0484 & 2.4 &   631   & 0.8604  &  0.0426  & 2.140(4) &  \err{1.358}{35}{-13}  \\  
		    & $16^3 \times 48$  &  6.586 & 0.0194 & 0.0484 & 2.4 &   631   & 0.8609  &  0.0426  & 2.129(3)  &  \err{1.352}{35}{-13} \\  
		    & $16^3 \times 48$  &  6.600 & 0.0290 & 0.0484 & 2.4 &   440   & 0.8614  &  0.0426  & 2.126(3)  &  \err{1.350}{35}{-13} \\  
		\hline \hline
	\end{tabular*}
	\label{Tbl:vacuum}
\end{table*}

\subsection{Meson Correlation Functions}  \label{sec:correlators}

Table~\ref{Tbl:mesons} lists the values of parameters used in the
valence-quark actions.
For the light valence quark, we again use the asqtad
action~\cite{stag_fermion}  and masses $am'_q$ close to the
physical value of the strange-quark mass, cf.\ Table~\ref{Tbl:vacuum}.
%
From Eqs.~(\ref{eq:L0})--(\ref{eq:L2}), one can see that with $m_2$
tuned to the physical mass, the leading mismatch between lattice and
continuum physics is in the hyperfine term in 
$\mathcal{L}_{\rm HQET}^{(1)}$.
In principle, one can tune $m_B$ to its continuum counterpart yielding
a match between lattice and continuum actions for both terms in Eq.~(\ref{eq:L1}).  
Here, we use the tree-level expression for $m_B$, which leaves the leading mismatch at
$O(\alpha_s a \Lambda)$.  
By setting $c_E = c_B$ we obtain the Sheikholeslami-Wohlert,
$O(a)$-improvement of discretization errors in the action~\cite{Sheikholeslami:1985ij}.
From the HQET perspective, this leaves $m_E\neq m_2$ in 
Eq.~(\ref{eq:L2}), but the effects of this mistuning are 
at $O(a^2 \Lambda^2)$ and $O(\alpha_sa\Lambda^2/m_Q)$.
Implementing the improvements above and using tree-level tadpole
improvement in the perturbative expressions~\cite{Lepage:1992xa,
Alford:1995hw}, we use $c_E = c_B = u_0^{-3}$.

The values of $u_0$ used in the heavy-quark and light-valence actions
are given in Table~\ref{Tbl:mesons}.  
For the fine and medium-coarse ensembles, they are the plaquette values
used to generate the MILC gauge configurations.  
For the coarse ensembles, the Landau-gauge link value was used.
The use of different $u_0$ definitions results in a slight
mismatch between the light valence- and sea-quark actions.  
In part because the meson mass is 
relatively insensitive to the strange sea-quark mass, 
we do not expect any
significant systematic errors from this mismatch.
Changes in $u_0$ result in changes to the bare mass of the heavy quark
as well, but this effect is partly absorbed by the nonperturbative
tuning of $\kappa$ and $\kc$.
Table~\ref{Tbl:mesons} also lists the nominal values of the light
valence-quark mass and sets of $\kappa$ values for bottom and charm
mesons.  These sets of $\kappa$ values, and mesons created from them,
are referred to as  charm-type or bottom-type.

With the parameters of the actions set, we now turn to the construction of the two-point correlators.
Contributions from excited states can be significantly reduced by using
a spatially smeared source, sink, or both, for the heavy-quark
propagator.
For the correlators in this work, we use two types of source-sink
combinations for the heavy quarks.
One is simply a delta function for both the source and sink; we refer
to this as the local correlator.
The other smears the field
$\psi(t,\bm{x})$ with a discretized 
version~\cite{Menscher:2005kj}
of the $1S$ charmonium wavefunction, $S(\bm{y})$, based on the Richardson
potential~\cite{Richardson:1978bt}:
\be  \label{eq:interpolating_op-smeared}
 	\phi(t, \bm{x}) = \sum_{\bm{y}} S(\bm{y}) \; \psi(t,
\bm{x}+\bm{y}) ,
\ee
and the smearing wavefunction is applied after fixing to Coulomb gauge.
Correlators using $\phi(t, \bm{x})$ are referred to as smeared
correlators.
All light valence quarks have a local source and sink.
The meson correlator is
\be \label{eq:correlator-smeared}  \label{eq:correlator}
	C_{i,j}(t,\p) = \sum_{\bm{x}}  \langle  \mathcal{O}^\dag_j (t,
\bm{x}) \; \mathcal{O}_i(0, \bm{0}) \rangle   e^{ i \bm{p} \cdot
\bm{x}} ,
\ee
where $i,j$ denote the source, sink smearing of the heavy-quark field;
for this work $i=j$.
%
$\mathcal{O}_i(t, \bm{x})$ is a bilinear interpolating operator with a
gamma-matrix structure that yields quantum numbers appropriate for
either pseudoscalar or vector mesons.
To construct this operator, we combine a one-component, staggered
light-quark spinor with a four-component, Wilson-type heavy-quark
spinor in a manner similar to Ref.~\cite{Wingate:2002fh},  
\be  \label{eq:interpolating_op}
 	\mathcal{O}_{\Xi} (t, \bm{x}) = \overline{\psi}_{\alpha}(t, \bm{x})
\, \Gamma_{\alpha \beta} \, \Omega_{\beta \Xi}(t, \bm{x}) \, \chi(t,
\bm{x}) ,
\ee
where $\Gamma = \gamma_5$ or $\gamma_\mu$;  ${\alpha, \beta}$ are spin
indices; and $\Omega(x) \equiv \gamma_1^{x_1}  \gamma_2^{x_2}
\gamma_3^{x_3} \gamma_4^{x_4}$.
The fields $\bar{\psi}$ and $\chi$ are the Wilson-type and staggered fields,
respectively, and the smeared correlator is constructed in the 
same way, but with $\bar{\phi}$ instead of~$\bar{\psi}$.  
The transformation properties of  $\mathcal{O}_{\Xi} (x)$ under
shifts by one lattice spacing are such that $\Xi$ can be viewed as playing the role of the (fermionic)
taste index~\cite{Golterman:1986jf, Kronfeld:2007ek}.
In our correlation functions, $\mathcal{O}_{\Xi} (x)$ is summed over $2^4$ hypercubes, and so $\Xi$ can be interpreted as
a taste degree of freedom in the sense of Refs.~\cite{Gliozzi:1982ib,
KlubergStern:1983dg}.

\begin{table*}
	\caption{Parameters used in the valence-quark actions.  
	The bare masses of the light and strange sea quarks $(am'_l, am'_s)$ label the ensemble.
	The mass of the light (staggered) valence quark is given by $am'_q$.
	$c_E$ and $c_B$ are the coefficients of the chromoelectric and chromomagnetic contributions to the Lagrangian.
	With $c_E=c_B$, they are the ususal Sheihkoleslami-Wohlert coupling.
	$u_0$ is the tadpole-improvement factor from measurements of the average plaquette for the fine and medium-coarse ensembles and from the Landau-gauge link on the coarse ensembles.
	Hopping parameter values \kp\ used for bottom-like and charm-like heavy quarks are given in the final two columns.
	}
	\label{Tbl:mesons}
	\begin{tabular*}{\textwidth}{c*{6}{@{\extracolsep{\fill}}c}}
		\hline \hline
		Lattice & $(am'_l, am'_s)$   &   $am'_q$         & $c_E=c_B$       &    $u_0$                    &  bottom-type $\kappa$  &   charm-type  $\kappa$ \\
		\hline
		Fine  
		& (0.0031, 0.031)  & 0.0272, 0.031 & 1.478 & 0.8779 & 0.0923 & 0.127         \\
		& (0.0062, 0.031)  & 0.0272, 0.031 & 1.476 & 0.8782 & 0.090, 0.0923, 0.093   & 0.1256, 0.127         \\
		& (0.0124, 0.031)  & 0.0272, 0.031 & 1.473 & 0.8788 & 0.0923 & 0.127         \\
		Coarse
		& (0.005, 0.050)   & 0.030, 0.0415 & 1.72   & 0.836  & 0.086 & 0.122 \\
		& (0.007, 0.050)   & 0.030, 0.0415 & 1.72   & 0.836  & 0.074, 0.086, 0.093        & 0.119, 0.122, 0.124           \\
		& (0.010, 0.050)   & 0.030, 0.0415 & 1.72   & 0.8346 & 0.074, 0.086, 0.093        & 0.119, 0.122, 0.124           \\
		& (0.020, 0.050)   & 0.030, 0.0415 & 1.72   & 0.8369 & 0.074, 0.086, 0.093        & 0.122, 0.124           \\
		& (0.030, 0.050)   & 0.030, 0.0415 & 1.72   & 0.8378 & 0.086 & 0.122 \\
		Medium-coarse
		& (0.0097, 0.0484) & 0.0387, 0.0484 & 1.570 & 0.8604 & 0.070, 0.080 & 0.115, 0.122\footnote{Used only with $am'_q=0.484$.}, 0.125 \\   
		& (0.0194, 0.0484) & 0.0387, 0.0484 & 1.567 & 0.8609 & 0.070, 0.076, 0.080 & 0.115, 0.122, 0.125            \\   
		& (0.0290, 0.0484) &             0.0484  & 1.565 & 0.8614 & 0.070, 0.080 & 0.115, 0.125            \\   
		\hline \hline
	\end{tabular*}
\end{table*}

%% file: Analysis.tex
In this section, we describe the components of our analysis.
Section~\ref{sec:twopnts} discusses the two-point correlator fits used to determine the meson energies $aE(\bm{p})$.
Section~\ref{sec:disprel} describes how we fit the meson dispersion relation to obtain $M_2$.
Finally, Sec.~\ref{sec:kappa_hfs} explains how $\kappa$ is tuned and how the hyperfine splitting is determined.

\subsection{Two-point Correlator Fits: $E(\bm{p})$}
\label{sec:twopnts}

To determine $E(\bm{p})$, 
we simultaneously fit the local and smeared heavy-light-meson two-point correlators to the function
\begin{equation}
	\label{fit_function}
	C_{i,i}(t, \bm{p}) \;
	 	= \; \sum_{\eta=0}^{N-1} 
				 \bigg[
	                            Z^2_{i, \eta}    
	                            	\left(   e^{-E_{\eta}(\bm{p}) t} + e^{-E_{\eta} (\bm{p}) (N_T-t) }    \right) \\
	                             + 
	                             (-1)^{t+1} (Z^{\rm p}_{i,{\eta}})^2    
	                             	\left(   e^{-E^{\rm p}_{\eta} (\bm{p}) t} + e^{-E^{\rm p}_{\eta} (\bm{p}) (N_T-t)}   \right) 
				\bigg], 
\end{equation}
where $N_T$ is the temporal extent of the lattice, and terms proportional to $e^{-E_{\eta} (\bm{p}) (N_T-t)}$ are due to periodic boundary conditions.
To simplify notation in this subsection, the lattice spacing $a$ is not written out explicitly.
Correlation functions containing staggered  
light quarks have contributions from both desired- and opposite-parity states with the opposite-parity states having the temporally-oscillating prefactor $(-1)^{t+1}$~\cite{Wingate:2002fh}.
We take each energy level $\eta$ in Eq.~(\ref{fit_function}) to include a pair of states consisting of one desired- and one opposite-parity state; the number of pairs of states in a fit is given by $N$. 
Quantities associated with the tower of opposite-parity states are denoted by the superscript ``p.''

Equation~(\ref{fit_function}) contains $2N$ exponentials, and the number of time slices in our data set is finite.
Although it is straightforward to separate the two different parities---because of the $(-1)^{t+1}$---it is difficult to separate states within each tower.
Rather than relying solely on taking $t$ large enough, we use the technique of constrained curve fitting~\cite{Bayes, Lepage:2001ym, Wingate:2002fh}.  
We thus minimize an augmented $\chi^2$~\cite{Lepage:2001ym},  
\be
	\chiaug \equiv \chi^2 + \sum_k \frac{\left(P_k - \tilde{P}_k  \right)^2}{\sigma^2_{\tilde{P}_k}} , 
    \label{Eqn:aug_chisq}
\ee
which means each fit parameter $P_k$ is provided a prior Gaussian probability distribution function with central value and width $(\tilde{P}_k, \sigma_{\tilde{P}_k})$.
The central value for fitted quantities comes from minimizing \chiaug\ on the whole ensemble.
%
We take the parameters to be $E_0^{\rm (p)}$, $\ln(Z^{\rm (p)}_{i, \eta})$, and (for $\eta>0$) $\ln(\Delta E_\eta^{\rm (p)})$,
where $\Delta E_\eta^{\rm (p)}=E_\eta^{\rm (p)}-E_{\eta-1}^{\rm (p)}$, thereby enforcing a tower of states with increasing energy.

In general, one considers a quantity to be determined by the data only if the statistical error, discussed next, is smaller than the corresponding prior width.
In this work, we are most concerned with the lowest-lying desired 
parity state, and the data---not the priors---always determine $E_0$ and $Z_{i,0}$.
For parameters that are poorly constrained by the data, such as those describing excited states, 
these priors prevent the fitter from searching fruitlessly along flat directions in parameter space. 
Because of the freedom in choosing the prior, we test whether the ground-state results are prior-indpendent, and stable.
When testing the stability of fit results, we use the Hessian error, defined as 
 \begin{equation}
	\sigma_{P_i} = \sqrt{  2 \left(  \frac{  \partial^2 \chi^2_{\rm aug} } { \partial P_i \partial P_j} \right)^{-1}_{ii}  } ,
\end{equation}
because its straightforward definition allows it to be quickly calculated for a single fit.

When using \chiaug\ to measure the goodness of fit, we count the degrees of freedom as the number of data points; 
the number of fit parameters is not subtracted since there are an equal number of extra terms in \chiaug.
In some cases, this could result in misleadingly low values of \chiaugdof.
For example, if the prior width $\sigma_{\tilde{P}_k}$ is 
much larger than
$(P_k - \tilde{P}_k )$, the associated term in \chiaug\ will be 
much smaller than the others.
This could be adjusted {\em a posteriori} by reducing the degrees of freedom, but it would require 
devising a criterion for ``large $\sigma_{\tilde{P}_k}$''.
We do not make such adjustments in our analyses.  
Instead, to determine goodness of fit, we monitor the values of 
\chiaugdof\ from constrained fits, but rely equally on the stability of 
fit results.

We estimate statistical uncertainties by generating pseudo-ensembles via the bootstrap method.  
When fitting a pseudo-ensemble, the central value of each prior is drawn randomly from its Gaussian probability distribution while
the prior width is kept the same~\cite{Lepage:2001ym, Wingate:2002fh}.
To prevent large, simultaneous but uncorrelated fluctuations among prior central values, which could destabilize
a fit, we restrict the randomized prior central values to $\pm1.5 \sigma_{\tilde{P}}$.
Final errors quoted for meson energies and functions thereof, such as the spin-averaged mass, 
are obtained from their bootstrap distributions.
We define the upper (lower) 68\%-distribution point as the value at which 16\% of the distribution has a higher (lower) value.
We refer to half of the distance between these two points as the average 68\% bootstrap error.


\subsection{Dispersion Relation Fits: The Kinetic Mass}    \label{sec:disprel}
Having determined $E(\bm{p})$, we use the dispersion relation to determine the kinetic meson mass, which we then use to tune the hopping parameter $\kappa$.
The low-momentum expansion for $E(\bm{p})$ is 
\be	\label{Eq:lowp}
	E(\p) = M_1 + \frac{\p^2}{2 M_2} - \frac{a^3 W_4}{6} \sum_i p_i^4  - 
   \frac{(\p^2)^2}{8 M_4^3}  + \cdots ,
\ee
where $W_4$ and the deviation of $M_4$ from $M_2$  capture lattice artifacts.  
(In the continuum limit $a^3 W_4 = 0$ and $M_4 = M_2$.)
 The vector \n~is defined by
 \be
	\label{Eq:vecn}
	a\p = (2\pi/N_L)  \, \n  ,
\ee
where $N_L$ is the spatial extent of the lattice, given in Table~\ref{Tbl:vacuum};
data are generated for $|\n|  \le 3 $.
Noise in $E(\p)$ increases with increasing momentum, though, and is substantial by the time $O(\p^4)$ effects become significant.   
For charm-type mesons, squaring the energy yields a substantial cancellation in the $\mathcal{O}(\p^4)$ contribution because  $aM_1 \approx aM_2 \approx aM_4$.  While this is not true for bottom-type mesons, the mass of these mesons is large enough to cause suppression via the $1/M$ factors whether  \Ep~or $E^2(\p)$ is used.
By fitting to $E^2(\p)$ then, the contributions from $O(\p^4)$ effects are reduced, and we are able to do a linear fit to low-momentum data, $|\n| \le 2$.
Setting $M_1=E(\bm{0})$ from the zero-momentum correlator, 
we square Eq.~(\ref{Eq:lowp}) and fit  
\begin{equation}
	\label{eq:disp_func_used}
    	E^2(\bm{p}) - M_1^2 = C \bm{p} ^2  
\end{equation}
to obtain $C$.  Finally, we set $M_2=M_1/C$.
The largest $\p$ is chosen so that the $O(\p^4)$ effects are 
expected to be negligible, based on tree-level values of 
the analogous quark quantities $w_4$ and $1/m_4^3$. 
We confirm the negligibility of these terms by inspecting plots of the 
data and monitoring $\chisq$.
(We do not use constrained curve fitting here and so we minimize the usual $\chi^2$.)
This procedure is 
repeated for each bootstrap-generated pseudo-ensemble, yielding bootstrap distributions for $aM_1$ and $aM_2$.

\subsection{The Hopping Parameter $\kappa$ and the Hyperfine Splitting $\Delta_1$}  \label{sec:kappa_hfs}

For tuning $\kappa$, it is helpful to remove the leading discretization errors from spin-dependent terms.
Let the spin-averaged kinetic meson mass be
\be  \label{eq:Mbar}
	\overline{M}_2 = \frac{1}{4}(M_2 + 3 M_2^*) ,
\ee    
where $M_2$ and $M_2^*$ are determined as described in Sec.~\ref{sec:disprel}.
This leaves the second, spin-independent term in Eq.~(\ref{eq:L2}) as the leading source of discretization error at $O(a^2 \Lambda^2)$.
Our goal then is to determine the value of $\kappa$ that will result in a value of $\overline{M}_2$  
that agrees with the experimental value taken from the Particle Data Group (PDG).
 
For each lattice spacing, we use the following procedure to tune $\kappa$.  
Using three or more ensembles, 
we study the light sea-quark mass dependence of $a\overline{M}_2$ for at least one combination of $\kappa$ and $m'_q$.  
This gives us some insight into the behavior of $a\overline{M}_2$ in the physical--sea-quark--mass limit and allows us to assign an uncertainty to $a\overline{M}_2$ due to non-physical sea-quark masses.
Next, on at least one ensemble, we determine $a\overline{M}_2$ at two staggered, valence-quark masses near the strange-quark mass.
This allows us to determine the dependence of $a\overline{M}_2$ on the staggered, valence-quark mass and interpolate linearly to the physical value if no simulated mass is close enough to the tuned strange-quark mass.
Having dealt with the staggered-valence and light sea-quark masses, we take $a\overline{M}_2$ at the physical, strange valence-quark mass at two values of $\kappa$ and 
interpolate linearly in $\kappa$ to the spin-averaged value of the 
meson masses, given by the Particle Data Group (PDG)~\cite{PDG06}, 
converted to lattice units with $a$ from Table~\ref{Tbl:vacuum}.
Finally, we combine the uncertainties in the 
tuned value of $\kappa$ from statistical and discretization errors in 
the meson mass, staggered-valence mass mistuning, non-physical 
sea-quark masses, and errors from the lattice-spacing conversion of the 
PDG mass.

To determine the hyperfine splitting, we start with the results for $ M_1 = E(\bm{0})$.
For each lattice spacing, we use values of \ahf~at, or linearly interpolated to, the tuned charm and bottom $\kappa$ values. 
We then consider uncertainties from statistics, the tuning of \kp~and \ams, non-physical sea-quark masses, and discretization.
The value of $a\Delta_1$ on the fine lattice is taken as our central value and results on the coarse and medium coarse lattices are used in the error analysis.
In the final value, we also include an uncertainty due to the conversion to physical units.

%% file: Fit_Details_Ep_M1_M2.tex
In this section, we describe the details of our fitting procedure for the meson energy $E(\bm{p})$ and the meson rest and kinetic masses, $M_1$ and $M_2$.
Our objective here is to document thoroughly our fitting procedures, including values for the priors, and tests.
Readers who are more interested in a summary can skip to Sec.~\ref{Sec:fitsum}.

Section~\ref{sec:2pt} discusses the parameters used in our two-point correlator fits for $E(\bm{p})$ (Sec.~\ref{sec:fitparams}) 
and the evaluation of goodness of fit via \chiaugdof~and tests of stability (Sec.~\ref{sec:stability_tests}).
In most tests discussed here, Hessian errors were used, because they 
are fast and straightforward.
Our complete data set, exhibited in Table~\ref{Tbl:mesons}, contains several ensembles at each of the three lattice spacings.
As explained in Sec.~\ref{sec:fitparams}, one ensemble at each lattice 
spacing is chosen for the purpose of setting priors in Eq.~(\ref{Eqn:aug_chisq}).
For tuning $\kappa$, we need data over a range of $\kappa$ and $\amq$ on a fixed ensemble.
At the fine lattice spacing, such data were generated on only one ensemble, (0.0062, 0.031), so we set priors and tune $\kappa$ on that same ensemble. 
For the coarse and medium-coarse lattice spacings, we have data for a range of $\kappa$ and $\amq$ on several ensembles.
We take the coarse (0.010, 0.050), and medium-coarse (0.0194, 0.0484) 
ensembles to set priors and then the ensembles with the smallest $am'_l$ (and a range of $\kappa$ and $\amq$) to tune $\kappa$.
We compute the hyperfine splittings from the same ensembles on which $\kappa$ was tuned.
These choices are summarized in Table~\ref{tbl:dataused}.
Data from other ensembles listed in Table~\ref{Tbl:mesons} are used to estimate uncertainties.

Fits of the dispersion relation to determine $M_2$ from $E(\bm{p})$ are comparatively simple, 
and Sec.~\ref{sec:kinmass} provides details that may be of interest.

\bt{Specific ensembles used in steps of the analyses.
	Setting priors is discussed in Sec.~\ref{sec:fitparams}.  
	Stability and goodness-of-fit tests done for $E(\p)$ results are described in Sec.~\ref{sec:stability_tests}.
	$\kappa$-tuning and hyperfine-splitting results are given in Secs.~\ref{sec:tuning} and~\ref{sec:hfs}, respectively.
	}{l  c  c  c  c}
		Lattice               &  setting priors         & $E(\p)$ tests,  tuning $\kappa$, and the hyperfine splitting \hf    \\
		\hline
		Fine                  &  (0.0062, 0.031)     &  (0.0062, 0.031)   \\
		Coarse              & (0.010, 0.050)        & (0.007, 0.050)       \\
		Medium-coarse & (0.0194, 0.0484)    & (0.0097, 0.0484)     \\  
\et{tbl:dataused}

\subsection{Two-point fits: $E(\bm{p}), M_1$}  \label{sec:2pt}

The number of gauge configurations in each ensemble is given in Table~\ref{Tbl:vacuum}.
To improve statistics, we generate data at four time sources on each of the fine and coarse gauge configurations and at eight time sources for medium-coarse configurations.
We also average the correlator points $C(t)$ and $C(N_T-t)$.
In order to reduce the effect of correlations between data points from sequential configurations, we bin the data by groups of $N_{\rm bin}$ configurations.  
Because fits for this project were done in concert with other projects,  $N_{\rm bin} = 4$ was adopted.
Comparisons of results using $N_{\rm bin} = 2, 4$, and 6 on the ensembles used here show no significant change in the fit-result error bars or the bootstrap distributions.
To account for correlations in the two-point correlator data, the fitter uses the normalized, data-sample covariance matrix as an estimate of the correlation matrix.
This matrix is remade for each bootstrap sample.

\subsubsection{Priors, time ranges, $N$}  \label{sec:fitparams}

We consider the setting of priors for the ground state parameters, excited-state amplitudes, and  energy splittings separately.
Ground-state ($\eta=0$) parameters are well-determined by the data;   
thus, the ground-state priors can, and should, be negligibly constraining.
In contrast, energy splittings and excited state amplitudes are not well determined by the data,  
and the related  priors are chosen such that they put reasonable bounds on the parameters.
The next paragraphs describe how the priors are set.
Note that the same set of priors is used for all ensembles at a given lattice spacing, for all momenta in the range  $|\n| = $~0 to~2, and for all \kp~and $am'_q$~of a given meson type, e.g., charm pseudoscalars.
The priors used are tabulated in Tables~\ref{priors-fine}--\ref{priors-medcoarse}.

We use information from a subset of our data, one ensemble per lattice spacing, to set the priors for the two-point--correlator fits.
This is necessary because we do not have enough external knowledge to set them independently.
The ensembles used to help set the priors are listed in Table~\ref{tbl:dataused}.
Other ensembles are statistically independent of these ensembles and so the prior information can be viewed as external to fits on those ensembles.
If possible, though, we do not want to exclude any data from our analysis, including the ensembles used in the setting of priors.
For this reason, our procedure for setting priors keeps the amount of 
information we take from these ensembles to a minimum.
Specifically, 
for a parameter $P$, we use averages over ranges of parameters, like the momentum, for the prior central value $\tilde{P}$
and chose prior widths $\sigma_{\tilde{P}}$ that are broad enough to 
cover the expected results for an entire subset of fits;
e.g., the same priors are used for fits  with $|\n| = $~0 to~2.

 To set  ground-state priors, we first fit to large-time data with $N=1$ in order to get a general idea of the ground-state parameter values.  
We then set $N> 1$ and fit correlators at low and high momenta to ascertain the range of values the ground state parameters may take.
We set prior central values for the ground-state energy of the desired- and opposite-parity states, $aE_0(\p)$ and $aE^p_0(\p)$, at about the midpoint of the range seen in these fits.  

To understand our logic for setting the prior widths for $aE_0(\p)$ and $aE^p_0(\p)$, recall that we use a Gaussian distribution for the prior $\tilde{P}$ with a width $\sigma_{\tilde{P}}$.  
We set $\sigma_{a\tilde{E}_0}$ and $\sigma_{a\tilde{E}_0^p}$ large enough so that results across the entire momentum range used in the analysis should fall well within the 1-$\sigma_{a\tilde{E}_0}$, or 1-$\sigma_{a\tilde{E}_0^p}$, range of the distribution.
After priors for the remaining parameters are set, we 
perform a complete set of fits and, 
for at least one ensemble at each lattice spacing, 
verify that, indeed, the final fit results for $aE_0$ and $aE^{\rm p}_0$ fit well within their respective prior distributions. 

Priors for the ground-state amplitudes are loosely based on the preliminary $N>1$ fits described above.  In most cases, the central value is the nearest whole number to the average of these results.  For the desired-parity state, the widths $\sigma_{\tilde{P}}$ are chosen such that they easily span the range of values seen in the fits.  For the opposite parity states, which are substantially noisier, the widths span the distance between the prior central value and the observed range in the results by about 1-$\sigma_{\tilde{P}}$.

Priors for all excited-state amplitudes were set to have a relatively small central value and a wide width.
%
To set the prior for the energy splitting, we note that experimentally measured meson splittings are a few hundred MeV.
We also bear in mind that the sum of a series of exponentials with a very small  energy splitting is not a well-posed problem.
Therefore, we chose the central value of the splitting to be several hundred MeV, slightly large, with a generous prior width.
For example, on the fine lattice the prior for the splitting, $\ln(a\Delta E) = -1.45(1.0)$ is equivalent to $ \Delta E \approx 550^{+950}_{-350}$~MeV.

In the charm sector,  the opposite-parity partner of the $D_s(0^-)$, the $D_{s0}^*(0^+)$, is close to the $D K$ threshold.
In this case, the energy splitting should not be viewed as a meson mass 
splitting, and our choice of prior for the $D_{s0}^*(0^+)$ energy 
splitting may be inappropriate.
The parity-partner signal is noisy, 
though, and in tests of the priors widths we see no change in the 
non-oscillating ground state energy $aE(\p)$, which is our main interest.
For details, see Sec.~\ref{sec:stability_tests}.

To choose the time ranges for the fits, $(t_{\rm min}, t_{\rm max})$, we first look at the data   
to determine the time by which the error in the data, e.g. the relative error in the correlator, has increased substantially.  
This gives us a potential value for $t_{\rm max}$.  
From effective mass plots we can also see at what time slice the majority of the excited-state contamination has died off, giving us a potential value for $t_{\rm min}$.
Constrained curve fitting is designed to reduce excited-state contamination of the lower-state fit parameters. 
Nevertheless, we do not see a significant reduction in the error from fitting to the smallest possible time slice, which requires including a larger number of states in the fit.
For simplicity, we chose final time ranges that are the same for similar sets of data.
These can be found in Table~\ref{tbl:NstatesTime}.

\begin{figure}
	\begin{tabular}{c c}
		\includegraphics[width=0.47\textwidth]{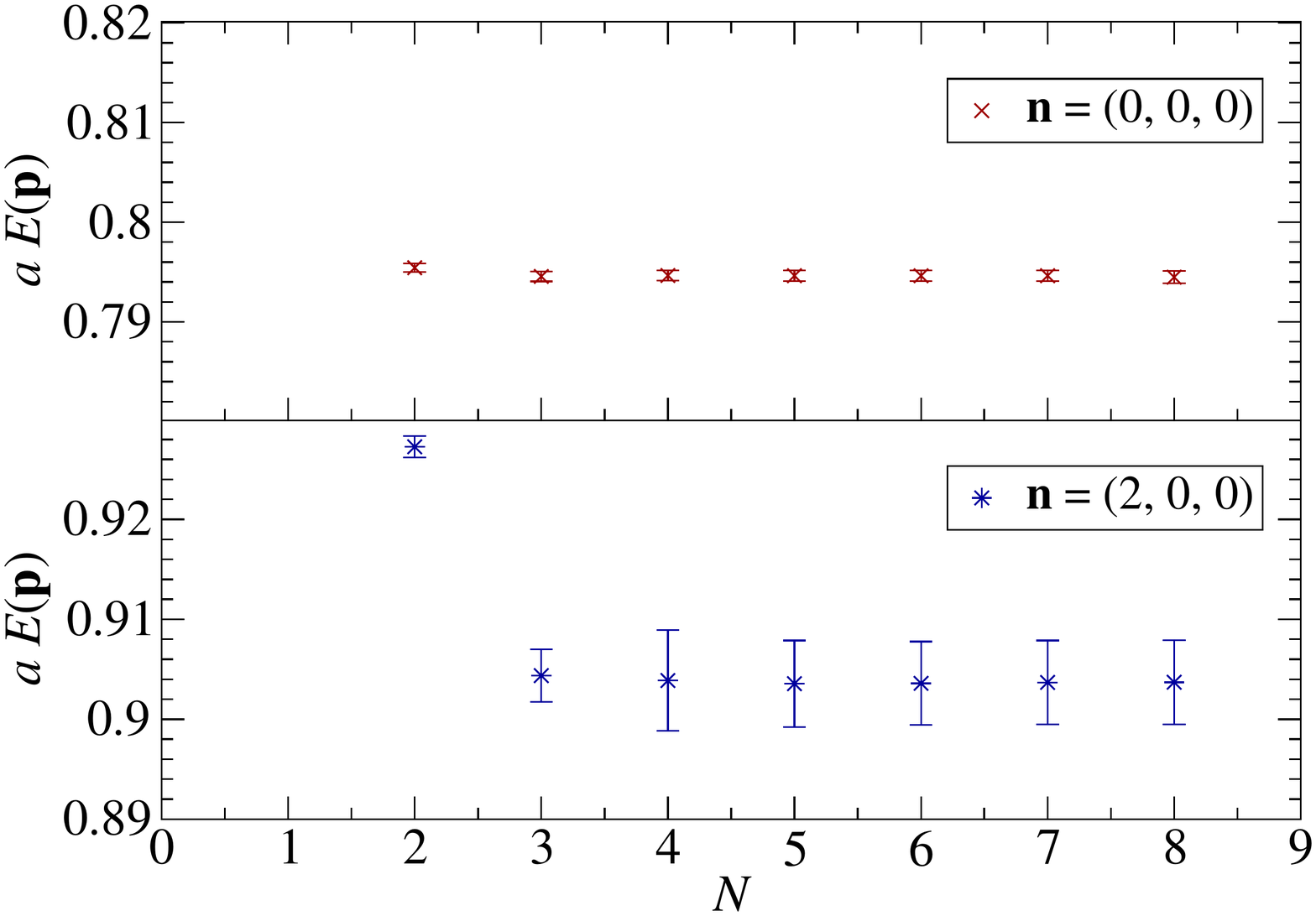}   
		&
		\includegraphics[scale=0.3]{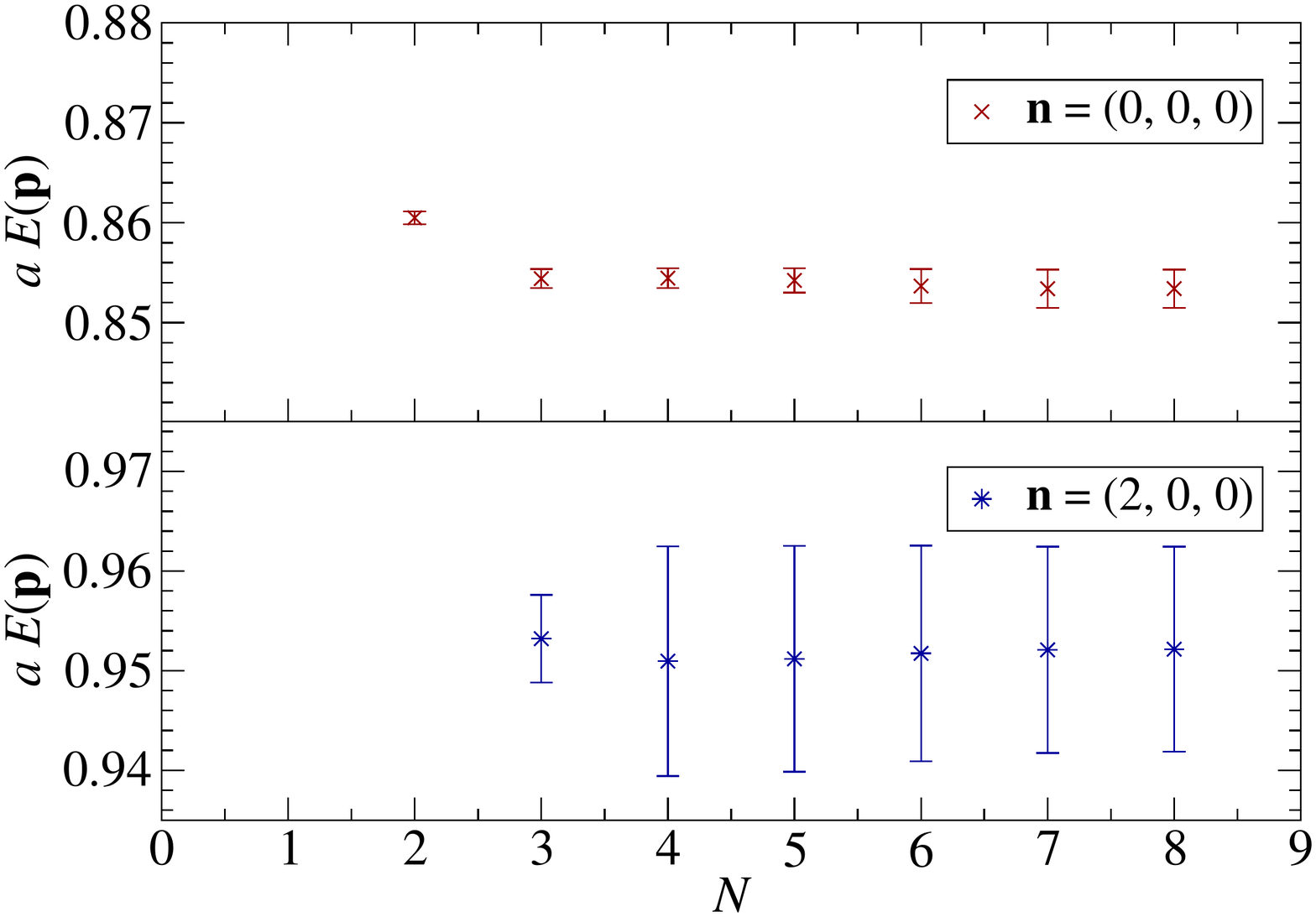}   
		 \\
	(a)  &  (b)  \\
%
%
		\includegraphics[scale=0.3]{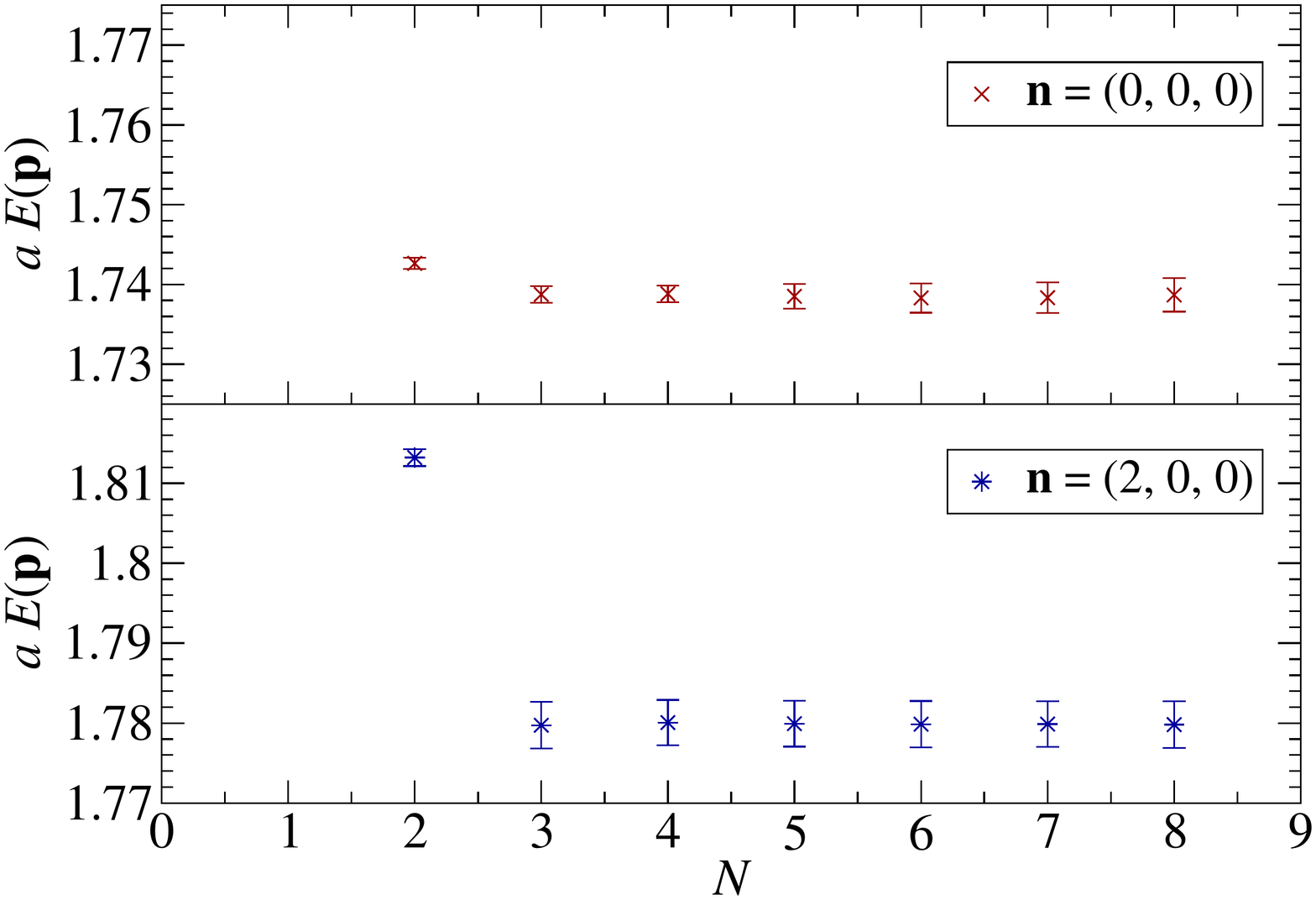}   
		&
		\includegraphics[scale=0.3]{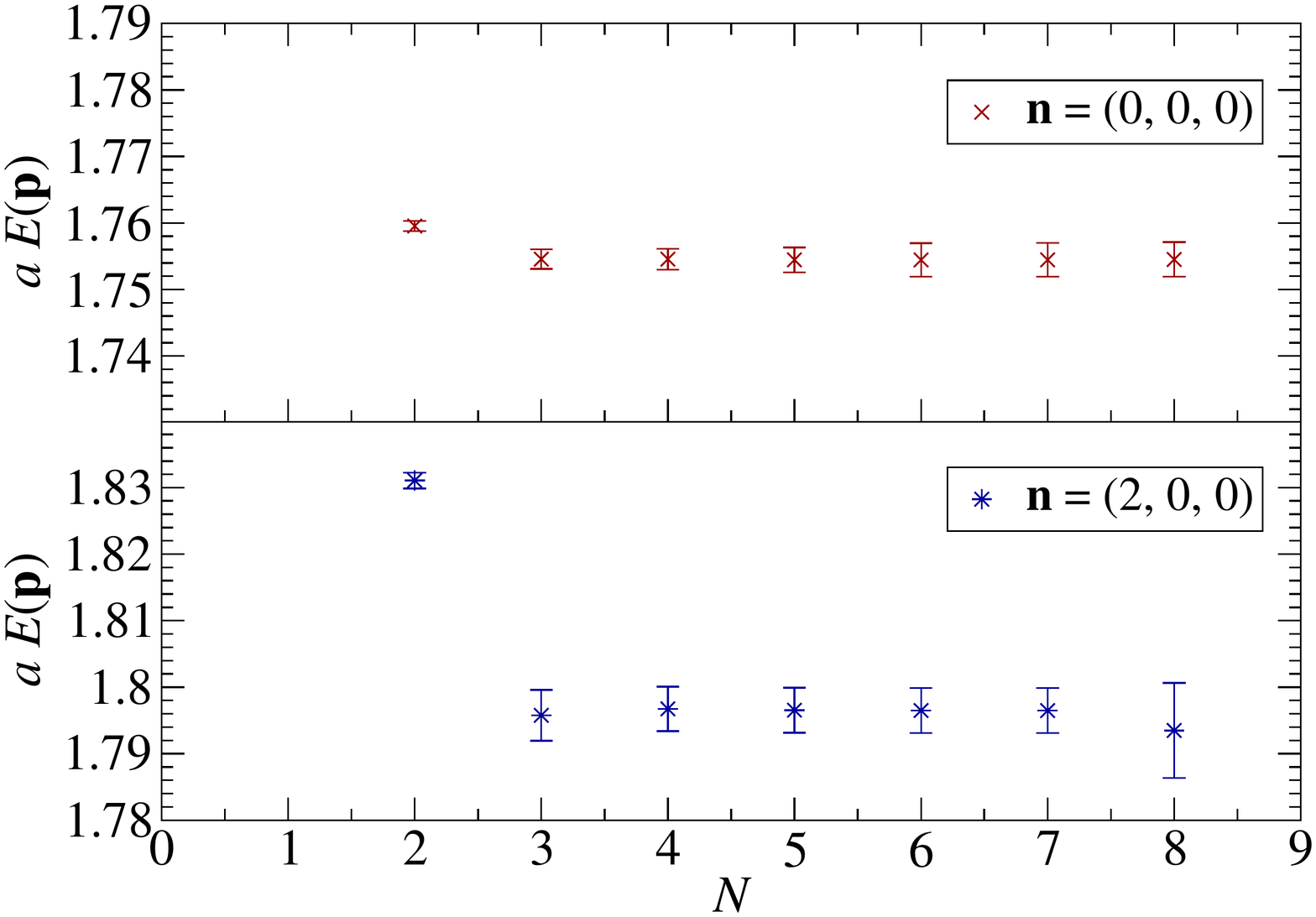}   
		 \\
		(c)  &  (d)  \\
	\end{tabular}
	\caption{Fitted values of \aEp~vs.~the number of (pairs of) states $N$ for  \kp~= 0.127, charm-type (a) pseudoscalar and (b) vector mesons and \kp~= 0.090, bottom-type (c) pseudoscalar  and (d) vector mesons on the (0.0062, 0.031) fine ensemble.   
	Results shown are for mesons with momenta $\n=(0,0,0)$ and $(2,0,0)$.
	Errors are Hessian.}
	\protect\label{fig:Ns-fine}
\end{figure}

With the time range set, we do fits for increasing values of the number of (pairs of) states $N$ and look for the ground-state energy to stabilize.  
We choose the final values of $N$ to be the minimum value needed to be in the stable region; these are given in Table~\ref{tbl:NstatesTime}.
Figure~\ref{fig:Ns-fine} shows representative plots of $aE(\p)$ versus $N$ from fits on the (0.0062, 0.031) fine ensemble.
It is clear that for the minimum-value $N$, the central value of the fit result is always well within the stable region.
In some cases, though, the (Hessian) error from the minimum-$N$ fit is smaller than that in the stable region. 
One could remedy this by choosing to fit with more states.  Unfortunately, an increase in the number of states leads to non-gaussian bootstrap distributions with a significant number of outliers --- clearly non-physical fit results that contain ground states with low energies and very small amplitudes.  
Using the minimum possible number of states, no outliers have been seen in the distributions.

\bt{Priors used for fine-ensemble two-point correlator fits for pseudoscalar and vector mesons.  
	Priors for all higher amplitudes and splittings are the same as those for the first excited state.
	The fit-parameter numbers 15--20 label the second excited state and so on.
	A prior of $\ln(a \Delta E) = \err{-1.45}{1.0}{-1.0}$ on the fine 
    ensembles corresponds approximately to $\Delta E = 
    \err{550}{950}{-350}$~MeV. }
	{l        c           r@{.}l     r@{.}l           r@{.}l     r@{.}l              }
                                                                   &&  \mc{4}{c}{Charm Mesons}                                                   &  \mc{4}{c}{Bottom Mesons}      \\ [0.5em]
	       fit parameter   & fit-parameter number  &  \mc{2}{c}{pseudoscalar}       &   \mc{2}{c}{vector}               &  \mc{2}{c}{pseudoscalar}       &   \mc{2}{c}{vector}   \\
	         \hline 
		 $E_0$                                              &  1  &    0&90(40)                            & 0&90(40)             &    1&75(60)                               & 1&75(60)            \\
		 $E^{\rm p}_0$                                  &  2  &    1&0(40)                              & 0&95(40)             &    1&85(60)                               & 1&85(60)           \\
		 $\ln(Z_{\rm1S, 0})$                          &  3  &    1&0(2.0)                             & 1&0(2.0)              &    1&0(3.0)                                & 1&0(3.0)            \\
		 $\ln(Z^{\rm p}_{\rm1S, 0}) \qquad$  &  4  &    1&0(2.0)                             &  1&0(2.0)             &    1&0(3.0)                                &  1&0(3.0)           \\
		 $\ln(Z_{\rm d, 0})$                            &  5  &   $-2$&0(2.0)                        & $-2$&0(2.0)         &   $-2$&0(3.0)                            & $-2$&0(3.0)        \\
		 $\ln(Z^{\rm p}_{\rm d, 0}) \qquad$   &  6  &   $-2$&0(2.0)                         & $-2$&0(2.0)         &   $-2$&0(3.0)                             & $-2$&0(3.0)       \\
		 $\ln(\Delta E)$                                  &  8  &   $-1$&45(1.0)                       & $-1$&45(1.0)       &   $-1$&45(1.0)                           & $-1$&45(1.0)     \\
		 $\ln(\Delta E^{\rm p})$                     &  9  &   $-1$&45(1.0)                        & $-1$&45(1.0)       &   $-1$&45(1.0)                           & $-1$&45(1.0)     \\
		 $\ln(Z_{\rm1S,1})$                           &  10  &   $-1$&0(3.0)                        & $-1$&0(3.0)         &   $-1$&0(3.0)                             & $-1$&0(3.0)       \\
		 $\ln(Z^{\rm p}_{\rm1S,1}) \qquad$   &  11  &   $-1$&0(3.0)                        & $-1$&0(3.0)        &   $-1$&0(3.0)                             & $-1$&0(3.0)        \\
	        $\ln(Z_{\rm d,1})$                              &  12  &   $-1$&0(3.0)                        & $-1$&0(3.0)         &   $-1$&0(3.0)                             & $-1$&0(3.0)       \\
		 $\ln(Z^{\rm p}_{\rm d,1}) \qquad$    &  13  &   $-1$&0(3.0)                        & $-1$&0(3.0)         &   $-1$&0(3.0)                             & $-1$&0(3.0)        \\
\et{priors-fine}

\bt{Same as Table~\ref{priors-fine}, but for the coarse ensembles.
	A prior of $\ln(a \Delta E) = \err{-1.2}{0.5}{-0.5}$ on the coarse ensembles corresponds approximately to $\Delta E = \err{500}{300}{-200}$~MeV. 
	}
	{l         c              r@{.}l     r@{.}l          r@{.}l     r@{.}l             c}
                                                                   &&  \mc{4}{c}{Charm Mesons}                                                   &  \mc{4}{c}{Bottom Mesons}      \\ [0.5em]
	        fit parameter         & fit-parameter number       &  \mc{2}{c}{pseudoscalar}       &   \mc{2}{c}{vector}               &  \mc{2}{c}{pseudoscalar}       &   \mc{2}{c}{vector}                \\
	         \hline
		 $E_0$                                          &  1  &    1&10(40)                                & 1&2(40)              &    2&00(40)                           &  2&00(40)           \\
		 $E^{\rm p}_0$                              &  2  &    1&30(40)                                & 1&3(40)              &    2&10(40)                           & 2&10(40)            \\
		 $\ln(Z_{\rm1S, 0})$                      &  3  &    1&0(2.0)                                 & 1&0(2.0)             &   1&0(2.0)                             & 1&0(2.0)             \\
		 $\ln(Z^{\rm p}_{\rm1S, 0}) \qquad$ & 4 &   1&0(3.0)                                 &  0&1(3.0)            &    $-1$&0(2.0)                       & $-0$&1(2.0)       \\
		 $\ln(Z_{\rm d, 0})$                       &  5  &   $-1$&0(2.0)                             & $-1$&0(2.0)        &   $-2$&0(2.0)                        & $-1$&0(2.0)        \\
		 $\ln(Z^{\rm p}_{\rm d, 0}) \qquad$ & 6 &   $-1$&0(3.0)                             & $-2$&0(3.0)        &   $-2$&0(2.0)                         & $-2$&0(2.0)       \\
		 $\ln(\Delta E)$                             &  8  &   $-1$&2(0.5)                             & $-1$&2(0.5)          &   $-1$&2(0.5)                         & $-1$&2(0.5)       \\
		 $\ln(\Delta E^{\rm p})$                 &  9  &   $-1$&2(0.5)                             & $-1$&2(0.5)          &   $-1$&2(0.5)                         & $-1$&2(0.5)       \\
		 $\ln(Z_{\rm1S,1})$                       &  10  &   $-1$&0(3.0)                             & $-1$&0(3.0)         &   $-1$&0(3.0)                          & $-1$&0(3.0)       \\
		 $\ln(Z^{\rm p}_{\rm1S,1}) \qquad$ &  11 &   $-1$&0(3.0)                             & $-1$&0(3.0)      &   $-1$&0(3.0)                          & $-1$&0(3.0)       \\
	        $\ln(Z_{\rm d,1})$                         &  12   &   $-1$&0(3.0)                             & $-1$&0(3.0)       &   $-1$&0(3.0)                          & $-1$&0(3.0)       \\
		 $\ln(Z^{\rm p}_{\rm d,1}) \qquad$ &  13  &   $-1$&0(3.0)                             & $-1$&0(3.0)       &   $-1$&0(3.0)                         & $-1$&0(3.0)        \\
\et{priors-coarse}

\bt{	Same as Table~\ref{priors-fine}, but for the medium coarse ensembles. 
	A prior of $\ln(a \Delta E) = \err{-1.0}{0.5}{-0.5}$ on the medium-coarse ensembles corresponds approximately to $\Delta E = \err{500}{300}{-200}$~MeV.
	}
	{l         c              r@{.}l     r@{.}l           r@{.}l     r@{.}l             c}     
                                                                   &&  \mc{4}{c}{Charm Mesons}                                                   &  \mc{4}{c}{Bottom Mesons}      \\ [0.5em]
	         fit parameter     & fit-parameter number         &  \mc{2}{c}{pseudoscalar}       &  \mc{2}{c}{vector}   &  \mc{2}{c}{pseudoscalar}       &   \mc{2}{c}{vector} \\
	         \hline
		 $E_0$                                              &  1  &    1&38(50)                                & 1&46(50)              &    2&35(40)                           &  2&38(50)           \\
		 $E^{\rm p}_0$                                 &  2  &    1&50(60)                                & 1&58(60)              &    2&48(50)                           & 2&50(50)            \\
		 $\ln(Z_{\rm1S, 0})$                         &  3  &    0&48(1.0)                               & 0&95(1.0)             &   0&12(1.4)                           & 0&60(1.0)             \\
		 $\ln(Z^{\rm p}_{\rm1S, 0}) \qquad$ & 4  &   $-0$&65(1.0)                           &  0&20(1.0)            &    $-1$&0(2.0)                       & $0$&1(2.0)       \\
		 $\ln(Z_{\rm d, 0})$                          &  5  &   $-0$&90(1.0)                           & $-0$&74(1.0)        &   $-1$&15(1.0)                      & $-0$&8(1.0)        \\
		 $\ln(Z^{\rm p}_{\rm d, 0}) \qquad$  &  6  &   $-2$&4(1.4)                             & $-1$&8(2.0)         &   $-2$&5(3.0)                         & $-1$&8(3.0)       \\
		 $\ln(\Delta E)$                                 &  8  &   $-1$&0(0.5)                            & $-1$&0(0.5)           &   $-1$&0(0.5)                         & $-1$&0(0.5)       \\
		 $\ln(\Delta E^{\rm p})$                    &  9  &   $-1$&0(0.5)                             & $-1$&0(0.5)           &   $-1$&0(0.5)                         & $-1$&0(0.5)       \\
		 $\ln(Z_{\rm1S,1})$                          & 10 &   $-1$&0(3.0)                            & $-1$&0(3.0)           &   $-1$&0(3.0)                         & $-1$&0(3.0)       \\
		 $\ln(Z^{\rm p}_{\rm1S,1}) \qquad$ &  11 &   $-1$&0(3.0)                            & $-1$&0(3.0)          &   $-1$&0(3.0)                          & $-1$&0(3.0)       \\
	        $\ln(Z_{\rm d,1})$                           &  12 &   $-1$&0(3.0)                           & $-1$&0(3.0)           &   $-1$&0(3.0)                          & $-1$&0(3.0)       \\
		 $\ln(Z^{\rm p}_{\rm d,1}) \qquad$   &  13 &   $-1$&0(3.0)                           & $-1$&0(3.0)          &   $-1$&0(3.0)                          & $-1$&0(3.0)        \\
\et{priors-medcoarse}

\bt{Time range $t_{\rm min}$--$t_{\rm max}$ and number of (pairs of) states $N$ used in two-point correlator fits at each lattice spacing.
For the time range, the first (second) number in parenthesis is $t_{\rm min}$ for the 1S-smeared (local) correlator;
$t_{\rm max}$ is the same for both correlators.}{ccc}
	Lattice spacing    &   Time range   & $N$ \\
	\hline
	    Fine           & $(2, 4)$--$25$ & $3$ \\ 
	    Coarse         & $(2, 8)$--$15$ & $2$ \\ 
	    Medium-coarse  & $(5, 6)$--$15$ & $2$ \\ 
\et{tbl:NstatesTime}

\subsubsection{Tests of Stability and Goodness-of-fit} \label{sec:stability_tests} 

Having set the priors, time range, and number of states for the fits, we check the stability of the results and goodness of fit in several ways.
For result stability, we check the effects of the time range used, the number of (pairs of) states $N$, and changes to the prior widths;  
we also compare the priors to the fit results.
%
We look at a representative subset of fits for each lattice spacing: pseudoscalar and vector meson correlators at two different $\kappa$ values (one for charm and one for bottom) for a given light-valence mass, on one ensemble per lattice spacing, and with momenta $\n=(0,0,0)$ and $(1,1,1)$ or $(2,0,0)$.  The specific values of $\kappa$, $am'_q$, and $(am'_l, am'_s)$ vary from test to test, and in some cases tests are extended to other values.
A description of the data used in the tests discussed here can be found in Table~\ref{tbl:dataused_intests}.
\bt{Data used in stability and goodness-of-fit tests.}{l  c  c  c}
		Lattice               & ensemble               &  $\kappa$                      & \amq   \\
		\hline
		Fine                   &  (0.0062, 0.031)    & 0.127; 0.090 or 0.093    & 0.0272  \\
		Coarse              & (0.007, 0.050)        & 0.122; 0.086                  & 0.0415 \\
		Medium-coarse & (0.0097, 0.0484)    & 0.125; 0.070                  & 0.0484 \\  
\et{tbl:dataused_intests}

For the time-range tests,
we vary $t_{\rm min}$ over two to four time slices, increasing $N$ if appropriate, and vary $t_{\rm max}$ over five to ten time slices.
We verify that there are no changes in the fit results beyond expected fluctuations.\footnote{In one case, $\kappa = 0.086$, coarse (0.010, 0.005), although the ground-state energy is stable as $t_{\rm max}$ is varied, the value of  \chisq~becomes large as $t_{\rm max}$ is increased beyond the final value ($t_{\rm max} = 15$).
This ensemble is not used directly for $\kappa$ tuning or hyperfine splitting determinations as explained in the introduction to this section.}
%
For number-of-states tests, we verify that the result is stable as $N$ is increased.
Figure~\ref{fig:Ns-fine} shows example results for the (0.0062, 0.031) fine ensemble.  
Similar results are seen for the coarse and medium-coarse ensembles and for the ground-state amplitudes $Z_{\rm 1S}$ and $Z_{\rm d}$.

\begin{figure}
	\begin{tabular}{c c}
		\includegraphics[scale=0.3]{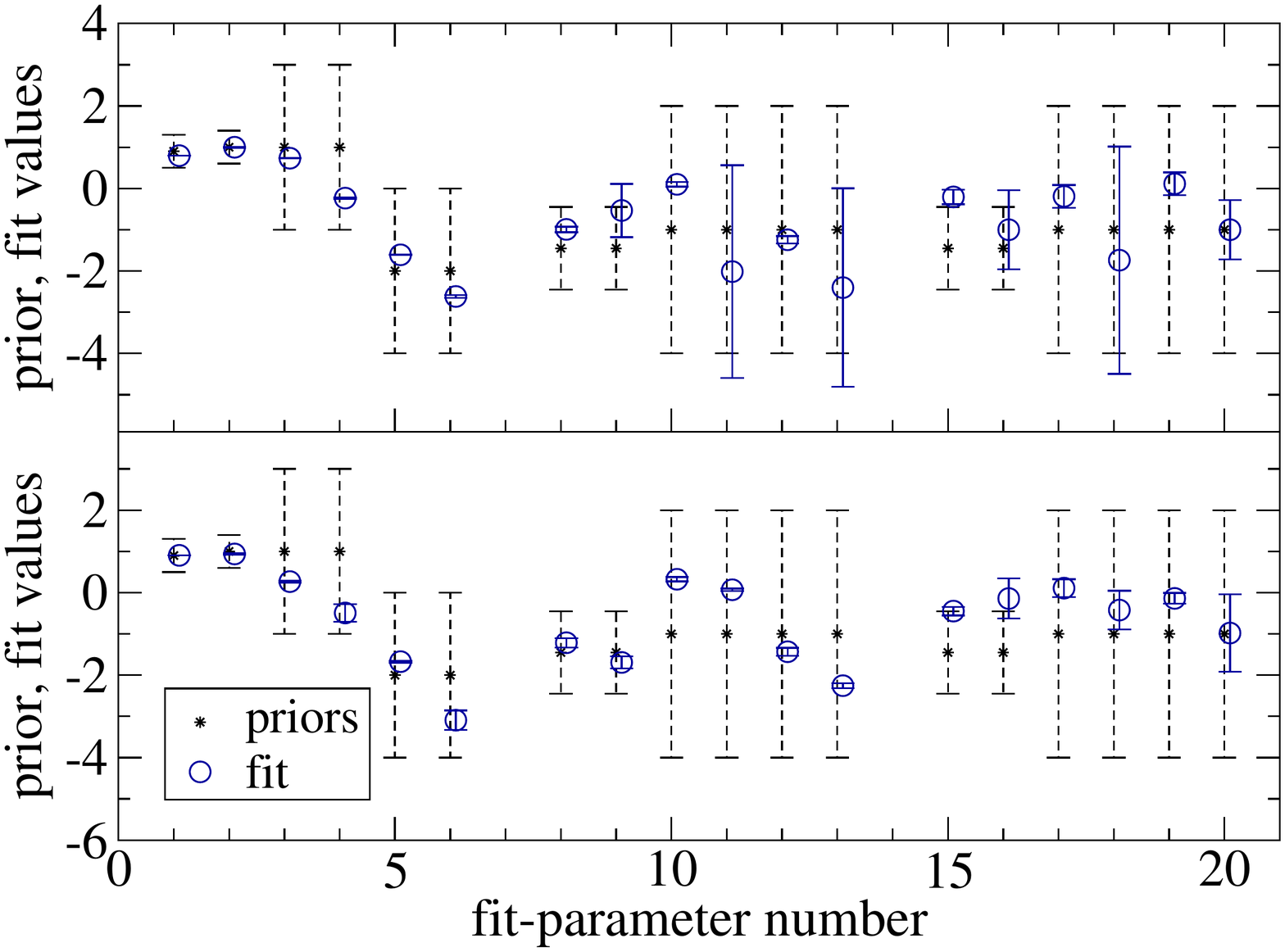}   
		&
		\includegraphics[scale=0.3]{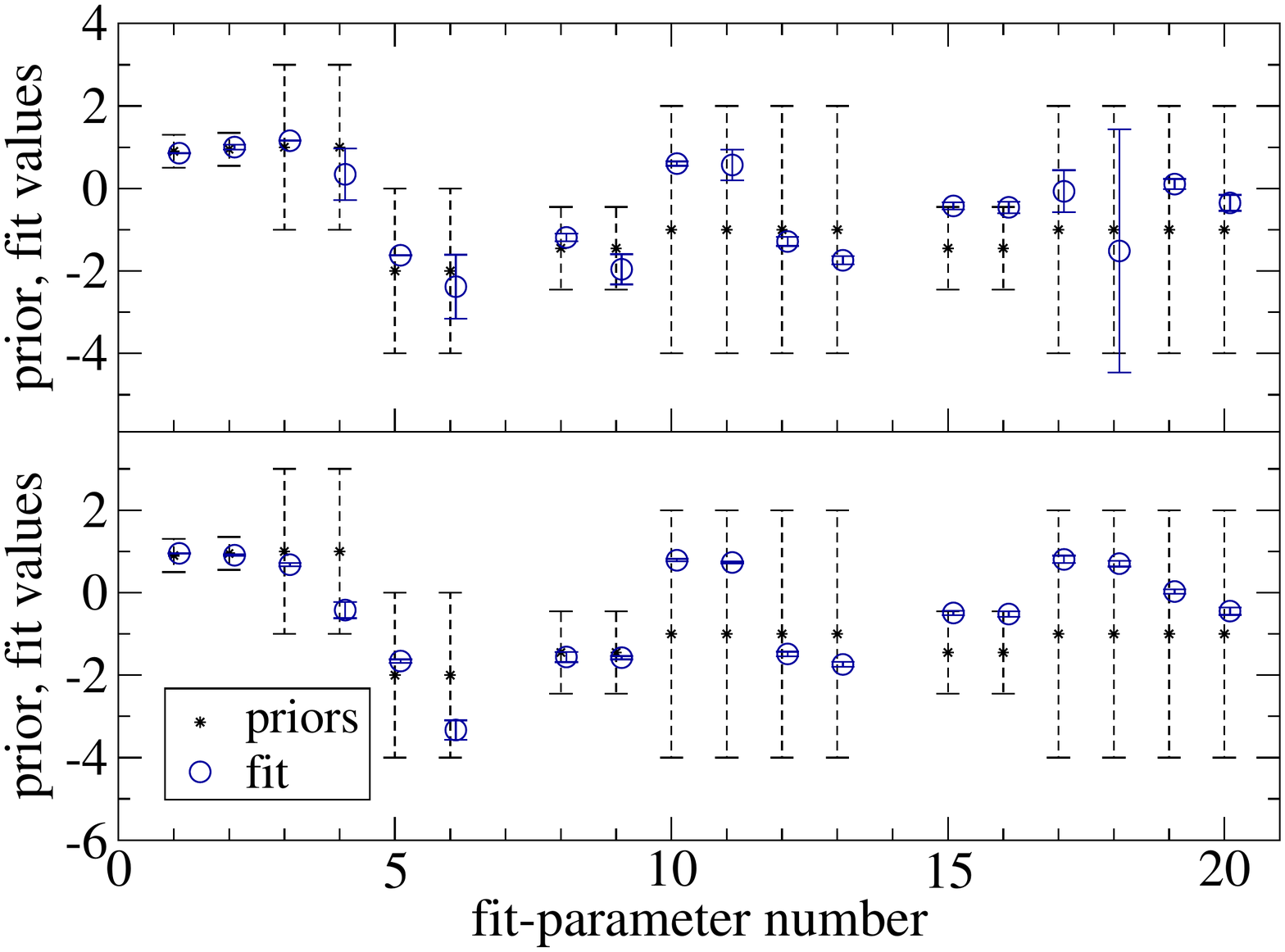}   
		 \\
		 (a)  &  (b)  \\
		\includegraphics[scale=0.3]{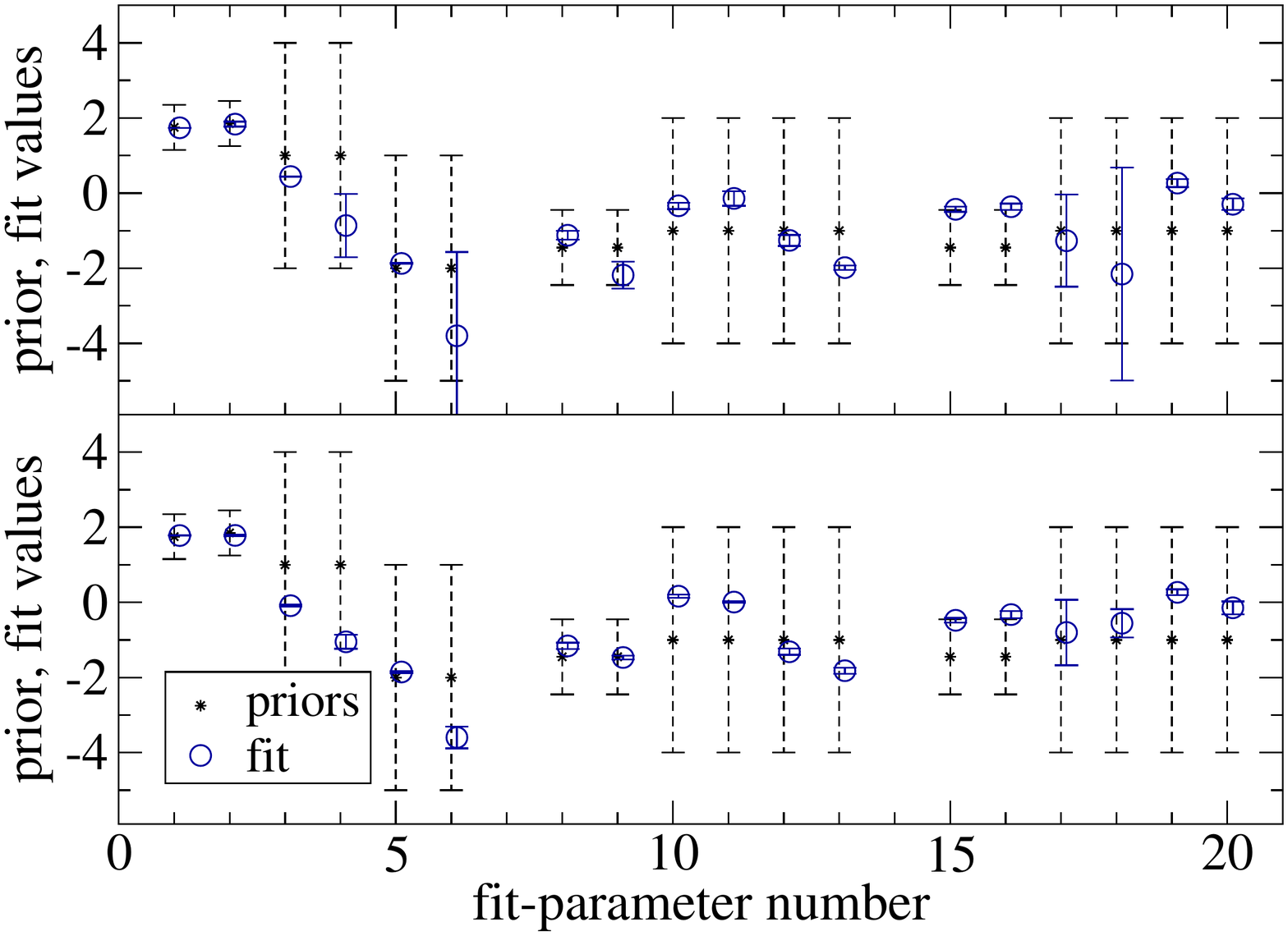}   
		&
		\includegraphics[scale=0.3]{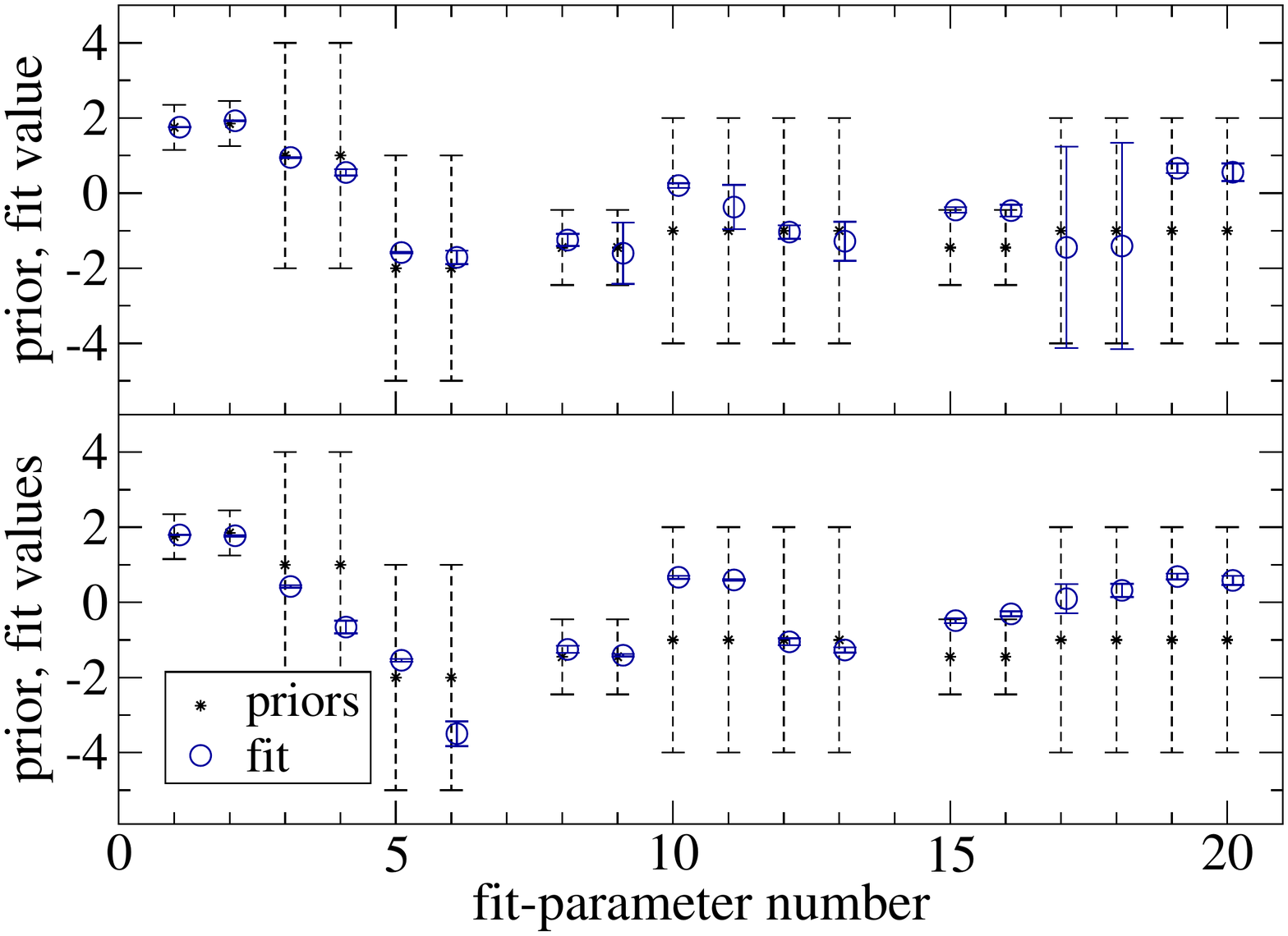}   
		 \\
		 (c)  &  (d)  \\
	\end{tabular}
	\caption{Fit results shown as open (blue) circles are overlaid on the priors, black dots with dashed widths, for charm-type (a) pseudoscalar and (b) vector mesons and bottom-type (c) pseudoscalar and (d) vector mesons on the (0.0062, 0.031) fine ensemble.   
	\kp~= 0.127 and 0.090 for charm- and bottom-type mesons, respectively; \amq~= 0.0272.
	The upper [lower] plot is from a fit where the meson has momentum of $\n=(0,0,0)$ [$(2,0,0)$].
	The fit-parameter numbers are defined in Table~\ref{priors-fine}.
	In each panel, the leftmost cluster corresponds to quantities from the ground state; the middle cluster corresponds to the first excited state; and the right most cluster to the second excited state. 
	Errors on the fit results are Hessian.
	For clarity, fit results are offset along the $x$-axis.}
	\protect\label{fig:PRIORS-fine}
\end{figure} 
%
\begin{figure}
	\includegraphics[scale=0.3]{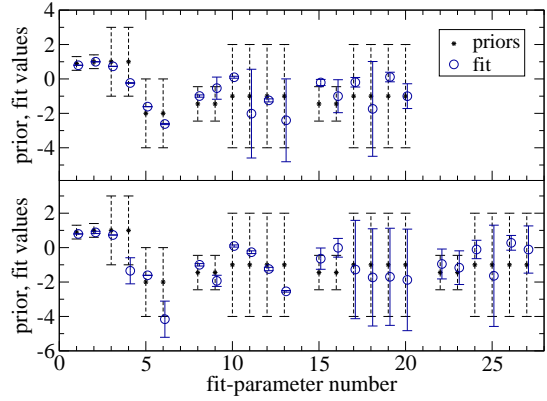}    
	\caption{Fit results shown as open (blue) circles are overlaid on the priors, black dots with dashed widths, for charm-type mesons on the (0.0062, 0.031) fine ensemble. \kp~= 0.127; \amq~= 0.0272; $\n=(0,0,0)$.
	The upper plot is the same as the upper left (pseudoscalar) panel of Fig.~\ref{fig:PRIORS-fine}~(a).
	The lower plot is from a fit which only differs by the use of $N=4$ pairs of states. 
	The fit-parameter numbers are defined in Table~\ref{priors-fine}.
	In each panel, the leftmost cluster corresponds to quantities from the ground state; the middle cluster corresponds to the first excited state; the next cluster corresponds to the second excited state and so on.
	The (desired-parity) ground-state quantities are stable to this change while other, excited-state, parameters are not.
	Errors on the fit results are Hessian.
	For clarity, fit results are offset along the $x$-axis. }
	\protect\label{fig:PRIORS-compare}
\end{figure}

For prior-width tests,  we reduce the widths by a factor of two for the non-oscillating ground state quantities and the energy splittings and repeat the fits. 
All changes observed are within statistical errors and, in most cases, the changes are substantially smaller than one $\sigma$.  
For charm, we also test for effects of the $DK$ threshold near the $D_{s0}^*(0^+)$ state.  
This splitting is 50 to 100 MeV, which is a several-$\sigma_{ \tilde{\Delta}_{aE_p} }$ deviation from our prior central value.
We ran separate tests on each lattice spacing using a prior width of $\sigma_{ \tilde{\Delta}_{aE_p} }= 2.5$ for the oscillating-state energy splitting.  
In units of MeV, this puts a 50-MeV splitting within 1$\sigma_{ \tilde{\Delta}_{aE_p} }$ of the prior central value.  
The ground and first-excited-state energies of the oscillating state are affected by this change but not in a systematic way.
This indicates that the oscillating-state signal is not strong in our data.
Our main interest, though, is the non-oscillating ground state energy $aE(\p)$;  this value is unaffected by the change in $\sigma_{ \tilde{\Delta}_{aE_p} }$.

In addition, we compare fit results with their priors.
Figure~\ref{fig:PRIORS-fine} gives examples of  these comparisons for fits on the (0.0062, 0.031) fine ensemble 
for charm- and bottom-type mesons.    
The $x$-axis labels the fit-parameter number, defined in Table~\ref{priors-fine}; the ground-state energy and amplitudes of the desired-parity state are at positions 1, 3, and 5.
We find that fit results for ground-state quantities are well within the prior widths.
For excited states, in some cases the fitter simply returns the prior value, indicating that the quantity is not constrained by the data.
In other cases, the results appear to be constrained by the data, indicating that some excited-state signal is in the correlator and the fitter adjusts the amplitudes to absorb it.
Although it may appear in Fig.~\ref{fig:PRIORS-fine} that a number of excited-state quantities are well-determined, this is an artifact of a minimum-$N$ fit; unlike the ground-state parameters, the excited state results are not stable as $N$ is increased.  
For example, Fig.~\ref{fig:PRIORS-compare} compares the fit results shown in the upper left (pseudoscalar) panel of Fig.~\ref{fig:PRIORS-fine}~(a), which uses $N=3$, with a fit which only differs by the use of $N=4$.
The comparison demonstrates that the (desired-parity) ground-state quantities are stable to the change in $N$ while other, excited-state, parameters are not.

\begin{figure}
	\begin{tabular}{c c}
		\includegraphics[scale=0.3]{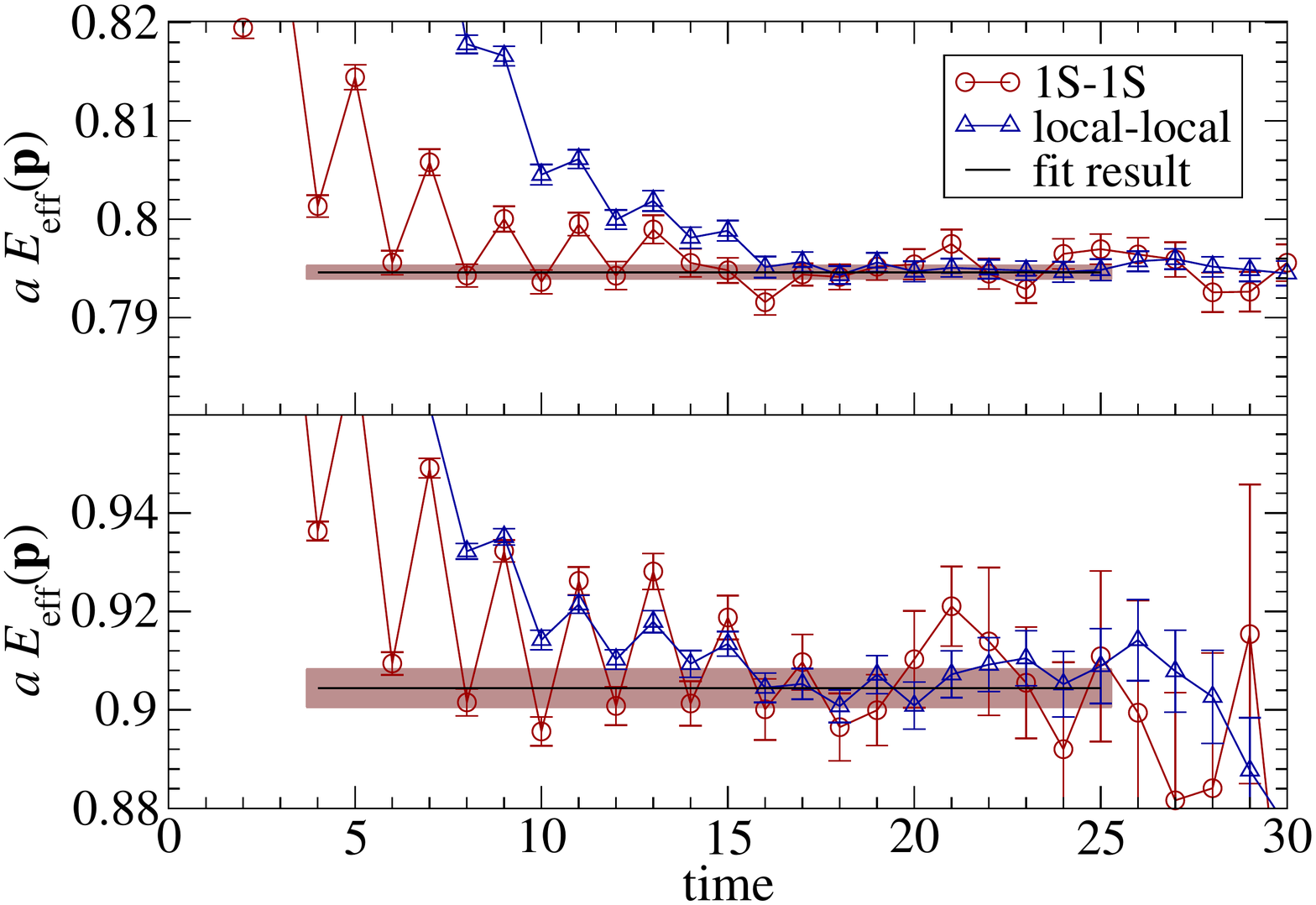}
		&
		\includegraphics[scale=0.3]{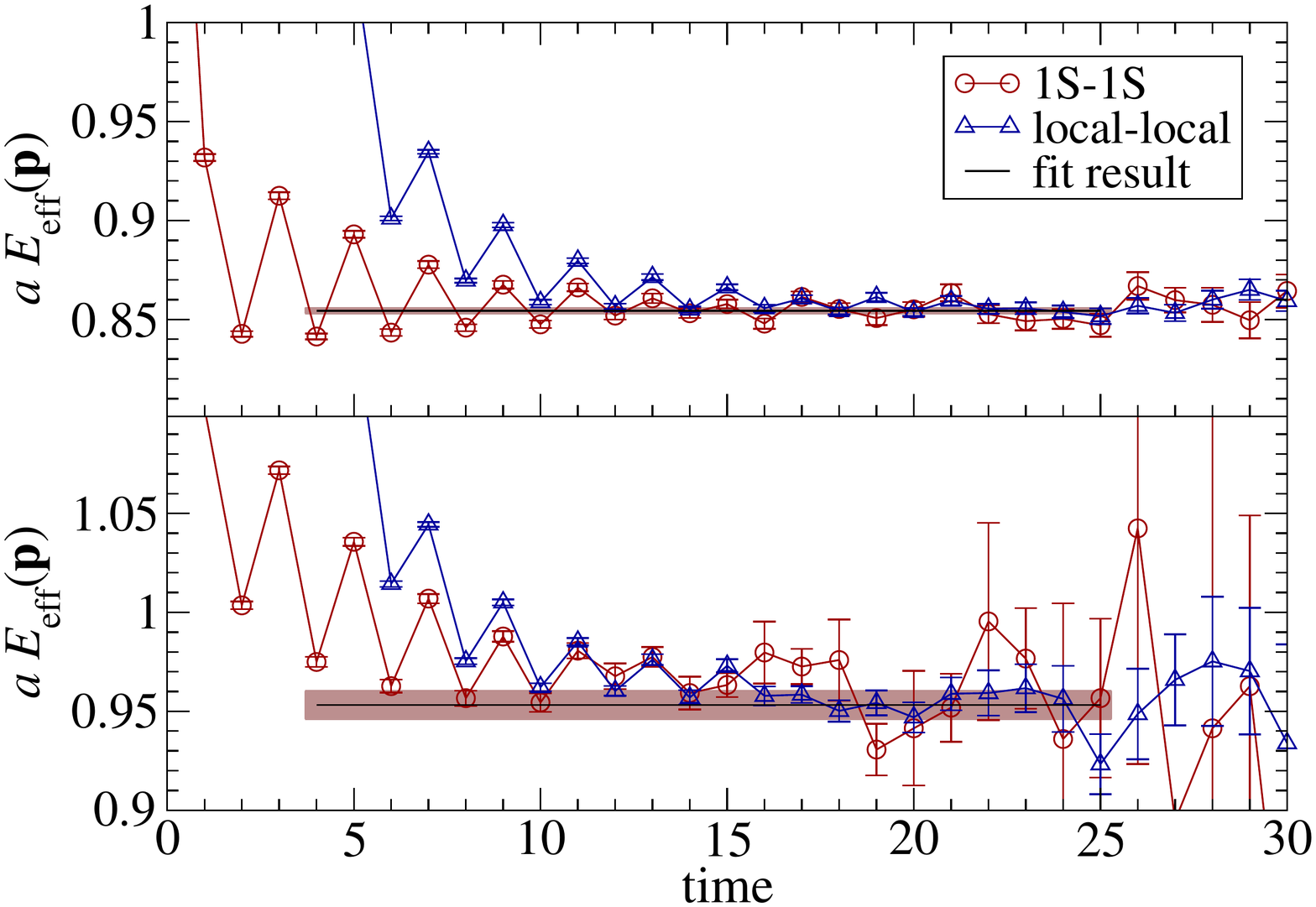}
		\\
		(a)  &  (b)  \\
		\includegraphics[scale=0.3]{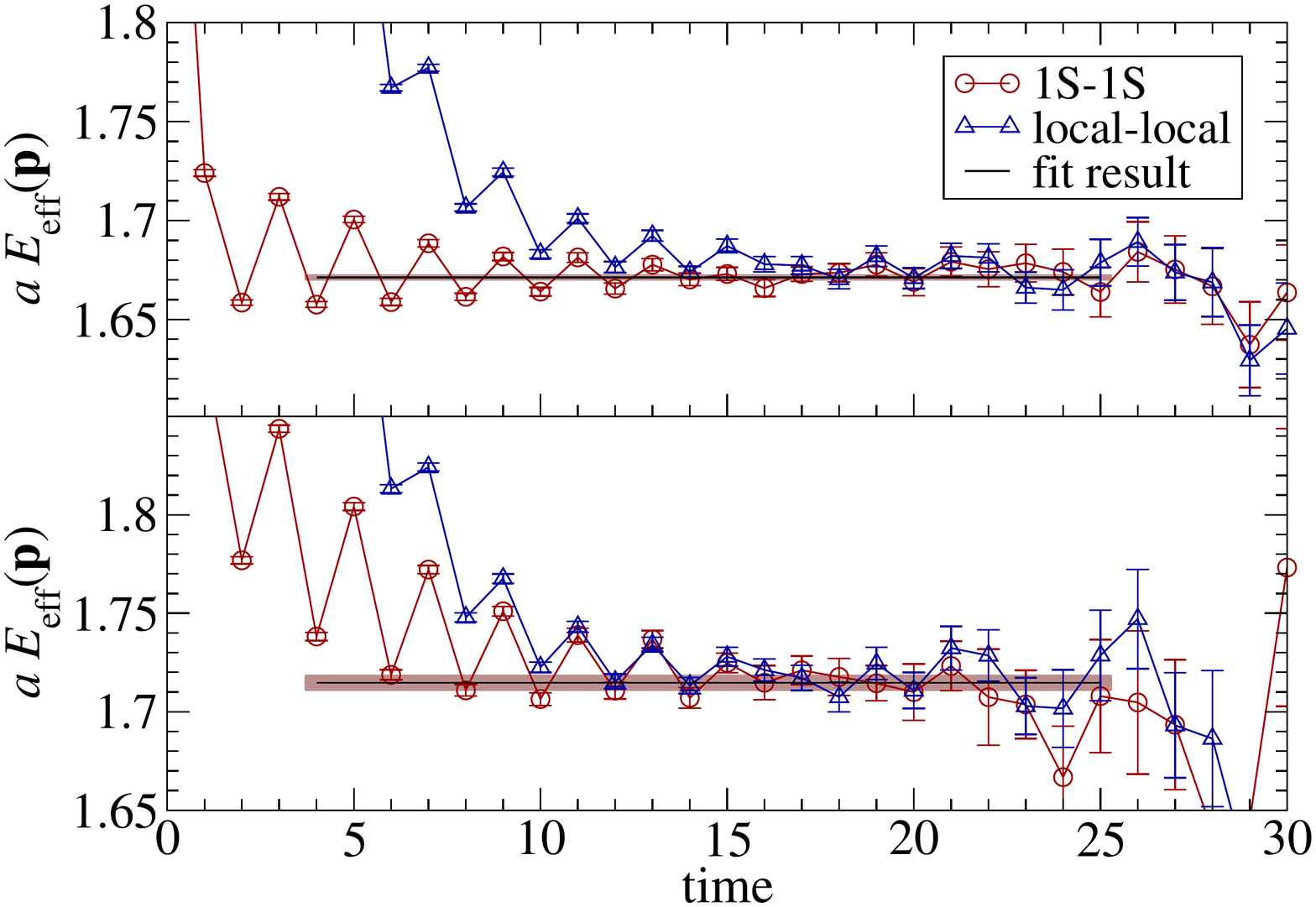}
		&
		\includegraphics[scale=0.3]{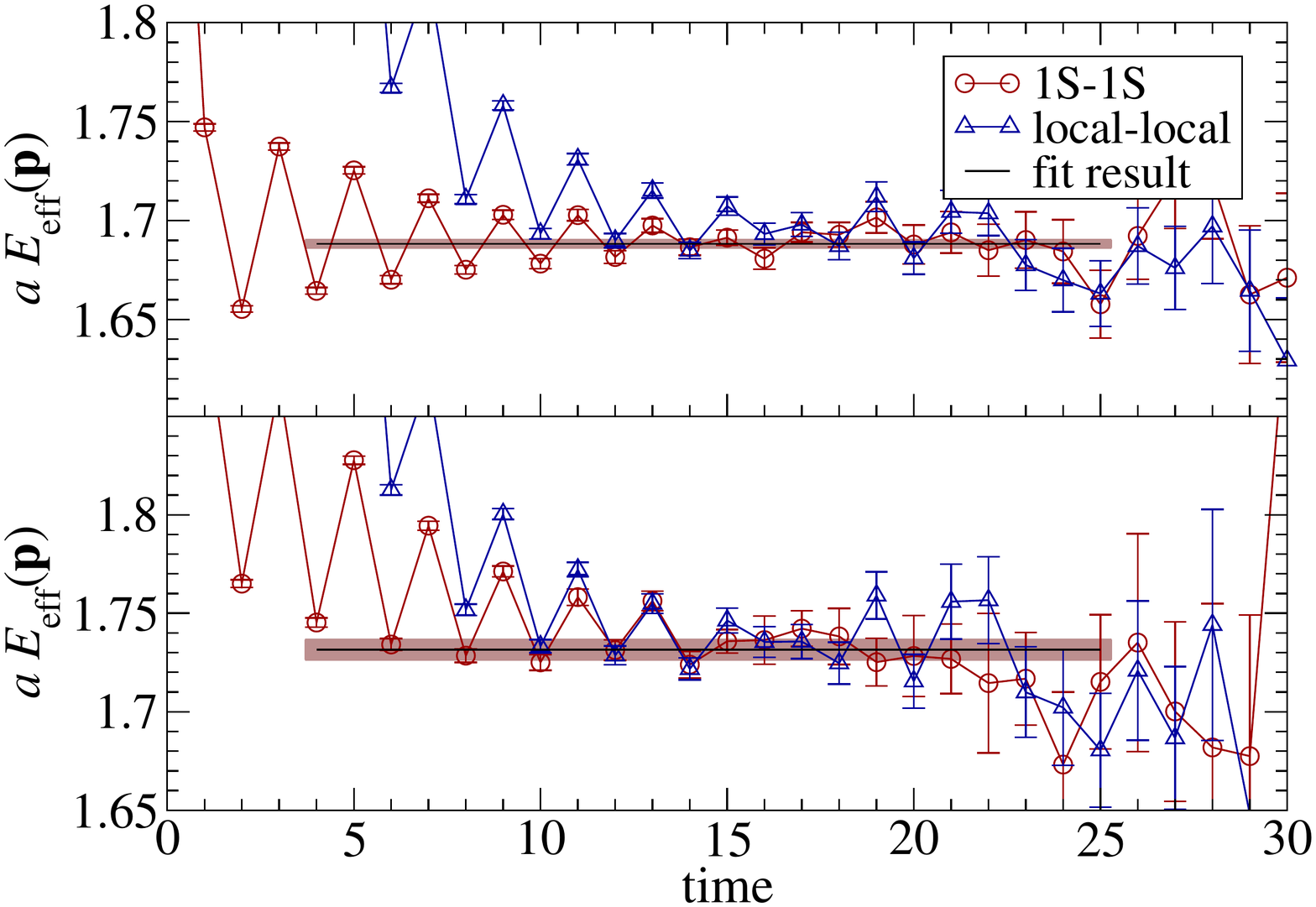}
		\\
		(c)  &  (d)  \\
	\end{tabular}
	\caption{Effective energy plots, $aE_{\rm eff}(\p)$, for  charm-type (a) pseudoscalar and (b) vector mesons and bottom-type (c) pseudoscalar and (d) vector mesons on the (0.0062, 0.031) fine ensemble.
	 \kp~= 0.127 and 0.093 for charm- and bottom-type mesons, respectively; \amq~= 0.0272.
	The upper [lower] plot is from a fit where the meson has momentum of $\n=(0,0,0)\, [\n=(2,0,0)]$.
	 Open (blue) triangles mark the local correlator and open (red) circles mark the 1S-smeared correlator.
	 Lines connecting the data points are simply to guide the eye; they are not a fit.
	The unadorned  black line is the multi-correlator fit result and the shaded band marks  the average 68\% bootstrap error.}
	\protect\label{EM-fine}
\end{figure}

For goodness-of-fit we begin by looking at the augmented \chisq~for each fit and verify that it is $\approx 1$ or smaller, where ``$\approx 1$'' is based on the 80\% range of the \chisq~distribution for a given number of degrees of freedom.
%
As a final check, we overlay the result on an effective-mass plot.  
We define the ``effective energy''
\be \label{eq:effmass}
	2aE_{\rm eff}(\p) = \ln \left[ C(t)/C(t+2a) \right] 
\ee
using a step of two time units in order to accommodate the oscillating contribution from the opposite-parity state.  
Figure~\ref{EM-fine} shows plots comparing $aE_{\rm eff}(\p)$ to the fit result on the  (0.0062, 0.031) fine ensemble.
The ground-state-energy result from the multiple-state fit is shown as a straight line segment over the time range fit.  
The band encompasses the average 68\% bootstrap error.
In each case, the fit result nicely matches the effective-energy plateau.

 
 \subsection{The kinetic mass $M_2$}  \label{sec:kinmass}

\begin{figure}
	\begin{tabular}{c c}
		\includegraphics[scale=0.3]{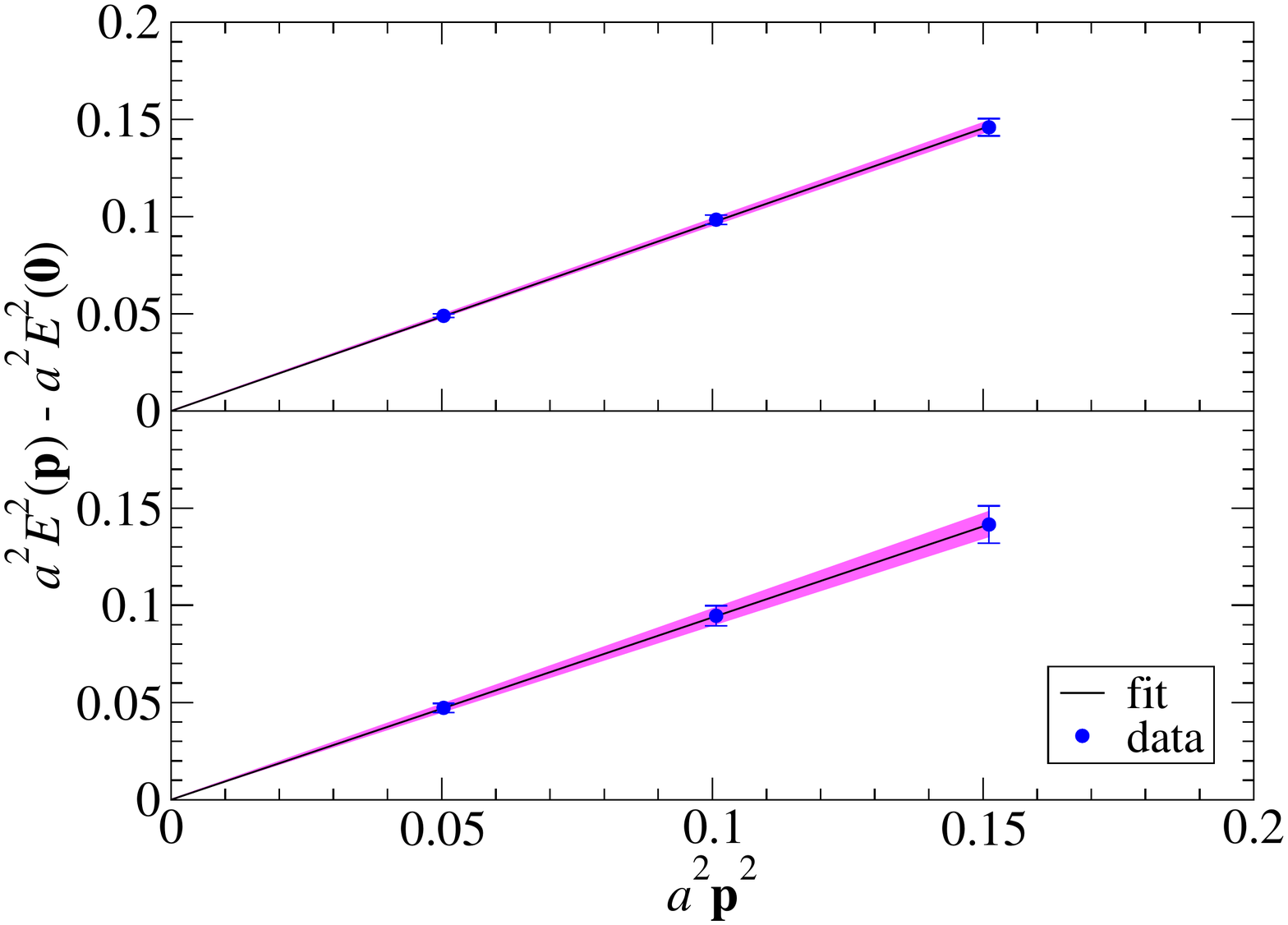}   
		&
		\includegraphics[scale=0.3]{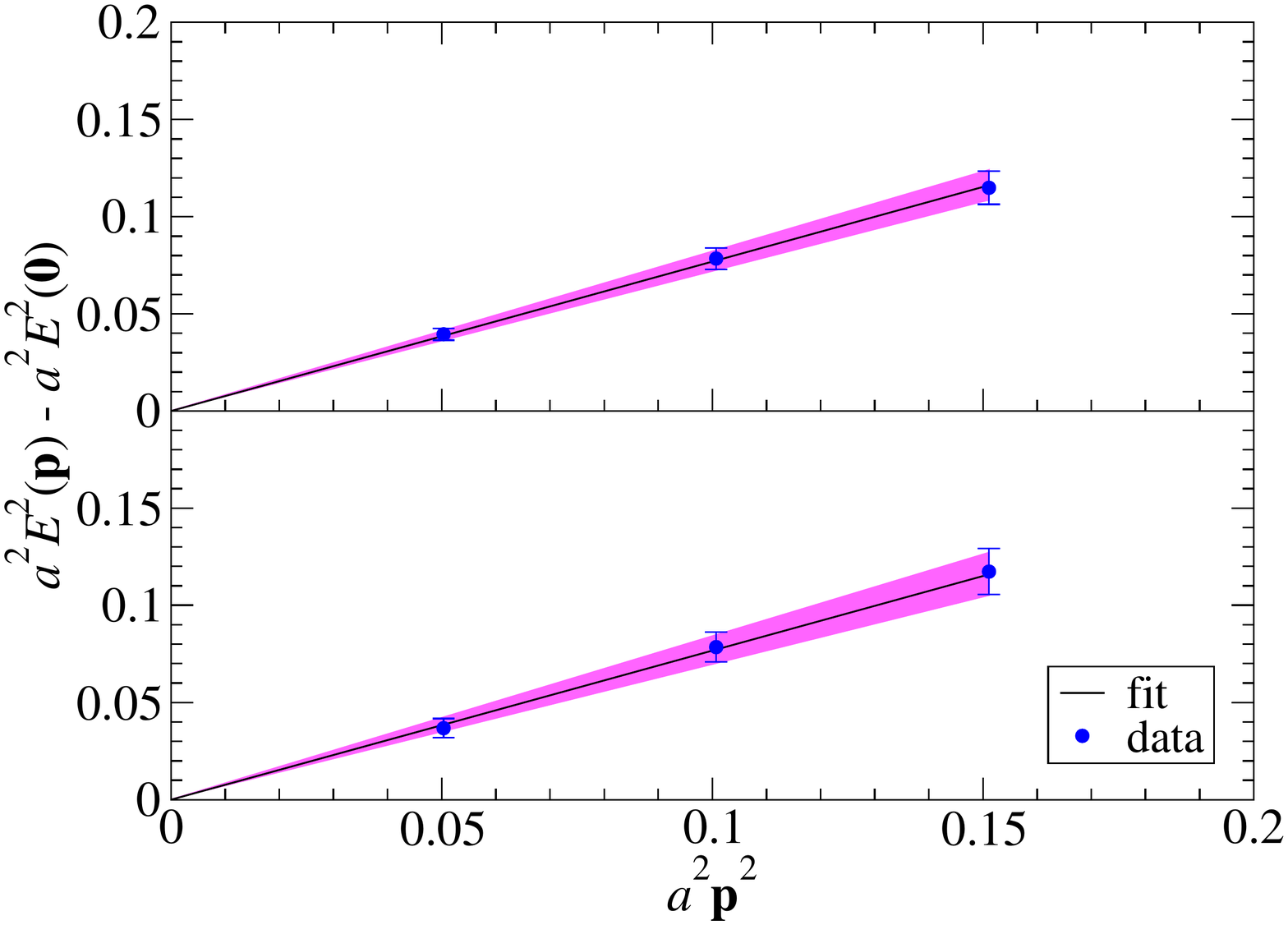}   
		 \\
		 (a)  &  (b)  \\
	\end{tabular}
	\caption{Results of fits to the dispersion relation for (a) charm-type (\kp~=~0.127) and (b) bottom-type (\kp~=~0.0923) mesons on the (0.0062, 0.031) fine ensemble. 
	(Blue) dots are the data.  A black line shows the fit result with the (pink) shaded band showing the one-sigma error from the fit.
	Upper panels show results for pseudoscalars and lower for vectors.}
	\protect\label{fig:DR-fine}
\end{figure}

Given results for \aEp, we fit data where $|\n| \le \sqrt{3}$ to Eq.~(\ref{eq:disp_func_used}) to determine the pseudoscalar and vector kinetic meson masses. 
Fits use a correlation matrix constructed from the bootstrap distributions.
The tables in Appendix~\ref{App_M2} give results for \aMkps, \aMkvm, and~\aMkbar~on the ensembles used for tuning, listed in Table~\ref{tbl:dataused}.
Included in the tables are the \chisq~and the probability that $\chi^2$ would exceed the value from the fit, known as the $p$ value~\cite{PDG06}.
Typical dispersion relation fits are shown for the (0.0062, 0.031) fine ensemble in Fig.~\ref{fig:DR-fine}.

In addition to statistical errors, we consider uncertainties from unphysical sea-quark masses, mistuning of the valence strange quark, and discretization. 
%
The noise in $\overline{M}_2$ makes it difficult to discern how $\overline{M}_2$ depends on the sea-quark masses. 
The $\overline{M}_1$ data is much cleaner, though, and we can use it to estimate the sea-quark error on $\overline{M}_2$, and hence $\kappa$.
To do this, we first note that, cf.~Eq.~(\ref{eq:HQETM1}),
\bea    
  	aM_1 &=& am_1 + a\bar{\Lambda}_{\rm lat} + O(1/m_Q)   \\
	aM_2 &=& am_2 + a\bar{\Lambda}_{\rm lat} + O(1/m_Q)    
\eea
where $am_1$ and $am_2$ capture the leading heavy-quark dependence and $\bar{\Lambda}_{\rm lat}$ depends only on the light degrees of freedom.
Taking $a\bar{\Lambda}_{\rm lat}$ to be the same for both $a\overline{M}_1$ and $a\overline{M}_2$ (see Appendix~\ref{App_M2DiscErr} and Ref.~\cite{Kronfeld:1996uy})
we can estimate the size of the effect of non-physical (light) sea quark masses on $a\bar{\Lambda}_{\rm lat}$, and hence $aM_2$, by studying the behavior of $aM_1$ as the light sea-quark masses are varied.

\begin{figure}
	\begin{tabular}{c c}
		\includegraphics[scale=0.3]{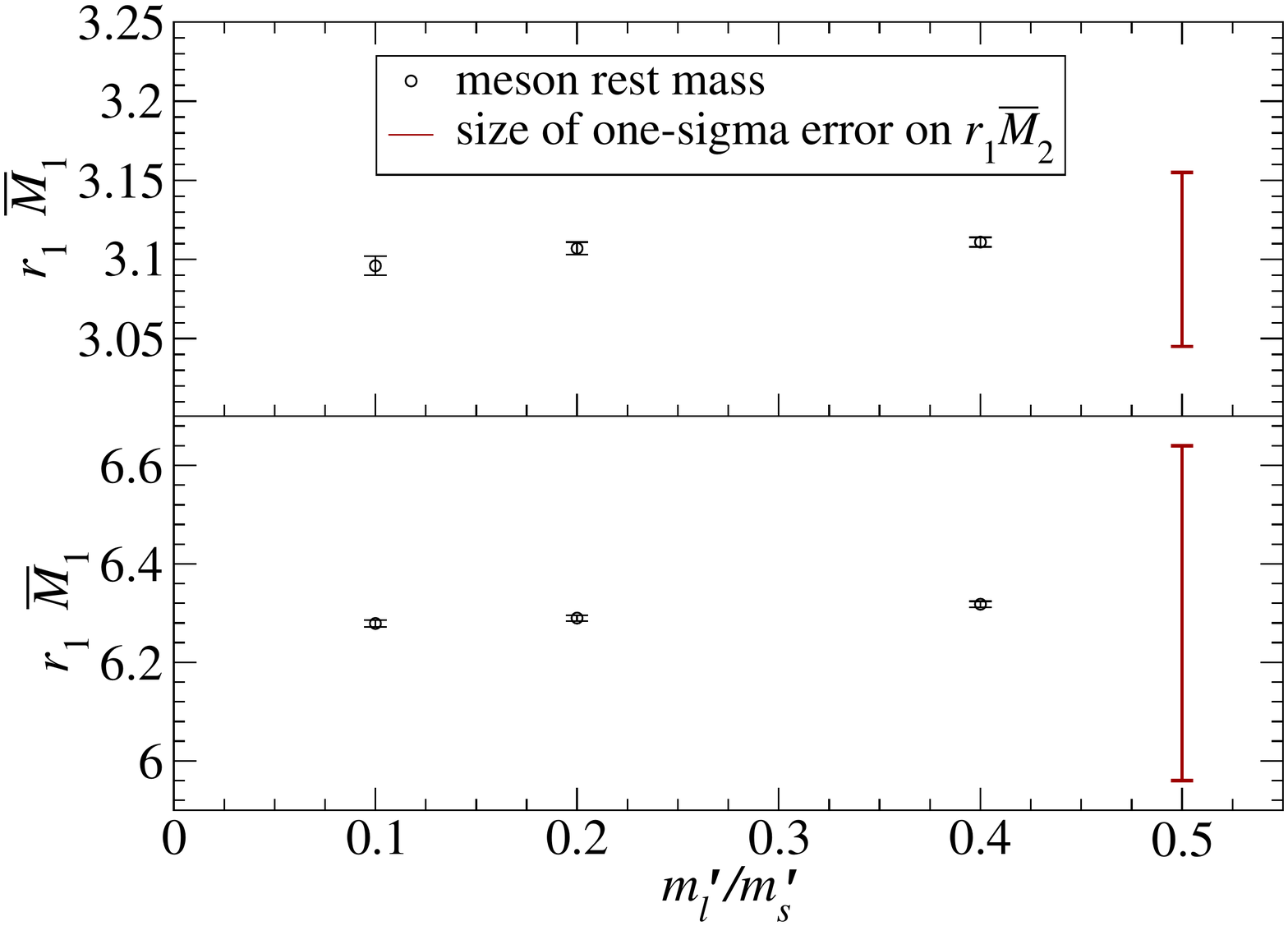}   
		&
		\includegraphics[scale=0.3]{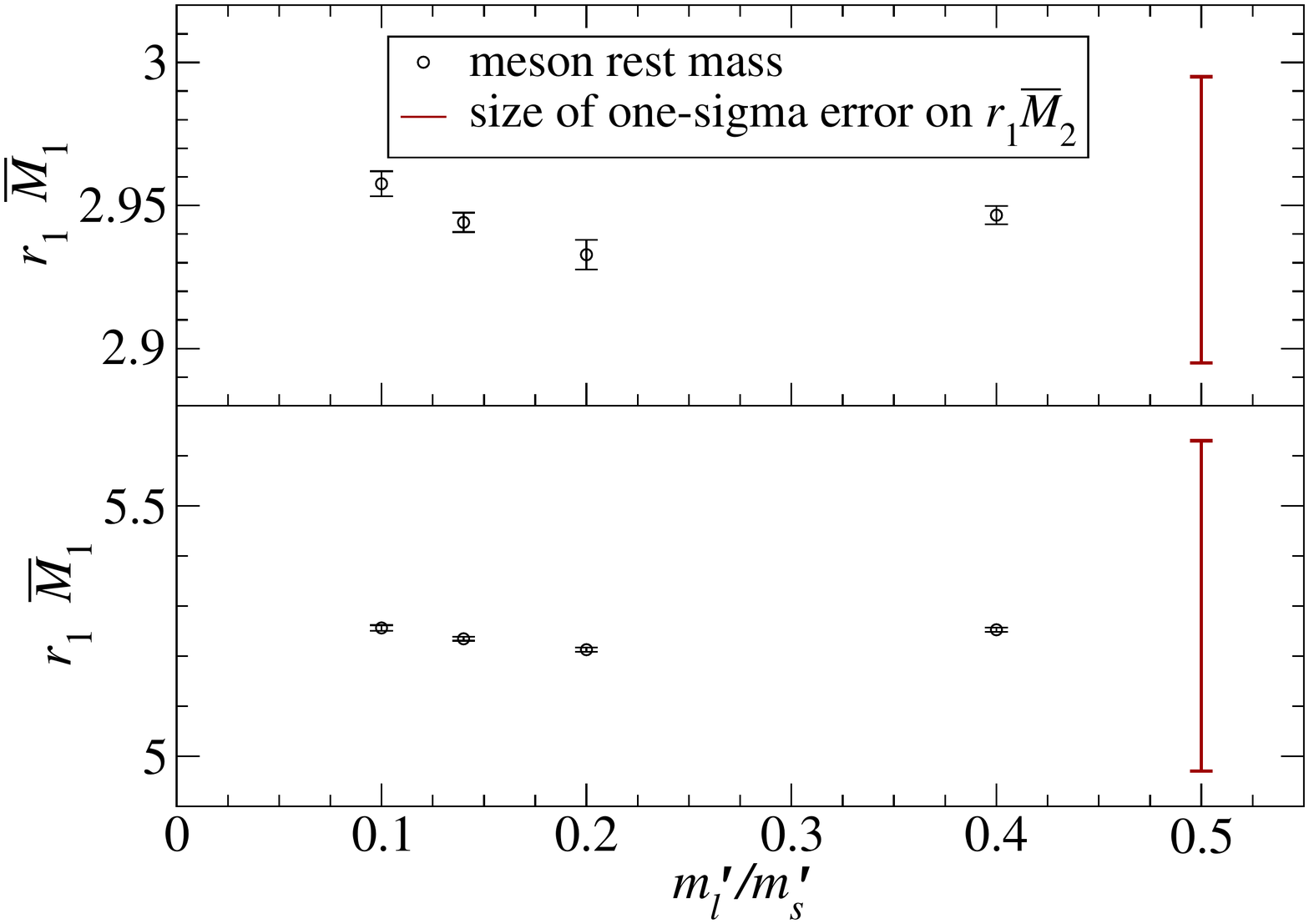}   
		\\
		(a)  & (b)  \\
	\end{tabular}
	\caption{The spin-averaged meson rest mass in physical units versus the ratio of the light to strange sea-quark masses $m'_l/m'_s$ for the (a) fine and (b) coarse ensembles.  
	Error bars are statistical only, from  the average 68\% bootstrap error.
	On the far right of the plot is a (red) bar indicating the size of the one-sigma statistical error on $r_1 \overline{M}_2$.
	The upper-panel plot is for charm-type mesons, lower is for bottom-type. 
	Values of  \kp~used are 0.127, 0.0923  on the fine ensembles with \amq~= 0.0272 and \kp~0.122, 0.086 on the coarse ensembles with \amq~= 0.0415.}
	\protect\label{fig:M2-sea}
\end{figure}

In Fig.~\ref{fig:M2-sea}, we plot the spin-averaged meson rest mass $r_1 \overline{M}_1$ versus the ratio of the light to strange sea-quark masses $m'_l/m'_s$ for the coarse and fine ensembles used here.
On the far right of each plot is a  bar indicating the size of the 1-$\sigma$ statistical error on $r_1 \overline{M}_2$; for fine this is from the (0.0062, 0.031) ensemble and for coarse the (0.007, 0.050) ensemble. 
The light sea-quark mass dependence is negligible compared to the statistical error on $r_1 \overline{M}_2$.  
We find similar behavior for the medium-coarse ensemble.

We must also consider how the non-physical value of the strange sea-quark mass affects $\overline{M}_2$.  
The strange sea-quark mass is mistuned by an amount $0.19 am'_s, 0.31 am'_s$ and $0.12 am'_s$ on the fine, coarse, and medium-coarse ensembles, respectively.    
The continuum chiral perturbation theory expression for the heavy-light spin-averaged mass~\cite{Jenkins:1992hx} shows that the leading sea-quark dependence of $\overline{M}_2$ is proportional to the sum over the sea-quark masses, $2m'_l + m'_s$. 
Hence, varying $am'_l$ tells us about the effect of varying $am'_s$.
Figure~\ref{fig:M2-sea} shows that a change of $0.3 am'_s$ in $am'_l$ has a negligible effect on $\overline{M}_2$, so we conclude that the mistuning of $am'_s$ has a negligible effect as well.

The tuned value of the strange-quark mass on each ensemble is given in Table~\ref{Tbl:vacuum}.
On the fine lattice, the valence-quark mass used in the simulation, \amq~ = 0.0272, differs from the physical value \ams~=~ 0.0252  by 0.0020.   
A comparison of our results for \aMkbar~in Table~\ref{M2-results-fine} shows that even a deviation in \amq~of twice this size does not discernibly affect  \aMkbar.  
The situation is similar for the coarse and medium-coarse results.  
For the coarse ensembles, the simulation mass $\amq = 0.03$ differs by 0.0044 from the tuned value of \ams.
Table~\ref{M2-results-coarse} shows that \aMkbar~is barely affected at the 1-$\sigma_{\aMkbar}$ level as \amq~changes by over twice this size.  
For the medium-coarse ensembles, the simulation mass of 0.0484 differs from the tuned strange-quark mass by 0.0058.   
A comparison of the values of \aMkbar~in Table~\ref{M2-results-medcoarse} shows that a deviation in \amq~just under twice this size yields, at most, a 1-$\sigma_{\aMkbar}$ variation in \aMkbar.
Therefore, we take our results of \aMkbar~at \amq~= 0.0272, 0.03, and 0.0484 as the masses of the $B_s$ and $D_s$ on the fine, coarse and medium-coarse ensembles, respectively, with no additional error for valence-mass mistuning.

In Appendix~\ref{App_M2DiscErr}, we derive an expression for the discretization error in $M_2$, $M_2 = M_{\rm continuum} + \delta M_2$.
The result, Eq.~(\ref{eq:error_nolightdisc}), can be written
\be  \label{eq:deltaM}
	\delta M_2  = \frac {\bar{\Lambda}^2} {6m_2}  \left[
	                                   5 \left( \frac {m_2^3} {m_4^3} - 1 \right)  + 4 w_4 (m_2 a)^3
	                                   \right] ,
\ee
replacing $\langle\bm{p}^2\rangle$ of Eq.~(\ref{eq:error_nolightdisc}) with $\bar{\Lambda}^2$.
Expressions for the short-distance coefficients $m_2$, $m_4$, and $w_4$ are given in Appendix~\ref{App_M2DiscErr}~\cite{El-Khadra:1996mp, Oktay:2008ex}.
To estimate the discretization error, we use values of the physical (pole) quark mass  (1.4~GeV for charm and 4.2~GeV for bottom) for $m_2$ in the prefactor of Eq.~(\ref{eq:deltaM}), and $\bar{\Lambda} = 0.7$~GeV.
Using these values,  $u_0$ from Table~\ref{Tbl:mesons}, and $\kc$ from Table~\ref{kcrit}
yields the values of $\delta M_2$ shown in Table~\ref{discerr-table}.  
The error estimate in Eq.~(\ref{eq:deltaM}) pertains to the kinetic mass, but the main focus here is the tuning of $\kappa$.
After tuning, we shall propagate this error from $M_2$ to $\kpch$ and $\kpbot$.

\bt{The relative error in the tuned hopping parameter $\delta \kappa / \kappa$  due to discretization effects in the kinetic meson mass.  
The ensembles used are (0.0062, 0.031),  (0.007, 0.050), and (0.0097, 00484) for the fine, coarse, and medium-coarse lattices, respectively.
Values of \kp~are 0.127 and 0.0923 on fine; 0.122 and 0.086 on coarse; and, 0.122 and 0.076 on medium-coarse.  
The $[\cdots]$ denotes the quantity in brackets in Eq.~(\ref{eq:deltaM}).
We use $(\bar\Lambda^2/6m_{\rm ch}) = 0.058\bar3$ and $(\bar\Lambda^2/6m_{\rm bot}) = 0.019\bar4$ to convert the $[\cdots]$ to $\delta M_2$.   
Values of $\delta \kappa / \kappa$ are given as fractions not a percentage.
}
{ l         c   c   c   c   c              c   c   c   c   c  }
	           &  \mc{5}{c}{charm}    & \mc{5}{c}{bottom}  \\ [0.25em]
   lattice spacing             & $m_0a$    & $[\cdots]$   &  $\delta M_2$   & $\frac{d m_2a}{d m_0 a}$ &  $\dkk$        & $m_0a$   & $[\cdots]$   &  $\delta M_2$   & $\frac{d m_2a}{d m_0 a}$ &  $\dkk$     \\  [0.25em]
   \hline
    fine                    &  0.391      & 1.31           & 0.0763              &   0.843                               & $-0.0086$    & 2.08         &16.8           & 0.327                &   0.880                             & $-0.0256$    \\ 
    coarse               &  0.565      & 2.37           & 0.1384              &   0.831                               & $-0.0203$    & 2.62         & 23.6          & 0.459                &   0.899                             & $-0.0440$    \\  
    medium-coarse  &  0.682     & 3.18            & 0.1857              &   0.830                               & $-0.0346$    & 3.56         & 37.2          & 0.724                &   0.922                             & $-0.0756$   \\   
\et{discerr-table}

\subsection{Fitting Summary}
\label{Sec:fitsum}

The preceding subsections contain many details intended for those engaged in similar analyses.
In this section, we re-emphasize the main features of the analysis.
Because, in this and related \cite{Aubin:2005ar,Aubin:2004ej,Evans:2008zz,Freeland:2007wk,Allison:2004be,Burch:2009az} work, 
we are interested in the ground state, we do not dwell on the excited states here.

Our priors are guided by the data, using one ensemble to set them and 
(generally) other ensembles for physical results.
We choose a time range such that the fit results for the ground state 
are stable, listed in Table~\ref{tbl:NstatesTime}.
We also test for stability as the number $N$ of (pairs of) exponentials 
grows---as shown in one example in Fig.~\ref{fig:Ns-fine}---and choose 
the minimum value of $N$ for which the central value is stable within errors.
The errors on the ground-state amplitudes and energies are always 
determined by the data, not the priors, as shown in 
Fig.~\ref{fig:PRIORS-fine} and~\ref{fig:PRIORS-compare}.
(In many cases, even excited-state information is data-determined, not 
prior-determined.)
Figure~\ref{EM-fine} shows that the fits agree with the effective energies.
(Note that the oscillations of $aE_{\rm eff}$ at small $t$ are to be expected with staggered quarks.)
In conclusion, the constrained curve fitting for $E(\bm{p})$ has worked as advertised, subsuming the subjectivity of fit ranges and different choices of $N$ 
into robust results for both central value and error bar.
Figures~\ref{fig:DR-fine} and~\ref{fig:M2-sea} show that, once $E(\bm{p})$ is well-determined, we can straightforwardly obtain the kinetic mass $M_2$ and the hyperfine splitting.

%% file: Results.tex
In this section, we present the main results of these calculations, including our error analysis.
Section~\ref{sec:tuning} focuses on the tuned values of $\kpch$ and $\kpbot$,
Sec.~\ref{sec:hfs} on the $D_s$ and $B_s$ hyperfine splittings, and
Sec.~\ref{sec:kcrit} on the critical value of the hopping parameter $\kc$. 

\subsection{The tuning of $\kpch$ and $\kpbot$}
\label{sec:tuning}

As discussed in Sec.~\ref{sec:kinmass}, effects from non-physical 
sea-quark masses and the mistuning of the valence strange-quark mass 
are negligible compared to the statistical error on \aMkbar.
In that section, we explain why taking \aMkbar\ at certain values of \amq\ is an acceptable approximation to \aMkbar\ at the tuned physical strange-quark mass.
We choose to tune~$\kappa$ at those same \amq, which are
$\amq = 0.0272$ on the (0.0062, 0.031) fine ensemble, 
$\amq = 0.03$  on the (0.007, 0.050) coarse ensemble, 
and  $\amq = 0.0484$ on the (0.0097, 0.0484) medium-coarse ensemble.

To obtain the tuned $\kappa$ for the charm (bottom) quark, $\kpch$ ($\kpbot$), we want to interpolate \Mkbar\ to the PDG value of the spin-averaged $D_s$ ($B_s$) mass~\cite{PDG06}.
In practice, it is simpler to do the interpolation with the meson mass in lattice units. 
Hence, we linearly interpolate \aMkbar\ to $a\overline{M}_{\rm PDG}$, the PDG value for the meson  mass converted to lattice units with $a$ from Table~\ref{Tbl:vacuum}.
This interpolation is repeated for the entire bootstrap distribution of $\aMkbar$.  We then estimate the statistical error on $\kappa$ as the
average 68\% bootstrap error described in Sec.~\ref{sec:twopnts}.
The discretization error in $M_2$, $\delta M_2$, is given by Eq.~(\ref{eq:deltaM}), and is always positive. 
This results in a single-sided, negative error bar on $\kappa$.
We convert $\delta M_2$ to the error, \dk, using $dM_2/d\kappa\approx dm_2/d\kappa$ and expressions for $m_0a$ and $m_2a$ given in Appendix~\ref{App_M2DiscErr}.  
The $\delta \kappa$ are given in Table~\ref{discerr-table}.
The experimental errors on the PDG values are negligible.
The remaining errors to consider are those which appear in the
conversion between lattice and physical units.
The error in the determination of $r_1/a$ is negligible, so we only need to consider the error in $r_1$, given in Eq.~(\ref{eq:r1}).

The error on $r_1$ is propagated to an error on $a^{-1}$ and then to an
error on $a\overline{M}_{\rm PDG}$, denoted $\sigma_{\rm PDG}$.
Table~\ref{tbl:PDG} gives the values of the PDG meson masses used in
this work and tabulates their spin-averaged mass and hyperfine
splitting.
Table~\ref{tbl:PDG-lat} gives the spin-averaged mass in lattice units.
The uncertainty $\sigma_{\rm PDG}$ is propagated to $\kappa$ using the
standard error formula $\sigma_{\kappa} = \sigma_{\rm PDG} / s$, where
$s$ is the slope used in the interpolation.
Table~\ref{tbl:kappa-error} gives the error budget for $\kpch$ and $\kpbot$,  and Table~\ref{tbl:kappa-final} lists the final tuned
results.

\bt{PDG values of the pseudoscalar and vector masses for the $D_s$ and $B_s$ mesons and the hyperfine splitting $\Delta$~\protect\cite{PDG06}.
	Also listed is the derived quantity $\overline{M}$, the spin-averaged mass.  
	}
	{ l    c  c  c  c  }
		& $M$ (GeV)                  &  $M^*$ (GeV)            & $\overline{M}$ (GeV)  &     $\Delta$ (MeV)     \\
		\hline  \\ [-0.5em]
		$D_s$
		& 1.96849(34)    &  2.1123(5)                   &  2.0763(4)             &   143.9(4)                           \\ [0.5em]     
		$B_s$
		& 5.3661(6)         &  5.4120(12)                 & 5.4005(9)         &     46.1(1.5)                          \\  [0.5em]  
\et{tbl:PDG}
\bt{Spin-averaged PDG masses converted to lattice units with an error
from the uncertainty in the lattice spacing $a$.
Values of $a$ used in the conversion can be found in
Table~\ref{Tbl:vacuum}.  }{ l      l          l      }
		  Ensemble                                      &
$a\overline{M}_{D_{\rm s}}$   &  $a\overline{M}_{B_{\rm s}}$    \\                    
		  \hline \\ [-0.5em]
		  Fine (0.0062, 0.031)                     &    \err{0.884}{0.009}{-0.023}     &  \err{2.299}{0.023}{-0.060}  \\ [0.5em]       
		  Coarse (0.007, 0.050)                   &    \err{1.242}{0.012}{-0.032}     &  \err{3.230}{0.031}{-0.083}   \\ [0.5em]      
		  Medium-coarse (0.0097, 0.0484)  &   \err{1.529}{0.015}{-0.039}      &  \err{3.977}{0.038}{-0.102}  \\ [0.5em]         
\et{tbl:PDG-lat}

\bt{Percent errors in the tuned $\kappa$ and the total error.
	For several sources of uncertainty, we determined that the error was
smaller than the precision of these calculations.  This is indicated by
an entry of ``0.0'' in the table.
	}
	{ l         c     c    c        c     c    c  }
       & \mc{3}{c}{Charm}    &   \mc{3}{c}{Bottom}                \\  [1.0em]
        Uncertainty      			&   Fine			&    Coarse                            & Medium-coarse      &Fine                            &    Coarse	       & Medium-coarse         \\
       \hline 
	Statistical                        & 1.26			&  0.57                                   &  0.53			&  5.0               		&  9.1			&  5.6  \\   
	Discretization                  & $(0, -0.86)$		&  $(0, -2.0)$                 	     &   $(0, -3.46)$		& $(0, -2.6)$             &  $(0, -4.4)$		& $(0, -7.56)$ \\
	Sea-quark masses         & 0.0				&  0.0                                     &  0.0				& 0.0                		 & 0.0			&   0.0    \\
	\ams~mistuning               & 0.0				&  0.0                                     &  0.0				& 0.0                 	& 0.0			&   0.0  \\  
	Unit conversion ($a$)      & $(+0.90, -0.35)$     &$(+0.49, -0.19)$                  & $(+0.77, -0.30)$       &$(+1.7, -0.64)$       & $(+1.9, -0.72)$	& $(+1.76, -0.66)$  \\
	\hline
	 Total                               & $(1.5, 1.6)$		&  ($+0.75, -2.1$)                   &  ($+0.93, -3.5$)       &($+5.3, -5.7$)         & ($+9.3, -10.1$)	& ($+5.9, -9.4$)  \\  
\et{tbl:kappa-error}

\bt{Final tuned results for $\kpch$ and $\kpbot$ with the total error.}
	{ l        c    c     c   }
                                             &   Fine              &    Coarse                           & Medium-coarse  \\ 
         \hline  \\ [-0.75em]
           $\kpch$               & 0.127(2)          &  $0.1219^{+9}_{-25}$           &  $0.122^{+1}_{-4}$      \\ [0.25 em]  
           $\kpbot$              & 0.090(5)           &  $0.082(8)$                          & $0.077^{+5}_{-7}$ \\ [0.25em]
\et{tbl:kappa-final}

\subsection{The rest mass and hyperfine splitting}  \label{sec:hfs}

In this section, we discuss the uncertainties in our calculation of the hyperfine splitting and compare our final results, for the $B_s$ and $D_s$ systems, with the PDG values.
To support the discussion, we tabulate our results for the pseudoscalar and vector meson rest masses  and the hyperfine splitting, $\aMps, \aMvm, \ahf, \ronehf$, in Tables~\ref{Tbl:M1-hyperfine-fine}--\ref{Tbl:M1-hyperfine-medcoarse} in Appendix~\ref{App_hyperfine}.
Statistical errors in these tables are the average 68\% bootstrap errors described in Sec.~\ref{sec:twopnts}. 
The other errors we consider are 
the mistuning of the valence strange-quark mass, 
unphysical sea-quark masses,  
the uncertainty in the tuning of $\kappa$,
discretization effects,
and the conversion to physical units.
For the central value, at each lattice spacing, we take $\ahf$ at the 
tuned values of $\kpch$ and $\kpbot$, linearly interpolating in $\kappa$ when necessary.

PDG results for the hyperfine splitting show a weak dependence on the light-quark valence mass, so we expect the mistuning in the simulated valence strange-quark mass to have a negligible effect.%
\footnote{For $X = B$ or $D$, the difference between the $M_{X_s^*}$-$M_{X_s}$ splitting and the $M_{X^*}$-$X_{X}$ splitting is measured to be about 1\% or less~\cite{PDG06}.}
The simulation valence masses  $\amq = 0.0272, 0.03, 0.0484$ for the fine, coarse, and medium-coarse
lattices, respectively, differ from the physical $am_s$ given in Table~\ref{Tbl:vacuum} by $0.0020, 0.0044, 0.0058$, respectively.
Tables~\ref{Tbl:M1-hyperfine-fine}--\ref{Tbl:M1-hyperfine-medcoarse} show that, indeed, these small mistunings have a negligible effect on the hyperfine splitting.  Hence, we do not interpolate to $am_s$; rather, we take $\ahf$ at the valence masses \amq~listed above as the result at the physical strange valence-quark mass and take the error for this approximation to be negligible.

To estimate the error due to the non-physical values of the sea-quark masses we use partially-quenched chiral perturbation theory.
The needed expression is derived in Appendix~\ref{App_hfsPQChiPT} and we repeat Eq.~(\ref{eq:pqchipt}) here for convenience.   
The hyperfine splitting $M^*_x-M_x$ of a heavy-light meson with light-valence quark $x$ is 
\begin{equation}
M^*_x-M_x = \Delta -\frac{\Delta g_\pi^2}{8\pi^2 f^2} \delta_{\rm log} + 2\Delta^{(\sigma)}(2m_l+m_s) +  2\Delta^{(a)}m_x \ ,
\end{equation}
where $\delta_{\rm log}$ contains the chiral logs, $m_l$ and $m_s$ are the light and strange sea-quark masses, and $\Delta^{(\sigma)}$ and $\Delta^{(a)}$ are counter terms which must be determined from the lattice data.
Working at a fixed value of $m_x$, we can use the difference of splittings at different values of $m_l$  to determine  $\Delta^{(\sigma)}$.
Given $\Delta^{(\sigma)}$, we can find the difference between the splitting at simulation values of $(m'_l, m'_s)$ and the physical values $(m_{l, \rm phys}, m_{s, \rm phys})$.
We take this difference as the error due to the non-physical sea-quark masses.

We have tabulated values of the hyperfine splitting in physical units, \ronehf, in Appendix~\ref{App_hyperfine:r1}.  
Figure~\ref{fig:rhf-sea} shows how \ronehf~varies with the light sea-quark mass on fine and coarse lattices.
From Fig.~\ref{fig:rhf-sea}, it is clear that, due to statistical variation in the splitting, using the difference in the central values of splittings from any two points will yield  different values for $\Delta^{(\sigma)}$.  
For the fine and coarse ensembles, we look only at the $am_l/am_s =$ 0.4 to 0.1 and $am_l/am_s =$ 0.4 to 0.2 differences and take the one that gives the larger error; for medium coarse, we have no $am_l/am_s = 0.1$ data and so take the error from the $am_l/am_s =$ 0.4 to 0.2 difference.

For the error estimate, we take $f = 131$~MeV and $g_\pi = 0.51$~\cite{Arnesen:2005ez}.
We relate meson to quark masses by 
\be
	M_{xy}^2 = B_0 (m_x + m_y)
\ee
where $B_0$ is determined empirically with $r_1 B_0 = 6.38, 6.23, 6.43$ on the fine, coarse, and medium-coarse lattices, respectively.   
These values of $B_0$ come from tree-level fits to MILC light-meson data, as described in Refs.~\cite{Aubin:2004ck, Bazavov:2009bb,Bernard:2007ps}.
We calculate $\Delta^{(\sigma)}$ for each meson type, $B_s$ and $D_s$, at each lattice spacing. 
We then calculate the difference 
\be
	(M^*_x-M_x)_{\rm  sim} - (M^*_x-M_x)_{\rm phys}
\ee
where the subscript ``sim'' (``phys'') denotes simulation (physical) sea-quark mass inputs $(am_l, am_s)$.
For the physical masses, we use $(am_{l, \rm phys}, am_{s, \rm phys}) = (0.00092, 0.0252), (0.00125, 0.0344), (0.00154, 0.0426)$ for the fine, coarse, and medium-coarse lattices, respectively.  These values of the quark masses are taken from Ref.~\cite{Bazavov:2009bb}, after adjustment for the $r_1$ scale used here.  
The simulation masses are those on the  the (0.0062, 0.031) fine, (0.007, 0.050) coarse, and (0.097, 0.0484) medium-coarse ensembles.  
The error calculated in this manner is labeled  ``sea-quark masses'' in Tables~\ref{Tbl:hyperfine_error_Ds} and~\ref{Tbl:hyperfine_error_Bs}.    
 
For the uncertainty in \ahf\ due to the error in \kp, recall that the non-negligible sources of error in \kp, from Table~\ref{tbl:kappa-error} in Sec.~\ref{sec:tuning}, are statistics, units conversion, and discretization error in $M_2$.  
Because we want to consider discretization errors separately from all others, 
we start by considering only the \kp-tuning error that comes from statistics and units-conversion.
To  convert  the error in \kp\ to an error in \ahf, we look at the change in \ahf between two values of \kp\ on the (0.0062, 0.031) fine, (0.007,
0.050) coarse, and (0.0097, 0.0484) medium-coarse ensembles; specific values can be found in Tables~\ref{Tbl:M1-hyperfine-fine}--\ref{Tbl:M1-hyperfine-medcoarse}.
This is the error labeled  ``\kp\ tuning'' in Tables~\ref{Tbl:hyperfine_error_Ds} and~\ref{Tbl:hyperfine_error_Bs}.

\begin{figure}
	\begin{tabular}{c c}
		\includegraphics[scale=0.32]{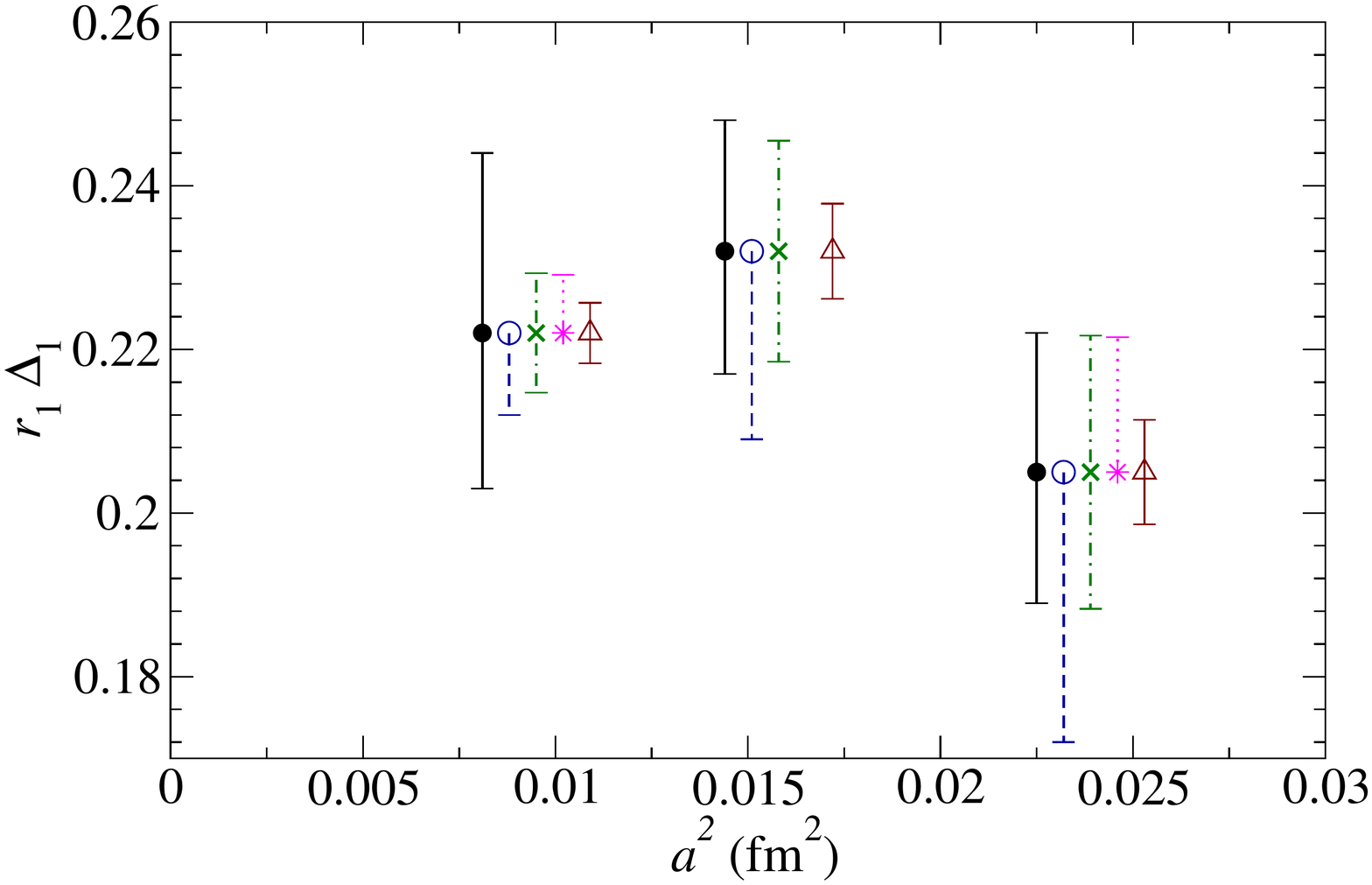}
		&
		\includegraphics[scale=0.32]{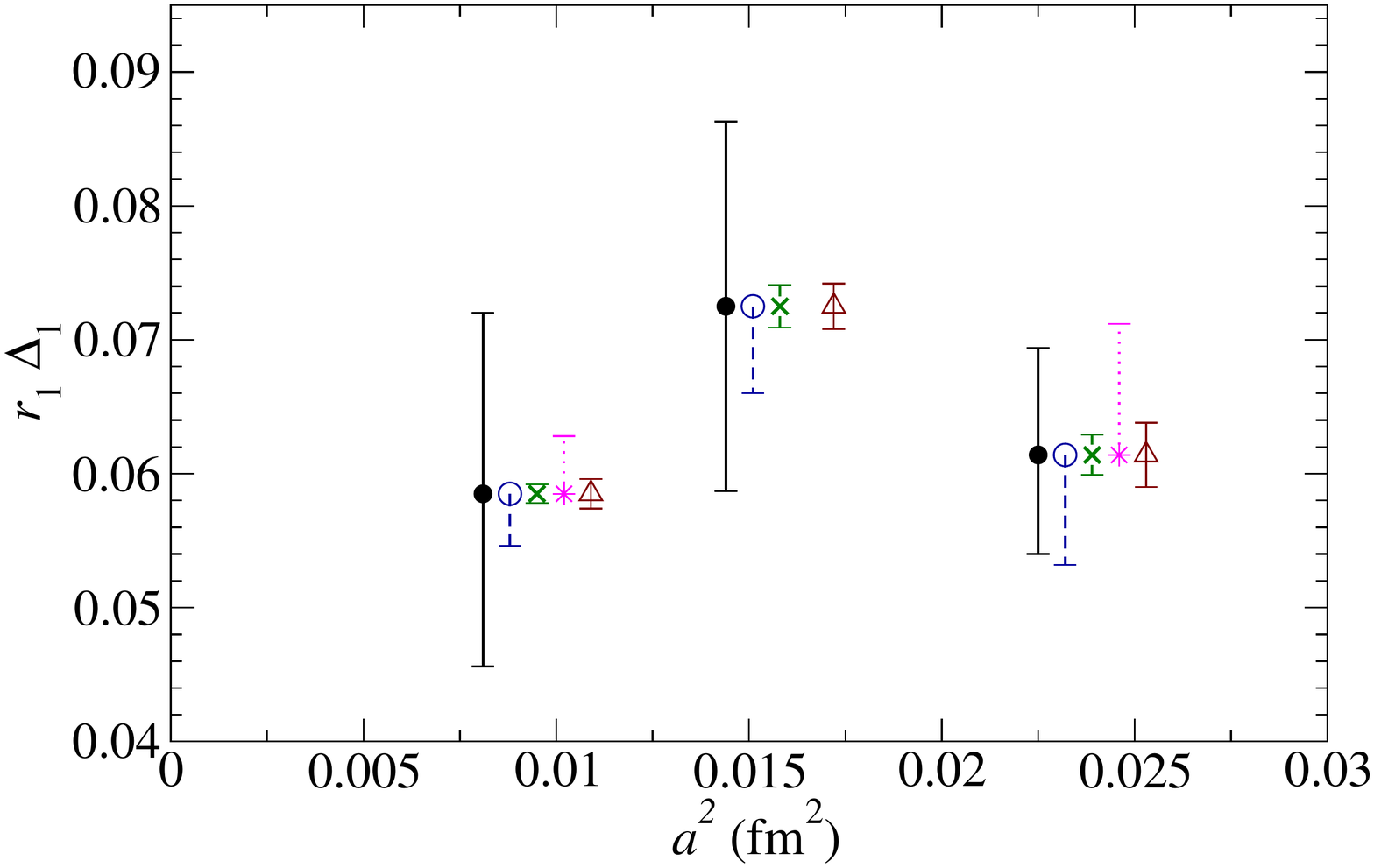}
		\\
		(a)  &  (b)  \\
	\end{tabular}
	\caption{Hyperfine splittings in $r_1$ units versus the squared lattice spacing $a^2$ (fm$^2$) for the (a) $D_s$ meson and (b) $B_s$ meson.
	Filled (black) dots with a solid error bar show the splitting with an error from all sources \emph{except} discretization.
	Open (blue) circles with a dashed error bar show the splitting with an error that also includes discretization error effects in \kp.
	(Green) X's with dash-dotted error bars show the estimated size of 
    discretization effects from the lattice-continuum mismatch of the 
    dimension-7 operator $\{i\bm{\sigma} \cdot \bm{B}, \bm{D}^2\}$ --- the errors are barely visible for the $B_s$ system.
	(Pink) stars with a dotted error bar show the $O(\alpha_s)$ discretization error from the 1-loop mismatch between $m_2$ and $m_B$.
	For the difference between the $O(\alpha_s)$ discretization effects on the coarse lattice versus the fine and medium-coarse lattices, see the text.}
	\protect\label{fig:hfs-discerr}
\end{figure}

For the $D_s$ ($B_s$) meson, Table~\ref{Tbl:hyperfine_error_Ds} (\ref{Tbl:hyperfine_error_Bs}) gives the error budget for \ahf\ at each lattice spacing, from all sources \emph{except} discretization.
These are statistics, valence-mass mistuning, unphysical sea-quark masses, and \kp\ tuning.  
In Fig.~\ref{fig:hfs-discerr}, these values are plotted as black, filled dots.

We now consider the three, distinct sources of discretization error in \ahf.
The first is indirect, coming from the discretization error in $aM_2$, which is propagated to an error on \kp\ as discussed in Sec.~\ref{sec:tuning}. 
This error can be traced to a mismatch between the spin-independent 
$O(\bm{p}^4)$ terms in Eq.~(\ref{eq:LHQET}) (not given explicitly) and the corresponding terms in the effective Lagrangian for continuum QCD.
These terms contribute to $\aMkbar$ as discussed in Appendix~\ref{App_M2DiscErr}.
The second source of discretization error is a direct result of the lattice-continuum mismatch of the dimension-seven operator $\{ i\bm{\sigma}\cdot \bm{B} , \bm{D}^2 \}$~\cite{Oktay:2008ex}.\footnote{Other dimension-six and -seven operators are either redundant, loop-suppressed, or known to have small coefficients~\cite{Oktay:2008ex}.} 
The third source of discretization error is the $O(\alpha_s)$ mismatch in the coefficient of the $i\bm{\sigma}\cdot \bm{B}$ operator in Eq.~(\ref{eq:L1}).  
For the discussion of error estimates below, it is useful to recall that the heavy-quark dynamics associate $m_2$ with the physical quark mass.  
Mismatches between $m_2$ and the generalized masses associated with other operators capture the heavy-quark discretization effects. 
We now give numerical estimates of the error from each source.

Our estimate of discretization error in $\aMkbar$ and its inclusion in the error on \kp\ is discussed in Sec.~\ref{sec:tuning}.
In Fig.~\ref{fig:hfs-discerr}, the value of \ronehf\ with an error that includes \emph{only} the uncertainty due to the discretization error on \kp\ is shown as an open (blue) circle with a dashed error bar.  
Note, as described in Sec.~\ref{sec:kinmass}, this uncertainty estimate depends on one's choice of $\Lambda_{\rm QCD}$.
In this paper, we use $\Lambda_{\rm QCD} = 0.7$~GeV.  Choosing $\Lambda_{\rm QCD} = 0.5$~GeV would cut the error on \kp\ in half and decrease the error on \ronehf.

Next we estimate the contribution from the dimension-seven operator $\{i\bm{\sigma}\cdot \bm{B} , \bm{D}^2 \}$.  
Using the notation of Ref.~\cite{Oktay:2008ex}, summarized in Sec.~\ref{sec:quarkmasses}, this operator's contribution to the hyperfine splitting has a coefficient
\be
	\frac{1}{(m_{B^{'}}a)^3} = \frac{1}{(m_{4}a)^3} ,
\ee
where the equality holds at the tree level for the choices of parameters in our action.
The difference between $am_4$ and $am_2$ captures the discretization error.
The fractional error in the hyperfine splitting due to this mismatch is
\be  \label{eq:dim7err}
	(a \Lambda_{\rm QCD})^2 2am_2 \left[ \frac{1}{(2am_4)^3}  - \frac{1}{(2am_2)^3} \right]  .
\ee
This error is plotted as a (green) dash-dot line on an X in Fig.~\ref{fig:hfs-discerr}.  
It would be added in quadrature with the error on the filled dot, if it were to be included in the total error.
Again we take $\Lambda_{\rm QCD} = 0.7$~GeV, but choosing $\Lambda_{\rm QCD} = 0.5$ would cut these error bars in half.  
The error from Eq.~(\ref{eq:dim7err}) is small for the $D_s$ splitting at the fine lattice spacing, but increasingly large and non-negligible at the coarse and medium-coarse lattice spacings; for the $B_s$ splitting, the error is negligible.

Finally, we turn to the effects of the $O(\alpha_s)$ mistuning in $c_B$, which leads to an $O(\alpha_s)$ mismatch between $m_B a$ and $m_2 a$.
Ideally, $c_B$ should be adjusted so the coefficient of $\bar{h}^{(+)}i\bm{\sigma}\cdot\bm{B}h^{(+)}$ equals $Z_B/2m_2$, where 
$Z_B$ is a coefficient with an anomalous dimension, such that $Z_B\bar{h}^{(+)}i\bm{\sigma}\cdot\bm{B}h^{(+)}$ is scale and scheme independent~\cite{Eichten:1990vp}.
In practice, $c_B$ is chosen in some approximation, in our case the tadpole-improved tree level of perturbation theory.

Given a value of $c_B$, our simulations produce
\be
	M_1^*-M_1=\Delta_1 =    \frac{4\lambda_2}{2m_B(c_B)}  .
\ee
From Eq.~(\ref{eq:mBa_app}), we see that $1/am_B$ has a contribution $c_B/(1+m_0a)$.
Hence, to include the leading correction to the hyperfine splitting, we shift
\be  \label{eq:Delta1cBmismatch}
	4\lambda_2 a  \left[ \frac{1}{2am_B(c_B)} \right]  \to   4\lambda_2 a  \left[ \frac{1}{2am_B(c_B)} + \frac{c_B^{\rm ideal}-c_B}{2(1+m_0a)} \right]
\ee
where $c_B^{\rm ideal}$ is the ideal choice.
(Because loop corrections to $1/am_B$ depend on $c_B$, subleading corrections also exist.)
To estimate the error in $\Delta_1$, we have to estimate $c_B^{\rm ideal}-c_B$.
In fact, Eq.~(\ref{eq:Delta1cBmismatch}) can also be used to shift the central value of the hyperfine splitting.

Reference~\cite{MNobes} describes preliminary work on a calculation of the one-loop corrections to $c_{B}^{[1]}$, as a function of the bare quark mass.
For all relevant values of $m_0a$, the one-loop effects are a small correction to the tadpole-improved Ansatz $c_B=u_0^{-3}$, provided that $u_0$ is the average link in Landau gauge.
On the coarse ensembles, we chose $u_0$ this way, and we can estimate the remaining correction directly from the calculation in Ref.~\cite{MNobes}.
Given further uncertainties from higher orders, we take this small correction as an uncertainty estimate.
On the medium-coarse and fine ensembles, however, we chose $u_0^4$ to be the average plaquette.
In those cases, the leading correction to $c_B$ comes from,   
\be
	 c_B^{\rm ideal}-c_B = u_{0, \rm LL}^{-3} - u_{0, \rm plaq}^{-3}
    \label{eq:u0swap}
\ee
where the labels refer to ``Landau-gauge link'' and ``plaquette.''
Equation~(\ref{eq:u0swap}) leads to significant corrections to the hyperfine splitting, so we shift $\Delta_1$ on the medium-coarse and fine ensembles by the amount corresponding to Eq.~(\ref{eq:Delta1cBmismatch}) and~(\ref{eq:u0swap}).
These shifts put $\Delta_1$ at the medium-coarse and fine lattice spacings on the same footing as those at the coarse spacing.
Empirically, they flatten the lattice-spacing dependence.

For the medium-coarse and fine data, we use the values of $u_0$ given in Table~\ref{tbl:u0} to calculate the shift described above.
It is displayed in Fig.~\ref{fig:hfs-discerr}  as a (pink) star with a single-sided, positive error bar.  
To obtain an error bar corresponding to the one-loop correction to $c_B$ in Ref.~\cite{MNobes}, 
we take $\alpha_s(0.09~\rm{fm}) = 1/3$ 
and use one-loop running to obtain values of $\alpha_s$ for the coarse and medium-coarse lattices.
These corrections are shown in in Fig.~\ref{fig:hfs-discerr} as a (red) triangle with a solid error bar.

In summary, discretization errors in the hyperfine splitting are small at the fine lattice spacing; therefore, we take as our final results the splittings calculated on the fine lattice.
In addition, since the effect of the leading $O(\alpha_s)$ mistuning of $c_B$ can be quantified, we shift our final central values by this amount.
All other discretization errors are included in our final error.
%
We convert our results to physical units using the values of $r_1/a$ and $r_1$ as listed in Table~\ref{Tbl:vacuum}.
After including the error from the units conversion in the total, our final results for the hyperfine splittings are 
\bea  
	\Delta_{D_s}    &=&  145 \pm15 \text{ MeV}	\label{eq:hf-final-D} \\    
	\Delta_{B_s}    &=&  \;\; 40 \pm 9  \text{ MeV}	\label{eq:hf-final-B}    
\eea
These results are in good agreement with the PDG values of 
$143.9\pm0.4$~MeV and $46.1\pm1.5$~MeV, respectively.

\begin{figure}
\begin{tabular}{c c}

\includegraphics[scale=0.3]{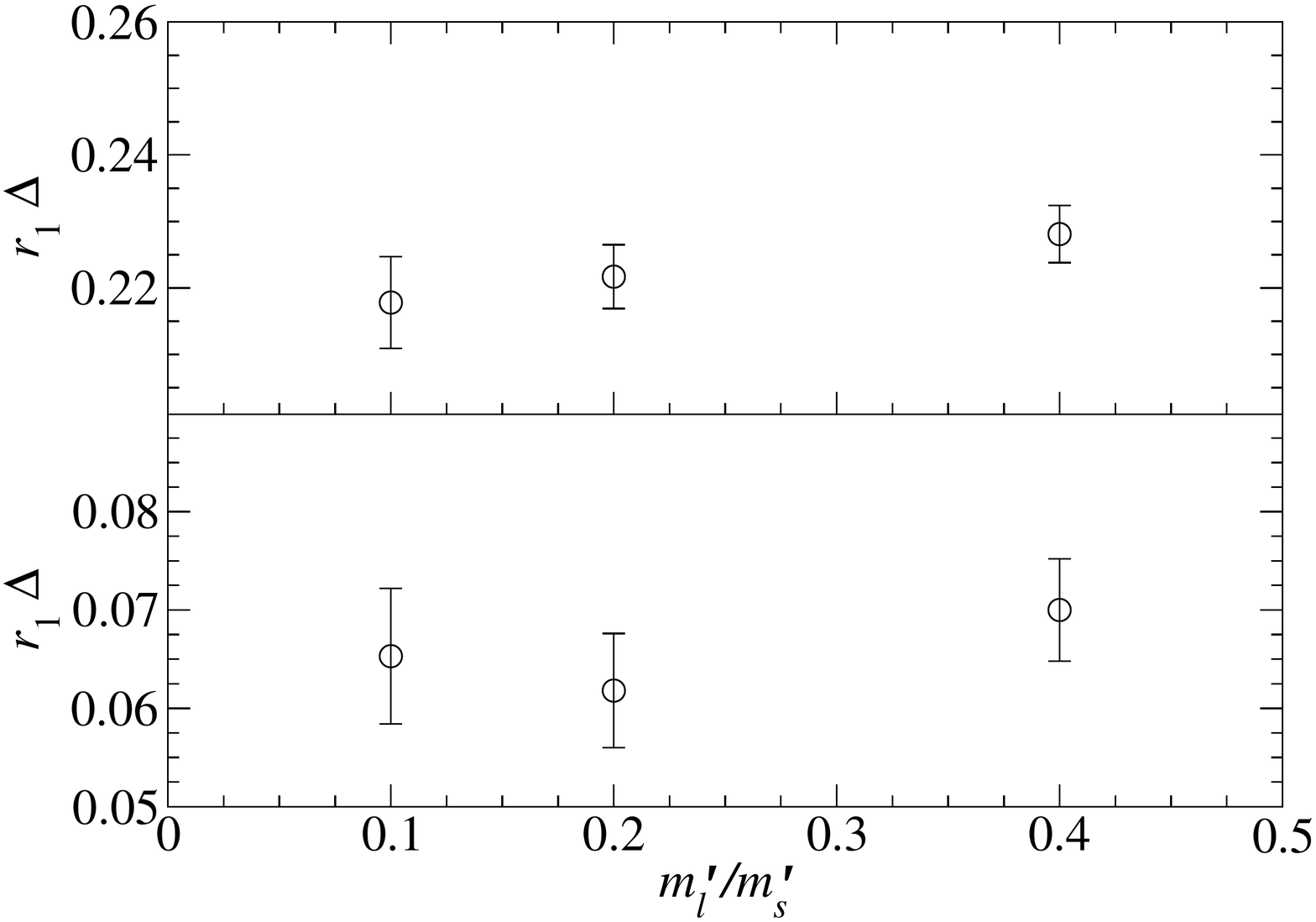}
	&

\includegraphics[scale=0.3]{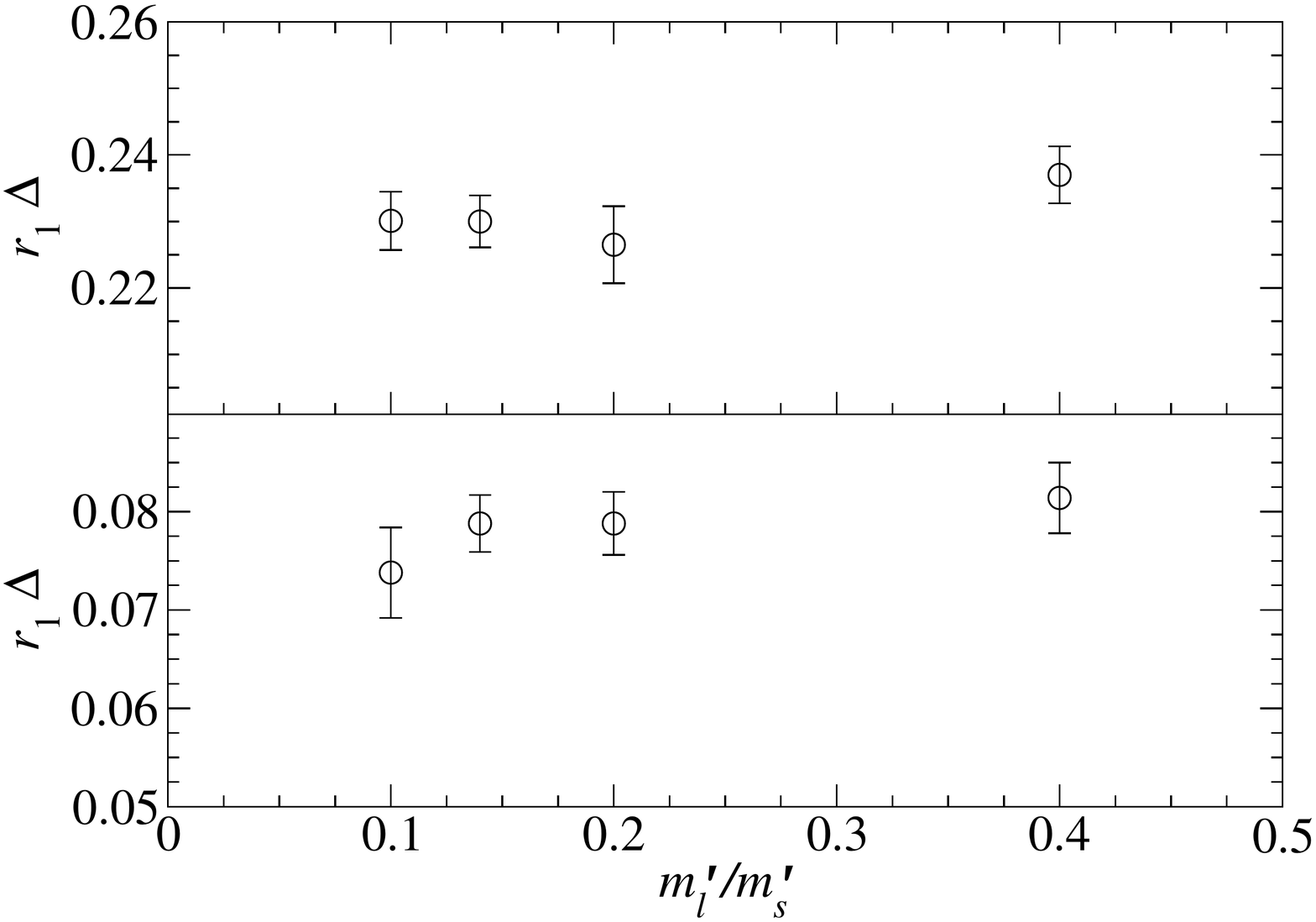}
	 \\
	 (a)  & (b) \\
\end{tabular}
\caption{The hyperfine splitting, in units of $r_1$, versus the ratio
of the light to strange sea-quark masses $m'_l/m'_s$ on (a) fine and
(b) coarse ensembles. 
Errors are  the average 68\% bootstrap error.
The upper panel in each plot is for charm-like splittings and the lower
panel is for bottom-like splittings.
Values of  \kp~are 0.127, 0.0923 for the fine ensembles and 0.122,
0.086 for coarse ensembles.
Values of \amq~are 0.0272 and 0.0415 for the fine and coarse ensembles,
respectively.}
\protect\label{fig:rhf-sea}
\end{figure}

\bt{Percent errors in the hyperfine splitting, \ahf, of $D_s$ \emph{not} including discretization effects.}{ l   c c c      }    
	 Uncertainty                              &   Fine 	                &    Coarse                             &    Medium-coarse     \\ [0.25em]
	\hline
	Statistical                                  &   2.2	                &   1.9                                     &    1.9    \\
	$\kappa$ tuning                        &   $(8.8, -7.5)$	&   $(4.0, -3.1)$                       &    $(4.0, -2.7)$  \\
	Valence $m_s$                         &       0		        &       0                                     &       0       \\
	Sea-quark masses                    &  3.6		        &   5.4                                      &       6.9   \\    
	\hline 
	Total                                          & $(10, -9)$		& $(7, -7)$                                &   $(8, -8)$   \\ 
\et{Tbl:hyperfine_error_Ds} 
\bt{Percent errors in the hyperfine splitting, \ahf, of $B_s$ \emph{not} including discretization effects.}{ l   c c c    }   

	 Uncertainty                               &   Fine          &    Coarse                     &    Medium-coarse     \\
[0.25em]
	 \hline
	Statistical                                   &   9.5	       &   4.0                             &   5.6     \\
	$\kappa$ tuning                         &   $(12, -11)$	&   $(17, -17)$               &  $(11, -10)$    \\ 
	Valence $m_s$                          &       0		&       0                           &       0       \\
	Sea-quark masses                     &      17		&    7.8                           &   2.6   \\ 
	\hline
	Total                                           &  $(23, -22)$	& $(19, -19)$                 & $(13, -12)$ \\
\et
{Tbl:hyperfine_error_Bs}

\bt{Tadpole-improvement factors for the estimate of the  $O(\alpha_s)$ discretization error shown in Fig.~\ref{fig:hfs-discerr}.   }
       {l  c   c}
	ensemble                                           &      $u_{0, \rm plaquette}$   &   $u_{0, \rm Landau}$  \\
	\hline
	fine  (0.0062, 0.031)                          &      0.878                              &   0.854   \\
	medium-coarse (0.0097, 0.0484)      &      0.860                              &  0.822  \\
\et{tbl:u0}


\subsection{The critical hopping parameter $\kc$}
\label{sec:kcrit}

In principle, it is possible to carry out a suite of nonperturbative heavy-quark calculations without knowing \kc, but in practice \kc\ is useful.
In particular, it enters the construction of improved bilinear and 4-quark operators via $m_0a$ Eq.~(\ref{eq:baremass}).
It also enters the computation of matching factors such as $Z_V$ and $Z_A$ \cite{HQETradiative}.
Note that these all amount to small corrections, so we do not need a very precise determination of \kc.
Equation~(\ref{eq:baremass}) shows that it does not have to be much better determined than \kpch\ and \kpbot.

A~nonperturbative definition of \kc\ is the value of $\kappa$ such that 
the mass of a pseudoscalar meson consisting of two Wilson quarks (with the clover action) vanishes.
The computation of these light-light pseudoscalar meson masses shares code with the work reported 
here and in Ref.~\cite{Burch:2009az}, and it is convenient to report the analysis here.
The value of \kc\ depends on $u_0$ via our choice of clover coupling, $c_B=c_E=u_0^{-3}$. 
In this and other work~\cite{Aubin:2005ar,Aubin:2004ej,Evans:2008zz,Freeland:2007wk,Allison:2004be,Burch:2009az}, $u_0$ has been set 
sometimes from the average plaquette and sometimes from the average link in Landau gauge.
The prescription for $u_0$ used in each \kc\ determination is given in column four of Table~\ref{kcrit}. 

\begin{table}
	\caption{Values of $\kc$ by ensemble.
	``$u_0$ used'' gives the origin of the $u_0$ value used in the $\kc$ determination.
	\kc\ values are given in two columns.  
	The first \kc\ column contains values which were determined by a fit.
	The second \kc\ column contains values which were estimated from fitted values at the same (approximate) lattice spacing. 
	The last column gives the fit method used in the determination, explained in the text.}
	\begin{tabular}{l l     c     ccc}
		\hline \hline 
	&                             &                                   &      \mc{3}{c}{\kc}       \\
	Lattice 
        & $(am'_l, am'_s)$  &  $u_0$ used             &  Iterated fit  & Direct fit   & Estimated   \\
		   \hline
		   Fine    
	& (0.0031, 0.031) & Landau-gauge link    &                  &                & $0.1372$     \\
	& (0.0062, 0.031) & Landau-gauge link       & $0.1372$   &                     &   \\   
	& (0.0062, 0.031) &    plaquette            & $0.1391$      &                  &   \\ 
	& (0.0124, 0.031) & Landau-gauge link       & $0.1372$  &                     &    \\
		   Coarse   
	& (0.005, 0.050)   &    Landau-gauge link      &                     &                & $0.1379$  \\
	& (0.007, 0.050)   &    Landau-gauge link      &                        &             & $0.1379$  \\
	& (0.010, 0.050)   & Landau-gauge link        & $0.1379$  &                    &         \\
	& (0.020, 0.050)   & Landau-gauge link        & $0.1378$    &                  &   \\
	& (0.030, 0.050)   & Landau-gauge link        & $0.1377$  &                    &         \\   
		   Medium-coarse
	& (0.0097, 0.0484) &   plaquette                &          & $0.1424$                        &    \\
	& (0.0194, 0.0484) &   plaquette                 &         & $0.1424$               &    \\   
	& (0.0290, 0.0484) &   plaquette                 &          & $0.1423$              &   \\    
		\hline \hline
	\end{tabular}
	\label{kcrit}
\end{table}

The determination of \kc\ is carried out on a subset of the available 
configurations, 50--100 configurations for the fine ensembles and 400--600 for the coarse and medium-coarse.
We compute two-point correlators for a range of $\kappa$ that yields meson masses 
of about $M_{\rm PS}=450$--$900$~MeV on the fine ensembles, 650--1100~MeV on 
the coarse ensembles, and 550--950~MeV on the medium-coarse ensembles.
It is impractical to push to lower $M_{\rm PS}$ due to exceptional configurations.
$M_{\rm PS}$ is a function of the quark mass, which we parametrize as the tree-level, tadpole-improved kinetic or rest mass.
In the relevant region, $m_{1,2}a=m_0a [1 -\frac{1}{2}m_0a] + O((m_0a)^3)$, so both pertain equally well.
The meson masses can be fit to a polynomial ansatz 
\be
	a^2 M_{\rm PS}^2 (\kappa) = A + B a m_2(\kp, \kc) + C a^2 m_2^2 (\kp, \kc)  
    \label{eq:ps4kcrit}
\ee
(or $m_1$ instead of $m_2$), where $A=0$ when \kc\ is correctly adjusted.

We use two techniques to determine \kc.
One method starts with a reasonable value of \kc\ and fits Eq.~(\ref{eq:ps4kcrit})
to obtain $A$, $B$, and $C$, which depend implicitly on $\kc$.
A better trial value of \kc\ is chosen, and the process is iterated until a \kc\ is found such that $A=0$.
We call this the ``iterated fit''.
The second method freezes $A$ to zero, and then $B$, $C$, and \kc\ are the fit parameters.
We call this the ``direct fit''.
On several ensembles the \kc\ values were simply estimated from the other ensembles with
the same (approximate) lattice spacing, these are labeled as ``estimated''.


Table~\ref{kcrit} contains our results for \kc, indicating the method used.
The table does not include error bars for \kc, but we believe that the results are correct to the number of significant figures shown, even though the range of $M_{\rm PS}$ is high.
We carried out several tests to verify this accuracy.  
We compared linear iterated fits [i.e., $C=0$ in Eq.~(\ref{eq:ps4kcrit})] to the baseline quadratic.
We also compared direct fits with and without the (continuum) chiral log.
These test show that higher order or log contributions do not alter our values of \kc\ significantly.
We fit comparable data with staggered valence quarks allowing $(m_0a)_{\rm crit}\neq0$, thereby testing whether a range of such large $M_{\rm PS}$ skews the results.
None of these tests suggests an error larger than a few in the fourth digit.
Such errors are negligible compared to those for 
\kpch\ and \kpbot---see Tables~\ref{tbl:kappa-error} and~\ref{tbl:kappa-final}---when forming $m_0a$ with Eq.~(\ref{eq:baremass}).

%% file: Future_Conclusions.tex

An accurate and precise determination of 
$\kpch$ and $\kpbot$ is important for all
calculations using the Fermilab action~\cite{Aubin:2005ar,Aubin:2004ej,%
Evans:2008zz,Freeland:2007wk,Allison:2004be,Burch:2009az}.
In this analysis, the error on $\kpbot$ is dominated
by statistics, and the error on $\kpch$  receives approximately equal 
contributions from statistics and discretization effects.
These errors play a significant role on quantities as diverse as $D$- 
and $B$-meson decay constants~\cite{Aubin:2005ar} and the quarkonium 
hyperfine splitting~\cite{Burch:2009az}.
Our final results for $\kpch$ and $\kpbot$ are given in 
Table~\ref{tbl:kappa-final}.

Another ingredient that is useful for matrix elements~%
\cite{Aubin:2005ar,Aubin:2004ej,Evans:2008zz} is the additive 
renormalization of the bare quark mass or, equivalently, \kc.
The improvement and matching of the operators needed to compute these 
matrix elements depends mildly on \kc\ via $m_0a$ \cite{HQETradiative}.
Our final results for \kc\ are given in Table~\ref{kcrit}.

The key ingredient needed to determine
$\kpch$ and $\kpbot$ is a computation of the pseudoscalar 
and vector heavy-strange meson masses.
These can be combined to yield the hyperfine splitting for $D_s$ and $B_s$ mesons.
Our final results for the hyperfine splittings are given in Eqs.~(\ref{eq:hf-final-D}) and~(\ref{eq:hf-final-B}). 
Both are in good agreement with the corresponding PDG averages.
These results bolster confidence in the tuning of $\kpch$ and $\kpbot$,
as well as the choice $c_B=u_0^{-3}$.
Further tests of these choices come from related calculations of the quarkonium spectrum~\cite{Burch:2009az}.
With detailed attention given to the connection between action parameters and mass splittings, those results are found to be consistent with experiment within the expected uncertainties.

Improved determinations of $\kpch$, $\kpbot$, and 
\kc\ for the medium-coarse, coarse, and fine ensembles are underway with higher statistics, as well as calculations on the new 
superfine ($a\approx 0.06$~fm), and ultrafine ($a\approx 0.045$~fm) lattices.   
The increased statistics will also allow us to use higher momentum data and
fit to the $O(p^4)$ terms in the dispersion relation.
Refinements in the determination and use of $r_1/a$ are allowing for a
better understanding of sea-quark effects which will be needed as the
statistical error on $aM_2$ decreases.
We are also investigating the use of twisted boundary conditions~\cite{twistedBC} which will
allow us to obtain data points at lower momenta.

As uncertainties in $M_2$ and $M_1$ decrease, there will be a need for
a better understanding of the chiral behavior of these masses.
One-loop, $O(\Lambda/m_Q)$ chiral perturbation theory results exist for continuum QCD~\cite{Jenkins:1992hx}.  
The extension to staggered chiral perturbation theory should be straightforward, and would allow us to extrapolate the light-valence mass to the physical up/down quark mass and determine the hyperfine splittings of the $B^\pm$ and $D^\pm$ mesons.
In this paper, we have included the partially quenched expression for the hyperfine splitting in Appendix~\ref{App_hfsPQChiPT}, since it is useful in estimating uncertainties from the unphysical sea-quark masses.

In addition, tuned values of $\kpch$, $\kpbot$, and $\kc$ combined with one-loop (lattice) perturbation theory can yield determinations of the pole masses $m_1$ and $m_2$ for both charmed and bottom quarks\footnote{The determination of the quark mass from $m_1$ requires a non-perturbative calculation of the binding energy as defined by $M_1 - m_1$~\cite{Freeland:2007wk}.}, 
which can be converted  to the potential-subtracted, $\overline{\rm MS}$, and other schemes~\cite{Freeland:2007wk}.
Quark masses combined with staggered chiral perturbation theory for the $B^\pm$ and $D^\pm$ mesons, can yield \emph{ab initio} calculations of HQET matrix elements~\cite{Kronfeld:2000gk, Freeland:2006nd}, which are used to calculate the Cabibbo-Kobayashi-Maskawa matrix element $|V_{cb}|$ via inclusive decay measurements.
Finally, improved determinations of the oscillating-state energy $E^p$ could make determinations of the experimentally accessible masses of the
positive parity states, $D^*_{s0}(2317)$ and
$D_{s1}(2460)$~\cite{Aubert:2006bk} a viable
option~\cite{diPierro:2003iw}.

%% file: App_M2DiscErr.tex
\section{Discretization Error in the Kinetic Meson Mass}
\label{App_M2DiscErr}

In this appendix, we present a semi-quantitative estimation of the
discretization error in the kinetic mass of heavy-light hadrons.
We use a formalism that applies when both quarks are non-relativistic,
even though this approximation is not good for the light quark in a
heavy-light meson.
A posteriori, we examine two ways to re-interpret the resulting formula
for a relativistic light quark.
Both estimates are numerically the same, so we proceed to use the
formula in Sec.~\ref{sec:kinmass}.

In what follows, the generalized masses $m_1, m_2, m_4$ and the
coefficient $w_4$ are used to describe the discretization errors.
Expressions for them when using the Fermilab action are in
Refs.~\cite{El-Khadra:1996mp} or~\cite{Oktay:2008ex} and are given at
the end of this appendix for convenience.
We assume that the light quark ($s$) has a mass in lattice units
$m_sa\ll1$ and makes no significant contribution to the discretization
error.

The bound state's kinetic mass can be read off from its kinetic energy
(by definition).
It will have a kinematic contribution, from the constituents' kinetic
energy, and a dynamical contribution, from the interaction that binds
the constituents.
We consider each in turn.

\subsection{Contributions from constituents' kinetic energy}

The hadron of interest is a heavy-strange meson, a bound state of a
heavy quark~$Q$ (momentum~$\bm{Q}$) and a strange antiquark $s$
(momentum~$\bm{s}$).  The non-relativistic kinetic energy is
\begin{equation}
	T =	m_{1Q} + \frac{\bm{Q}^2}{2m_{2Q}} - 
		\frac{(\bm{Q}^2)^2}{8m_{4Q}^3} - \sixth w_{4Q}a^3\sum_iQ_i^4 +
		m_{1s} + \frac{\bm{s}^2}{2m_{2s}} - 
		\frac{(\bm{s}^2)^2}{8m_{4s}^3} - \sixth w_{4s}a^3\sum_is_i^4 .
	\label{eq:T}
\end{equation}
The binding energy is communicated to the bound-state kinetic mass via
the terms quartic in the momenta 
and via corrections to the potential, given below.
In general, the lattice breaks relativistic invariance, so $m_1\neq
m_2\neq m_4$, $w_4\neq 0$.
Re-writing the kinetic energy in center-of-mass coordinates
\begin{eqnarray}
	\bm{Q} & = & \frac{m_{2Q}}{m_{2Q}+m_{2s}}\bm{P} + \bm{p} ,\\
	\bm{s} & = & \frac{m_{2s}}{m_{2Q}+m_{2s}}\bm{P} - \bm{p} ,\\
	\bm{P} & = & \bm{Q} + \bm{s} ,\\
	\bm{p} & = & \frac{m_{2s}\bm{Q} - m_{2Q}\bm{s}}{m_{2Q}+m_{2s}},
\end{eqnarray}
one finds
\begin{eqnarray}
	T & = & m_{1Q} + m_{1s} + \frac{\bm{P}^2}{2(m_{2Q}+m_{2s})} + 
		\frac{\bm{p}^2}{2\mu_2} -
		\frac{\bm{P}^2\,\bm{p}^2 +2(\bm{P}\cdot\bm{p})^2}{4(m_{2Q}+m_{2s})^2}
			\left[\frac{m_{2Q}^2}{m_{4Q}^3} + \frac{m_{2s}^2}{m_{4s}^3} 
			\right] \nonumber \\
	  & - & a^3 \sum_i \frac{P_i^2p_i^2}{(m_{2Q}+m_{2s})^2}
			\left(w_{4Q}m_{2Q}^2+w_{4s}m_{2s}^2\right) + \cdots, 
		\label{eq:BSkinetic} \\
	\frac{1}{\mu_2} & = & \frac{1}{m_{2Q}} + \frac{1}{m_{2s}} .
\end{eqnarray}
The only quartic terms shown are those quadratic in $\bm{P}$; the
omitted terms are not smaller; they just do not contribute to the bound
state's kinetic energy.
The objective is to collect all terms quadratic in $\bm{P}$, because
their overall coefficient will yield the bound state's kinetic mass.

\subsection{Contribution from the interaction: Breit equation}

To obtain the two-particle system's potential energy, one has to work
out the scattering amplitude from one-gluon exchange, obtaining an
expression called the Breit equation~\cite{Berestetskii:1971vol6,
Kronfeld:1996uy}.

In momentum space, for the color-singlet channel
\begin{eqnarray}
	V(\bm{K}) & = & -C_Fg^2 D_{\mu\nu}(K) \mathcal{N}_Q(\bm{Q}+\bm{K})
		\bar{u}(\xi',\bm{Q}+\bm{K}) \Lambda_Q^\mu(Q+K,Q) u(\xi,\bm{Q})
		\mathcal{N}_Q(\bm{Q}) \nonumber \\
		& & \times \mathcal{N}_s(\bm{s})
		\bar{v}(\xi,\bm{s}) \Lambda_s^\nu(s,s-K) v(\xi',\bm{s}-\bm{K})
		\mathcal{N}_s(\bm{s}-\bm{K}),
	\label{eq:V(K)}
\end{eqnarray}
where $D_{\mu\nu}$ is the (lattice) gluon propagator, $\Lambda_q^\mu$
is the lattice vertex function (for $q=Q,s$), and $\mathcal{N}_q$ is an
external-line factor needed with the normalization conditions on
spinors employed here~\cite{El-Khadra:1996mp,Oktay:2008ex}.
(In continuum field theory, $\mathcal{N}=\sqrt{m/E}$.)

To the accuracy needed here, the gluon propagator can be replaced with
the continuum propagator.
The heavy-quark line is
\begin{eqnarray}
	J^4_Q & = & \mathcal{N}_Q(\bm{Q}+\bm{K})
		\bar{u}(\xi',\bm{Q}+\bm{K}) \Lambda_Q^4(Q+K,Q) u(\xi,\bm{Q})
		\mathcal{N}_Q(\bm{Q}) \nonumber \\
		& = & 
		\bar{u}(\xi',\bm{0}) \left[1 - 
		\frac{\bm{K}^2 - 2i\bm{\Sigma}\cdot(\bm{K}\times\bm{Q})}{8m_{EQ}^2} 
		+ \cdots \right] u(\xi,\bm{0}), \label{eq:J4lat} \\
	\bm{J}_Q & = & \mathcal{N}_Q(\bm{Q}+\bm{K})
		\bar{u}(\xi',\bm{Q}+\bm{K}) \bm{\Lambda}_Q(Q+K,Q) u(\xi,\bm{Q})
		\mathcal{N}_Q(\bm{Q}) \nonumber \\
		& = & -i\bar{u}(\xi',\bm{0})\left[
			\frac{\bm{Q}+\half\bm{K}}{m_{2Q}} +
			\frac{i\bm{\Sigma}\times\bm{K}}{2m_{BQ}}
		+ \cdots \right] u(\xi,\bm{0}),  \label{eq:Jilat}
\end{eqnarray}
and, to the extent that the strange antiquark is non-relativistic, one
has a similar expression for the antiquark line 
$J^\nu_s=\mathcal{N}_s(\bm{s}) \bar{v}(\xi,\bm{s}) \Lambda_s^\nu(s,s-K)
v(\xi',\bm{s}-\bm{K}) \mathcal{N}_s(\bm{s}-\bm{K})$.

In Coulomb gauge,
\begin{equation}
	D_{44}(K) = \frac{1}{\bm{K}^2}, \quad\quad
	D_{ij}(K) = 
	\frac{1}{K^2}\left(\delta^{ij}-\frac{K^iK^j}{\bm{K}^2}\right),
\end{equation}
and the other components vanish.  Thus, noting that
$K_4=i[(\bm{Q}+\bm{K})^2-\bm{Q}^2]/2m_Q$ is subleading,
\begin{eqnarray}   
	V(\bm{K}) & = & - C_Fg^2
		\left[\frac{1}{\bm{K}^2} - 
			\left(\frac{1}{8m_{EQ}^2} + \frac{1}{8m_{Es}^2}\right) -
			\frac{1}{m_{2Q}m_{2s}} \left( \bm{Q}\cdot\bm{s} - 
				\frac{\bm{Q}\cdot\bm{K}\bm{K}\cdot\bm{s}}{\bm{K}^2} 
			\right) \frac{1}{\bm{K}^2} 
		\right] \nonumber   \label{eq:V(K)-nospin} \\ [1.0em]
		& + & \textrm{spin-dependent terms}. 
\end{eqnarray}
Let us discuss each part of the bracket in turn.
The leading term yields, after Fourier transforming to position space,
the $1/r$ potential.
The second yields a contact term proportional to~$\delta(\bm{r})$: it
is a relativistic correction to the bound state's \emph{rest} mass, so
it is of no further interest here.
Similarly, the spin-dependent terms do not contribute to the bound
state's kinetic energy, so they are not written out.
The remaining exhibited contributions do contribute to the bound
state's kinetic energy, when $\bm{Q}$ and $\bm{s}$ are eliminated in
favor of $\bm{P}$ and $\bm{p}$.

Next we Fourier transform from $\bm{K}$ to $\bm{r}$ using
\begin{eqnarray}  
	\int\frac{d^3K}{(2\pi)^3} \frac{e^{i\bm{r}\cdot\bm{K}}}{\bm{K}^2} 
		& = & \frac{1}{4\pi r}, 
	\label{eq:V} \\
\int\frac{d^3K}{(2\pi)^3}\frac{K_iK_je^{i\bm{r}\cdot\bm{K}}}{(\bm{K}^2)^2}
		& = & \half(\delta_{ij}+r_i\nabla_j)
		\int\frac{d^3K}{(2\pi)^3}\frac{e^{i\bm{r}\cdot\bm{K}}}{\bm{K}^2}.
	\label{eq:virial}
\end{eqnarray}
Following with the substitution of $\bm{P}$ and $\bm{p}$ for $\bm{Q}$
and $\bm{s}$ this yields
\begin{equation}
	V(\bm{r},\bm{P},\bm{p}) = - \frac{C_F\alpha_s}{r} \left[1 -
		\frac{\bm{P}^2}{2(m_{2Q}+m_{2s})^2} \right] -
			r_i\nabla_j\frac{C_F\alpha_s}{r}
			\frac{P_iP_j}{2(m_{2Q}+m_{2s})^2} + \cdots,
\end{equation}
where the omitted terms do not influence the bound state's kinetic
energy.  

Note that $\bm{K}$ changes $\bm{p}$ but not $\bm{P}$, so $\bm{r}$ is
conjugate to $\bm{p}$.
To take expectation values, we use the virial theorem
\begin{equation}
	\langle r_i\nabla_jV(\bm{r})\rangle = 
		\frac{\langle p_ip_j\rangle}{\mu_2},
\end{equation}
so the total energy of the bound state, $E(\bm{P})=\langle T+V\rangle$,
is
\begin{eqnarray}
	E(\bm{P}) & = & m_{1Q} + m_{1s} + 
		\frac{\langle\bm{p}^2\rangle}{2\mu_2} -
		\left\langle\frac{C_F\alpha_s}{r}\right\rangle 
	\nonumber \\
	 & + & \frac{\bm{P}^2}{2(m_{2Q}+m_{2s})} \left[1 -
		\frac{\langle\bm{p}^2\rangle}{2\mu_2(m_{2Q}+m_{2s})} +
		\frac{1}{(m_{2Q}+m_{2s})}
			\left\langle\frac{C_F\alpha_s}{r}\right\rangle \right]
	\nonumber \\
	 & + & \frac{\bm{P}^2}{2(m_{2Q}+m_{2s})^2}
		\frac{\langle\bm{p}^2\rangle}{2\mu_2} 
			\left[1 - \mu_2
			\left(\frac{m_{2Q}^2}{m_{4Q}^3} + \frac{m_{2s}^2}{m_{4s}^3} 
			\right) \right]
	\nonumber \\
	 & + & \frac{P_iP_j}{(m_{2Q}+m_{2s})^2} 
		\frac{\langle p_ip_j\rangle}{2\mu_2}
			\left[1 -\mu_2
				\left(\frac{m_{2Q}^2}{m_{4Q}^3} + \frac{m_{2s}^2}{m_{4s}^3} 
				\right)\right] \nonumber \\
	 & - & a^3 \sum_i \frac{P_i^2\langle p_i^2\rangle}{(m_{2Q}+m_{2s})^2}
			\left(w_{4Q}m_{2Q}^2+w_{4s}m_{2s}^2\right) + \cdots.
	\label{eq:E(P)}
\end{eqnarray}
The first line of Eq.~(\ref{eq:E(P)}) shows the binding energy adding
to the quarks' rest masses to form the bound state's rest mass,
\begin{equation}
	M_1 = m_{1Q} + m_{1s} + 
		\frac{\langle\bm{p}^2\rangle}{2\mu_2} -
		\left\langle\frac{C_F\alpha_s}{r}\right\rangle.
\end{equation}
The second line shows the same binding energy modifying the kinetic
energy.   The remaining terms are discretization errors.
In general they are a bit messy, but they simplify for the $S$-wave
states we use to tune~$\kappa$.
Then $\langle p_ip_j\rangle=\third\delta_{ij}\langle\bm{p}^2\rangle$,
whence
\begin{equation}
	E(\bm{P}) = M_1 + \frac{\bm{P}^2}{2M_2} + \cdots,
\end{equation}
where
\begin{eqnarray}
	M_2 & = & m_{2Q} + m_{2s} +
		\frac{\langle\bm{p}^2\rangle}{2\mu_2} -
			\left\langle\frac{C_F\alpha_s}{r}\right\rangle
	\nonumber \\
	 & + &  \frac{5}{3}
		\frac{\langle\bm{p}^2\rangle}{2\mu_2} 
			\left[\mu_2
			\left(\frac{m_{2Q}^2}{m_{4Q}^3} + \frac{m_{2s}^2}{m_{4s}^3} 
			\right) - 1 \right] +
		\frac{4}{3} a^3 \frac{\langle\bm{p}^2\rangle}{2\mu_2} \mu_2
			\left(w_{4Q}m_{2Q}^2+w_{4s}m_{2s}^2\right) + \cdots.
\end{eqnarray}
The last line exhibits the discretization errors, which would vanish if
$m_4=m_2$, $w_4=0$.

The error can be re-written 
\begin{equation}
	\delta M_2 = \frac{1}{3}
		\frac{\langle\bm{p}^2\rangle}{2\mu_2} \left\{
		5 \left[\mu_2
			\left(\frac{m_{2Q}^2}{m_{4Q}^3} + \frac{m_{2s}^2}{m_{4s}^3} 
			\right) - 1 \right] +
		4 a\mu_2
			\left[w_{4Q}(m_{2Q}a)^2+w_{4s}(m_{2s}a)^2\right]\right\},
	\label{eq:deltaM2}
\end{equation}
which is equivalent to Eq.~(14) of Ref.~\cite{Kronfeld:1996uy}.
Note that the error ends up being proportional to the internal kinetic
energy of the bound state, $\langle\bm{p}^2\rangle/2\mu_2$.

\subsection{Relativistic light degrees of freedom}


For asqtad light quarks, the discretization errors are $O(\alpha_s
m_s^2a^2)$ and $O(m_s^4a^4)$.
So, for a semi-quantitative estimate of the discretization error, it
should be safe to assume $m_{4s}=m_{2s}=m_{1s}=m_s$, $a^3 w_{4s}=0$.
Equation~(\ref{eq:deltaM2}) is then 
\begin{equation}
	\delta M_2 = \frac{1}{3m_{2Q}} \frac{\langle\bm{p}^2\rangle}{2 \mu_2 }
\mu_2
		\left[5 \left(\frac{m_{2Q}^3}{m_{4Q}^3} - 1 \right) +
		4 w_{4Q}(m_{2Q}a)^3\right].
	\label{eq:error_nolightdisc}
\end{equation}
To use this formula we need a value for $\langle\bm{p}^2\rangle$, and
we consider two possibilities.
The first is to replace $\langle\bm{p}^2\rangle$ with $\bar{\Lambda}^2$.
The reduced mass $\mu_2$ then cancels, yielding a sensible limit even
when $m_s\to0$.
The second is to replace the non-relativistic kinetic energy
$\langle\bm{p}^2\rangle/2\mu_2$ with a relativistic version, namely
$\bar{\Lambda}$.
If we take a constituent quark mass $m_s=\frac{1}{2}\bar{\Lambda}$,
then this discretization-error estimate equals that of the first
approach to $O(m_s/m_Q)$.

\subsection{The generalized masses and $w_4$}  \label{sec:quarkmasses}

General tree-level expressions for the quark masses and $w_4$ were
originally given in  Ref.~\cite{El-Khadra:1996mp} and succinctly
recapitulated in Ref.~\cite{Oktay:2008ex}.  For convenience we give
them here with parameters $\zeta = 1 = r_s$ as in our simulations 
\bea 
	m_0 a &=& \frac{1}{u_0} \left(  \frac{1}{2\kp} - \frac{1}{2\kc}
\right),      \label{eq:baremass_app} \\
	m_1 a &=& \ln (1 + m_0a)    \label{eq:m1_app} \\
	\frac{1}{m_2 a} &=&  \frac{2}{m_0a (2 + m_0a)}  + \frac{1}{1 + m_0a},
\label{eq:m2a_app}\\
	%
	\frac{1}{m_B a} &=&  \frac{2}{m_0a (2 + m_0a)}  + \frac{c_B}{1 +m_0a},
    \label{eq:mBa_app}\\
	\frac{1}{4m_E^2 a^2} &=&  \frac{1}{[m_0a (2+m_0a)]^2}  +
\frac{c_E}{m_0a(2+m_0a)}, \label{eq:mEa} \\
	\frac{1}{m_4^3 a^3} &=&  \frac{8}{[m_0a (2+m_0a)]^3}  + \frac{4 +
8(1+m_0a)}{[m_0a(2+m_0a)]^2} + \frac{1}{(1+m_0a)^2}, \label{eq:m4a} \\
\nonumber \\		    
	w_4     &=&  \frac{2}{m_0a(2+m_0a)} + \frac{1}{4(1+m_0a)} .
\label{w4}
\eea
These expressions and Eq.~(\ref{eq:error_nolightdisc}) are used to
obtain Table~\ref{discerr-table}.

%

%% file: App_M2.tex
 
\section{Tables of the kinetic mass}   \label{App_M2}

In this appendix, we tabulate values of the pseudoscalar, vector, and spin-averaged kinetic mass,  $ \aMkps,  \aMkvm$, and $\overline{M}_2 $, respectively.  
Values are given for all combinations of $\kappa$ and $\amq$ on the ensembles used for tuning \kp.
\chisq~and the $p$ value, one minus the $\chi^2$ cumulative distribution~\cite{PDG06}, from the dispersion relation fits are also given.

  
 \btd{ l c         ccc    cc }       
 	  &   		&			&			&					& \mc{2}{c}{$\chisq$~~($p$)}  \\
          &$\kappa$    &    \aMkps   &  \aMkvm     &  $a\overline{M}_2 $       &   \aMkps   &  \aMkvm            \\ 
         \hline   \\ [-0.5em]
          \amq~= 0.0272
          & 0.090     &  2.30(17)  &  2.31(25)  &  2.31(21)  &   0.21 (0.81)      &  0.24 (0.79)   \\ [0.5em]
         & 0.0923     &  2.19(15)  &  2.22(22)  &  2.21(19)  &   0.22 (0.80)      & 0.35 (0.71)  \\ [0.5em]
         & 0.093 &    2.16(14)  &  2.19(22)  &  2.18(18)  &   0.22 (0.80)    &    0.37 (0.69)  \\ [0.5em]
         \\
         & 0.1256        & 0.860(19)  &  0.936(42)  &  0.917(32)     &     0.09 (0.92)     &  0.14 (0.87)    \\ [0.5em]
         & 0.127         &   0.819(15)  &  0.912(39)  &  0.889(30)   &  0.36 (0.70)      &  0.00 (1.0)  \\ [0.5em]
	 \\
	 \\
          \amq~= 0.031
          & 0.0923     & 2.22(14)  &  2.22(21)  &  2.22(18)  &   0.18 (0.84)       &  0.31 (0.73)     \\ [0.5em]
          \\
          & 0.1256       & 0.871(18)  &  0.947(38)  &  0.928(30)    &     0.14 (0.87)     &  0.16 (0.85)       \\ [0.5em]
         & 0.127       &  0.828(15)  &  0.918(37)  &  0.895(29) &   0.40 (0.67)     &   0.00 (1.0)    \\ [0.5em]
\etd{The kinetic meson mass for bottom- and charm-type mesons on the (0.0062, 0.031) fine ensemble from fits to $E^2(\bm{p}) - E^2(\bm{0})$ using $|\n| \le \sqrt{3}$.  
	Fits are done to obtain \aMkps~and \aMkvm~and the results are then spin averaged. 
	Uncertainties are the average 68\% bootstrap error.
	\chisq~with the $p$ value in parentheses is also given.
	 The $p$ value is one minus the $\chi^2$ cumulative distribution~\cite{PDG06}.
	 }{M2-results-fine}

  
 \btd{ l  c    c c c   cc}       
 	&  			 &			&			&					& \mc{2}{c}{$\chisq$~~($p$)}  \\
         & $\kappa$         &   \aMkps   &  \aMkvm         &   $a\overline{M}$     &   \aMkps   &  \aMkvm       \\ 
         \hline   \\ [-0.5em]
         \amq~= 0.03
         & 0.074       &  3.78(49)  &  3.64(54)  &  3.67(50)   &    1.12 (0.33)      &  0.17 (0.84)   \\ [0.5em]
         & 0.086       &  2.93(21)  &  3.03(33)  &  3.01(29)    &  0.28 (0.76)        &   0.21 (0.81)   \\ [0.5em]
         & 0.093       &  2.50(14)  &  2.66(24)  &  2.62(21)   &   0.05 (0.95)      &  0.11 (0.90)     \\ [0.5em]
         \\
         & 0.119       &  1.263(16)  &  1.402(43)  &  1.368(34)    &  0.84 (0.43)     &  0.20 (0.82)   \\ [0.5em]   
         & 0.122       &    1.132(17)  &  1.270(46)  &  1.236(37) &    0.35 (0.70)   &     0.34 (0.71)         \\ [0.5em]   
         & 0.124       &    1.038(16)  &  1.161(43)  &  1.130(33)      &   0.28 (0.76)      &    0.52 (0.60)       \\ [0.5em]   
	 \\
	 \\ 
          \amq~= 0.0415
          & 0.074       &   3.66(35)  &  3.75(53)  &  3.73(48)  &    0.26 (0.77)     &   0.23 (0.79) \\ [0.5em]
         & 0.086       &  2.99(19)  &  3.09(28)  &  3.06(25)  &   0.46 (0.63)    &      0.48 (0.62)   \\ [0.5em]
         & 0.093       & 2.57(13)  &  2.75(25)  &  2.70(21)   &      0.14 (0.87)    &  0.31 (0.73) \\ [0.5em]
         \\
         & 0.119       & 1.292(15)  &  1.456(41)  &  1.415(33)     &    0.88 (0.41)   & 0.38 (0.69)   \\ [0.5em]   
         & 0.122       & 1.157(17)  &  1.310(44)  &  1.272(36)      &    0.19 (0.83)    &   0.24 (0.79)   \\ [0.5em]   
         & 0.124       & 1.065(15)  &  1.200(43)  &  1.166(34)    &  0.15 (0.86)       &  0.47 (0.63)    \\ [0.5em]   
\etd{Same as Table~\ref{M2-results-fine} but for mesons on the (0.007, 0.050) coarse ensemble.}{M2-results-coarse}




 \btd{ lc   ccc  cc } 
	&  			&			&			&					& \mc{2}{c}{$\chisq$~~($p$)}  \\
        & $\kappa$ &   \aMkps   &  \aMkvm         & $a\overline{M}$        &   \aMkps   &  \aMkvm    \\ 
         \hline   \\ [-0.5em]
          \amq~= 0.0484
          & 0.070       &  4.54(32)  &  4.53(45)  &  4.53(41)   &   0.55 (0.58)     &  0.78 (0.46) \\ [0.5em]
         & 0.080       &  3.79(19)  &  3.77(27)  &  3.78(24)   &   0.54 (0.58)     &  1.19 (0.31) \\ [0.5em]
         \\
          & 0.115       &   1.747(25)  &  1.825(47)  &  1.805(37)   &    1.32 (0.27)    &   0.36 (0.70) \\ [0.5em]
         & 0.125       &  1.304(12)  &  1.415(32)  &  1.387(26)   & 1.23 (0.29)      &  0.05 (0.95)  \\ [0.5em]
         \\
	\\
         \amq~= 0.0387
         & 0.070       &  4.47(35)  &  4.44(50)  &  4.44(46)   &   0.47 (0.63)     &  0.72 (0.49)  \\ [0.5em]
         & 0.080      & 3.73(21)  &  3.70(31)   &   3.71(28)  &    0.53 (0.59)    &  1.07 (0.34) \\ [0.5em]
         \\
         & 0.115       &    1.725(28)  &  1.804(58)  &  1.784(47)    &    1.06 (0.43)    &   0.04 (0.96)   \\ [0.5em]
         & 0.125       &   1.282(13)  &  1.388(37)  &  1.361(29)   &   0.89 (0.41)      &  0.07 (0.93) \\ [0.5em]         
\etd{Same as Table~\ref{M2-results-fine} but for mesons on the (0.0097, 0.0484) medium-coarse ensemble.}{M2-results-medcoarse}

%% file: App_hyperfine.tex
 
\section{Tables of $M_1 = E(0)$ and the hyperfine splitting}   \label{App_hyperfine}

In this appendix, we tabulate the hyperfine splitting \ahf\ and \ronehf\ discussed in Sec.~\ref{sec:hfs}.

\subsection{The hyperfine splitting in lattice units \ahf}
In this subsection, we tabulate values of \ahf\ relevant to the discussion in Sec.~\ref{sec:hfs} of the uncertainty in the hyperfine splitting due to statistics, $\kappa$ tuning, and the light valence mass.


\btd{ l  c  c c c c   }
	&\kp         &   ensemble           &      $\quad \aMrps$                           &     $\quad \aMrvm$                        &     $\quad \ahf $                                   \\  
	\hline  \\ [-0.5em]
	\amq = 0.0272
	&0.090     &   (0.0062, 0.031)   &  1.7387(13)  &  1.7546(19)  &  0.0158(15)                             \\  [0.75em]
	\\ 
	&0.0923   &   (0.0031, 0.031)  &    1.6877(21)  &  1.7054(25)  &  0.0177(19)    \\ [0.75em]
	&0.0923   &   (0.0062, 0.031)  &    1.6870(13)  &  1.7037(20)  &  0.0167(16)     \\  [0.75em]
	&0.0923   &   (0.0124, 0.031)  &    1.6835(16)  &  1.7024(19)  &  0.0188(14)     \\  [0.75em]
	\\   
         &0.1256    &   (0.0062, 0.031)     & 0.8408(8)  &  0.8968(16)  &  0.0561(15)      \\  [0.75em]
	\\ 
         &0.127     &   (0.0031 0.031)        & 0.7944(9)  &  0.8534(19)  &  0.0590(19)      \\  [0.75em]
         &0.127     &   (0.0062, 0.031)       & 0.7946(7)  &  0.8544(15)  &  0.0599(13)     \\  [0.75em]
         &0.127     &   (0.0124, 0.031)       &0.7901(7)  &  0.8514(11)  &  0.0613(12)    \\  [0.75em]
	\\ 
	\\
	\amq = 0.031
	&0.090          &   (0.0062, 0.031)     &1.7441(13)  &  1.7601(17)  &  0.0159(14)  \\  [0.75em]
	&0.0923        &   (0.0062, 0.031)     & 1.6926(12)  &  1.7093(18)  &  0.0167(14)  \\  [0.75em]
	 \\
         &0.1256     &   (0.0062, 0.031)     &  0.8470(8)  &  0.9030(14)  &  0.0560(13) \\  [0.75em]
         &0.127     &   (0.0062, 0.031)     &  0.8009(7)  &  0.8606(14)  &  0.0597(12)  \\  [0.75em]
\etd{Fine-ensemble values of the rest mass $\aMr=E(\bm{0})$ and hyperfine splitting \hf.  \amq~= 0.0272 and ~0.031.  
Uncertainties are the average 68\% bootstrap error.}{Tbl:M1-hyperfine-fine}


\btd{ l  c   c c c c      }
	&\kp         &   ensemble           &      $\quad \aMrps$                           &     $\quad \aMrvm$                        &     $\quad \ahf $                \\  
	\hline  \\ [-0.5em]
	\amq = 0.0415
	&0.074    & (0.007, 0.050)       &  2.2394(22)  &  2.2618(25)  &  0.0224(09)     \\   [0.75em]
	\\
	&0.086    & (0.005, 0.050)       & 1.9662(17)  &  1.9941(27)  &  0.0279(18)         \\   [0.75em]
	&0.086    & (0.007, 0.050)       &  1.9644(17)  &  1.9943(21)  &  0.0299(11)       \\   [0.75em]
	&0.086    & (0.010, 0.050)       &  1.9676(16)  &  1.9978(21)  &  0.0301(12)      \\   [0.75em]
	&0.086    & (0.020, 0.050)       & 1.9584(16)  &  1.9891(21)  &  0.0307(14)      \\   [0.75em]
	\\
	\\
	&0.122    & (0.005, 0.050)       &  1.0529(10)  &  1.1399(22)  &  0.0870(17)      \\  [0.75em]
        &0.122    & (0.007, 0.050)       &  1.0520(7)  &  1.1393(17)  &  0.0873(15)         \\  [0.75em]
        &0.122    & (0.010, 0.050)       &  1.0549(10)  &  1.1414(26)  &  0.0865(22)       \\  [0.75em]
        &0.122    & (0.020, 0.050)       &  1.0446(9)  &  1.1339(16)  &  0.0894(16)       \\  [0.75em]
	\\
	&0.124    & (0.007, 0.050)       &  0.9871(7)  &  1.0819(17)  &  0.0948(15)                  \\  [0.75em]
	\\
	\\
	\amq = 0.030
	&0.074      & (0.007, 0.050)     &  2.2241(26)  &  2.2466(29)  &  0.0225(11)     \\   [0.75em]
	&0.086      & (0.007, 0.050)     &  1.9488(21)  &  1.9787(25)  &  0.0299(14)     \\   [0.75em]
	\\
        &0.122      & (0.007, 0.050)      &  1.0339(08)  &  1.1220(20)  &  0.0881(17)     \\  [0.75em]
\etd{Same as Table~\ref{Tbl:M1-hyperfine-fine} but for the coarse ensembles with \amq~= 0.0415 and ~0.03.}{Tbl:M1-hyperfine-coarse}

\btd{ l cl   c c c c  }
	& \kp         &   ensemble           &      $\quad \aMrps$                           &     $\quad \aMrvm$                        &     $\quad \ahf $                   \\   
	\hline  \\ [-0.5em]
	\amq = 0.0484
	&0.076    & (0.0097, 0.0484) &  2.3192(27)  &  2.3472(36)  &  0.0280(17)         \\   [0.75em]
	&0.076    & (0.0194, 0.0484) &  2.3153(30)  &  2.3424(47)  &  0.0270(26)         \\   [0.75em]
	&0.076    & (0.0290, 0.0484) & 2.3137(23)  &  2.3445(21)  &  0.0308(16)         \\   [0.75em]
	\\
	&0.080    & (0.0097, 0.0484) &  2.2298(24)  &  2.2606(34)  &  0.0308(17)            \\   [0.75em]
	\\
	\\
	&0.122    & (0.0097, 0.0484) &  1.2427(8)  &  1.3390(21)  &  0.0963(19)         \\   [0.75em]
	&0.122    & (0.0194, 0.0484) &  1.2397(8)  &  1.3400(17)  &  0.1004(14)         \\   [0.75em]
	&0.122    & (0.0290, 0.0484) &  1.2364(7)  &  1.3402(17)  &  0.1038(14)         \\   [0.75em]
	\\
	&0.125    & (0.0097, 0.0484) &  1.1565(8)  &  1.2634(21)  &  0.1069(20)           \\   [0.75em]
	\\
	\\
	\amq = 0.0387
	&0.076    & (0.0097, 0.0484) &  2.3060(31)  &  2.3341(40)  &  0.0281(19)            \\   [0.75em]
	\\
	&0.122    & (0.0097, 0.0484) &  1.2271(9)  &  1.3237(25)  &  0.0966(24)           \\   [0.75em]
	\etd{Same as Table~\ref{Tbl:M1-hyperfine-fine} but for medium-coarse ensembles with \amq~= 0.0484 and ~0.0387.}{Tbl:M1-hyperfine-medcoarse}

\clearpage
\subsection{The hyperfine splitting in physical units \ronehf}    \label{App_hyperfine:r1}
In this subsection, we tabulate values of \ronehf\ relevant to the discussion in Sec.~\ref{sec:hfs} of the dependence of the hyperfine splitting on the sea-quark masses.


\btd{ l  c  c c c c c   }
	&   ensemble           &     $\quad \ronehf$            \\  
	\hline  \\ [-0.5em]
	$\kappa = 0.0923$
	&   (0.0031, 0.031)  &     0.0653(69) \\ [0.75em]
	&   (0.0062, 0.031)  &     0.0618(58)  \\  [0.75em]
	&   (0.0124, 0.031)  &     0.0700(52) \\  [0.75em]
	\\   
	$\kappa = 0.127 $
         &   (0.0031 0.031)        &     0.2178(69)  \\  [0.75em]
         &   (0.0062, 0.031)       &     0.2217(48)  \\  [0.75em]
         &   (0.0124, 0.031)       &     0.2281(43) \\  [0.75em]
\etd{Fine-ensemble values of the hyperfine splitting \hf~in units of $r_1$.  \amq~= 0.0272.  Uncertainties are  the average 68\% bootstrap error.}{Tbl:r1hfs-fine}

\btd{ l  c   c c c c c     }
	&   ensemble           &     $\quad \ronehf$            \\  
	\hline  \\ [-0.5em]
	$\kappa = 0.086  $
	& (0.005, 0.050)       &  0.0738(46)    \\   [0.75em]
	& (0.007, 0.050)       &   0.0788(29)   \\   [0.75em]
	& (0.010, 0.050)       &  0.0788(32)  \\   [0.75em]
	& (0.020, 0.050)       &  0.0814(36)  \\   [0.75em]
	\\
	$\kappa = 0.122  $
	& (0.005, 0.050)       & 0.2301(44)  \\  [0.75em]
        & (0.007, 0.050)       &  0.2300(39)    \\  [0.75em]
        & (0.010, 0.050)       &  0.2265(58)   \\  [0.75em]
        & (0.020, 0.050)       &  0.2370(43)   \\  [0.75em]
\etd{Same as Table~\ref{Tbl:r1hfs-fine} but for the coarse-ensembles with \amq~= 0.0415.}{Tbl:r1hfs-coarse}

\btd{ l cl   c c c c c }
	&   ensemble           &     $\quad \ronehf$         \\   
	\hline  \\ [-0.5em]
	$\kappa = 0.076 $
	& (0.0097, 0.0484) &  0.0616(37)    \\   [0.75em]
	& (0.0194, 0.0484) &  0.0603(57)    \\   [0.75em]
	& (0.0290, 0.0484) &  0.0699(37)    \\   [0.75em]
	\\
	$\kappa = 0.122 $
	& (0.0097, 0.0484) &  0.2117(41)    \\   [0.75em]
	& (0.0194, 0.0484) &  0.2244(30)    \\   [0.75em]
	& (0.0290, 0.0484) &  0.2357(39)   \\   [0.75em]
	\etd{Same as Table~\ref{Tbl:r1hfs-fine} but for medium-coarse--ensembles with \amq~= 0.0484.}{Tbl:r1hfs-medcoarse}

%% file: App_hfsPQChiPT.tex

\section{Partially quenched chiral perturbation theory for the heavy-light hyperfine splitting}
\label{App_hfsPQChiPT}

For full (unquenched) QCD,
Jenkins \cite{Jenkins:1992hx} has calculated the hyperfine splitting at one loop in heavy-meson
chiral perturbation theory.  
It is not difficult to take her result (Eq.~(A.10) of Ref.~\cite{Jenkins:1992hx}) and
extend it to partially quenched QCD.  The further step of including staggered taste-violations
({\it i.e.}, doing staggered chiral perturbation theory) would also be fairly straightforward,
but we do not take it here because the continuum partially quenched form is sufficient
for estimating the small systematic effect due to the mistuning of sea quark masses.
Unlike Jenkins, we neglect electromagnetic and isospin-violating effects.

At the quark-flow level, the relevant diagrams are the
self-energy diagrams 
shown in Fig.~5(a) [left] of Ref.~\cite{Aubin:2005aq} (the ``connected diagram'') and
in Figs.~5(b),(c)  [left] of Ref.~\cite{Aubin:2005aq} (the ``disconnected diagram'').\footnote{%
One should ignore the solid square in each of the figures from  Ref.~\cite{Aubin:2005aq} because it
represents a current insertion, not relevant here.} One simply needs to determine how much of
Jenkins's result comes from each of these two diagrams. This is accomplished by
noting that, when the light valence quark is a $u$ ($a=1$ in Jenkins's notation), an 
internal kaon only appears in the connected diagram, when the quark in the virtual
loop is an $s$.  This fixes the normalization of the connected diagram. 
Using the methods described in Refs.~\cite{Aubin:2003mg,Aubin:2003uc,Aubin:2005aq} (but dropping
the taste violations and indeed the taste degree of freedom itself), the disconnected diagram 
is  easily calculated. Its normalization can then be fixed so that it supplies the 
remainder of the $a=1$ result in Ref.~\cite{Jenkins:1992hx}.

There are ample checks of this reasoning. First, the same normalizations must apply for any
choice of the valence mass.  The $\eta$ contributes in each case only through the disconnected
diagram, while the pion contributions come from both connected and disconnected diagrams for
valence $u$ or $d$ ($a=1,2$), and must be absent for valence $s$ ($a=3$). Finally the 
contribution from the unphysical $s\bar s$ state, which appears in each diagram for $a=3$, 
should cancel.

It is then immediate to write down the partially quenched version. 
Let the light valence quark be $x$, with
mass $m_x$, and let the sea quarks be $u,d,s$ with masses $m_u=m_d=m_l$ and $m_s$.
With the light meson decay 
constant $f$ normalized so that $f\approx f_\pi\approx 130$ MeV, the
hyperfine splitting $M^*_x-M_x$ is given by
\begin{equation}  \label{eq:pqchipt}
M^*_x-M_x = \Delta -\frac{\Delta g_\pi^2}{8\pi^2 f^2} \delta_{\rm log} + 2\Delta^{(\sigma)}(2m_l+m_s) +  2\Delta^{(a)}m_x \ ,
\end{equation}
where $\Delta$ is the splitting in the (three-flavor) chiral limit, and $\Delta^{(\sigma)}$
and $\Delta^{(a)}$ are LECs that start at order $1/m_Q$ in the heavy quark expansion.
The non-analytic chiral logarithms $\delta_{\rm log}$ are 
\begin{equation}
\delta_{\rm log} = \sum_{F=u,d,s}\ell(M^2_{xF}) - \frac{1}{3} R_X^{[2,2]}(\{m\},\{\mu\})\;
\tilde \ell(M^2_X) 
- \frac{1}{3} \sum_{j=X,\eta} D_{j,X}^{[2,2]}(\{m\},\{\mu\})\; \ell(M^2_j) \ .
\end{equation}
Here $M_X$ is the mass of the valence $x\bar x$ meson, and 
$M_{xF}$ is the mass of the mixed valence-sea $x\bar F$ meson.
The residue functions $R_j^{[n,k]}$ and $D_{j,i}^{[n,k]}$, as well as the chiral logarithm
functions  $\ell(m^2)$ and $\tilde \ell(m^2)$, are defined in 
Refs.~\cite{Aubin:2003mg,Aubin:2003uc}. The term with the sum over $F$ comes from the
connected diagram, while those with the residue functions come from the disconnected diagram,
which has a double pole at $M_X^2$ in the partially quenched case. 
The denominator ($\{m\}$) and numerator ($\{\mu\}$) mass sets are
\begin{equation}
\{m\} =  \{M_X,M_\eta\}\ , \qquad \{\mu\} =  \{M_U,M_S\}
\end{equation}
with $M_U$ and $M_S$ the masses of the $u\bar u$ and $s\bar s$ mesons, respectively.


%% file: Draft_Mesons.bbl
\begin{thebibliography}{99}



%
\bibitem{Davies:2003ik}
  C.~T.~H.~Davies {\it et al.}\ 
  [HPQCD, MILC, and Fermilab Lattice Collaborations],
  Phys.\ Rev.\ Lett.\  {\bf 92}, 022001 (2004)
  [arXiv:hep-lat/0304004];

%
\bibitem{Aubin:2004ck}
  C.~Aubin {\it et al.}  [HPQCD, MILC, and UKQCD Collaborations],
  Phys.\ Rev.\ D {\bf 70}, 031504(R) (2004)
  [arXiv:hep-lat/0405022];
  C.~Aubin {\it et al.}  [MILC Collaboration],
  Phys.\ Rev.\  D {\bf 70}, 114501 (2004)
  [arXiv:hep-lat/0407028].



%
\bibitem{Aubin:2005ar}
  C.~Aubin {\it et al.} [Fermilab Lattice, MILC, and HPQCD Collaborations],
  Phys.\ Rev.\ Lett.\  {\bf 95}, 122002 (2005)
  [arXiv:hep-lat/0506030].
  C.~Bernard {\it et al.} [Fermilab Lattice and MILC Collaborations],
  PoS {\bf LATTICE2008}, 278 (2008)
  [arXiv:0904.1895 [hep-lat]].


%
\bibitem{Aubin:2004ej}
  C.~Aubin {\it et al.}  [Fermilab Lattice, MILC, and HPQCD Collaborations],
  Phys.\ Rev.\ Lett.\  {\bf 94}, 011601 (2005)
  [arXiv:hep-ph/0408306];
  M.~Okamoto {\it et al.} [Fermilab Lattice and MILC Collaborations],
  Nucl.\ Phys.\ B Proc.\ Suppl.\  {\bf 140}, 461 (2005)
  [arXiv:hep-lat/0409116];
  C.~Bernard {\it et al.} [Fermilab Lattice and MILC Collaborations],
  Phys.\ Rev.\  D {\bf 79}, 014506 (2009)
  [arXiv:0808.2519 [hep-lat]];
  J.~A.~Bailey {\it et al.} [Fermilab Lattice and MILC Collaborations],
  Phys.\ Rev.\  D {\bf 79}, 054507 (2009)
  [arXiv:0811.3640 [hep-lat]];
  C.~Bernard {\it et al.} [Fermilab Lattice and MILC Collaborations],
  Phys.\ Rev.\  D {\bf 80}, 034026 (2009)
  [arXiv:0906.2498 [hep-lat]].


\bibitem{Evans:2008zz}
  R.~T.~Evans {\it et al.} [Fermilab~Lattice and MILC collaborations], 
  PoS {\bf LATTICE 2008} 052 (2008);
  R.~T.~Evans, E.~G\'amiz, A.~El-Khadra and A.S.~Kronfeld  [Fermilab Lattice and
                  MILC Collaborations],
  PoS {\bf LATTICE 2009} 245 (2009)
  [arXiv:0911.5432 [hep-lat]].


\bibitem{Freeland:2007wk}
  E.~D.~Freeland, A.~S.~Kronfeld, J.~N.~Simone and R.~S.~Van de Water
                  [Fermilab Lattice and MILC Collaborations],
  PoS {\bf LAT2007}, 243 (2007)
  [arXiv:0710.4339 [hep-lat]].


%
\bibitem{Allison:2004be}
I.~F.~Allison \emph{et al.}\ 
  [HPQCD and Fermilab Lattice Collaborations],
  Phys.\ Rev.\ Lett.\  {\bf 94}, 172001 (2005)
  [arXiv:hep-lat/0411027].

%
\bibitem{Burch:2009az}
  T.~Burch {\it et al.} [Fermilab Lattice and MILC Collaborations],
    Phys.\ Rev.\ D {\bf 81}, 034508 (2010)
  [arXiv:0912.2701 [hep-lat]].



%
\bibitem{Bernard:2001av}
  C.~W.~Bernard {\it et al.},
  Phys.\ Rev.\ D {\bf 64}, 054506 (2001)
  [arXiv:hep-lat/0104002].
%
\bibitem{Aubin:2004wf}
  C.~Aubin {\it et al.},
  Phys.\ Rev.\ D {\bf 70}, 094505 (2004)
  [arXiv:hep-lat/0402030].
  

\bibitem{Bazavov:2009bb}
  A.~Bazavov {\it et al.},
  Rev.~Mod.~Phys., to be published
  [arXiv:0903.3598 [hep-lat]].
  



\bibitem{r1}
  C.~W.~Bernard {\it et al.},
  Phys.\ Rev.\  D {\bf 62}, 034503 (2000)
  [arXiv:hep-lat/0002028]; 
  R.~Sommer,
  Nucl.\ Phys.\  B {\bf 411}, 839 (1994)
  [arXiv:hep-lat/9310022].
%


  \bibitem{stag_fermion}
  T.~Blum {\it et al.},       
  Phys.\ Rev.\ D {\bf 55}, R1133 (1997)
  [arXiv:hep-lat/9609036];
%
  K.~Orginos and D.~Toussaint  [MILC Collaboration],
  Phys.\ Rev.\ D {\bf 59}, 014501 (1998)
  [arXiv:hep-lat/9805009];
%
  J.~F.~Laga\"e and D.~K.~Sinclair,
  Phys.\ Rev.\ D {\bf 59}, 014511 (1998)
  [arXiv:hep-lat/9806014];
%
  G.~P.~Lepage,        
  Phys.\ Rev.\ D {\bf 59}, 074502 (1999)
  [arXiv:hep-lat/9809157];
%
  K.~Orginos, D.~Toussaint and R.~L.~Sugar  [MILC Collaboration],      
  Phys.\ Rev.\ D {\bf 60}, 054503 (1999)
  [arXiv:hep-lat/9903032];
%
  C.~W.~Bernard {\it et al.}  [MILC Collaboration],
  Phys.\ Rev.\ D {\bf 61}, 111502(R) (2000)
  [arXiv:hep-lat/9912018].
  


\bibitem{El-Khadra:1996mp}
  A.~X.~El-Khadra, A.~S.~Kronfeld and P.~B.~Mackenzie,
  Phys.\ Rev.\ D {\bf 55}, 3933 (1997)
  [arXiv:hep-lat/9604004].

  

\bibitem{Sheikholeslami:1985ij}
  B.~Sheikholeslami and R.~Wohlert,
  Nucl.\ Phys.\  B {\bf 259}, 572 (1985).


\bibitem{wilson:1977}
  K.~G.~Wilson, 
  in {\em New Phenomena in Subnuclear Physics}, 
  edited by A.~Zichichi (Plenum, New York, 1977).




\bibitem{Kronfeld:2000ck}
  A.~S.~Kronfeld,
  Phys.\ Rev.\ D {\bf 62}, 014505 (2000)
  [arXiv:hep-lat/0002008].


\bibitem{HQETradiative}
  J.~Harada, S.~Hashimoto, K.~I.~Ishikawa, A.~S.~Kronfeld, T.~Onogi and N.~Yamada,
  Phys.\ Rev.\  D {\bf 65}, 094513 (2002)
  [Erratum-ibid.\  D {\bf 71}, 019903 (2005)]
  [arXiv:hep-lat/0112044];
  J.~Harada, S.~Hashimoto, A.~S.~Kronfeld and T.~Onogi,
  Phys.\ Rev.\  D {\bf 65}, 094514 (2002)
  [arXiv:hep-lat/0112045].
  
%
\bibitem{Symanzik:1979ph}
  K. Symanzik,
  in \emph{Recent Developments in Gauge Theories},
  edited by G. 't~Hooft \emph{et al}.\ (Plenum, New York, 1980);
  in \emph{Mathematical Problems in Theoretical Physics},
  edited by R. Schrader \emph{et al}.\ (Springer, New York, 1982);
  Nucl.\ Phys. B {\bf 226}, 187, 205 (1983).

    
%
\bibitem{Lepage:1992xa}
  G.~P.~Lepage and P.~B.~Mackenzie,
  Phys.\ Rev.\  D {\bf 48}, 2250 (1993)
  [arXiv:hep-lat/9209022].

    
 
%
\bibitem{Mertens:1997wx}
  B.~P.~G.~Mertens, A.~S.~Kronfeld and A.~X.~El-Khadra,
  Phys.\ Rev.\ D {\bf 58}, 034505 (1998)
  [arXiv:hep-lat/9712024].
  

  
\bibitem{Kronfeld:2000gk}
  A.~S.~Kronfeld and J.~N.~Simone,
  Phys.\ Lett.\  B {\bf 490}, 228 (2000)
  [Erratum-ibid.\  B {\bf 495}, 441 (2000)]
  [arXiv:hep-ph/0006345].



\bibitem{Oktay:2008ex}
  M.~B.~Oktay and A.~S.~Kronfeld,
  Phys.\ Rev.\  D {\bf 78}, 014504 (2008)
  [arXiv:0803.0523 [hep-lat]].
  
  

  
%

\bibitem{Alford:1995hw}
  M.~G.~Alford, W.~Dimm, G.~P.~Lepage, G.~Hockney and P.~B.~Mackenzie,
  Phys.\ Lett.\  B {\bf 361}, 87 (1995)
  [arXiv:hep-lat/9507010].
%
\bibitem{more_imp_glue}
  C.~W.~Bernard {\it et al.}  [MILC Collaboration],
  Phys.\ Rev.\  D {\bf 58}, 014503 (1998)
  [arXiv:hep-lat/9712010]; 
%
  M.~L\"uscher and P.~Weisz,
  Commun.\ Math.\ Phys.\  {\bf 97}, 59 (1985)
  [Erratum-ibid.\  {\bf 98}, 433 (1985)];
%
  M.~Luscher and P.~Weisz,
  Phys.\ Lett.\  B {\bf 158}, 250 (1985).





\bibitem{Shamir:2006nj}
  Y.~Shamir,
  Phys.\ Rev.\  D {\bf 71}, 034509 (2005)
  [arXiv:hep-lat/0412014];
  Phys.\ Rev.\  D {\bf 75}, 054503 (2007)
  [arXiv:hep-lat/0607007].



\bibitem{Bernard:2006zw}
  C.~Bernard,
  Phys.\ Rev.\  D {\bf 73}, 114503 (2006)
  [arXiv:hep-lat/0603011].

\bibitem{Bernard:2007ma}
  C.~Bernard, M.~Golterman and Y.~Shamir,
  Phys.\ Rev.\  D {\bf 77}, 074505 (2008)
  [arXiv:0712.2560 [hep-lat]].




\bibitem{staggered:reviews}
  S.~D\"urr,
  PoS {\bf LAT2005}, 021 (2006)
  [arXiv:hep-lat/0509026];
%
  S.~R.~Sharpe,
  PoS {\bf LAT2006}, 022 (2006)
  [arXiv:hep-lat/0610094];
  M.~Golterman,
  PoS {\bf CONFINEMENT8}, 014 (2008)
  [arXiv:0812.3110 [hep-ph]].


\bibitem{Kronfeld:2007ek}
  A.~S.~Kronfeld,
  PoS {\bf LAT2007}, 016 (2007)
  [arXiv:0711.0699 [hep-lat]].
  


\bibitem{MILC07}
For $m_s$ and $r_1/a$, we use the fitting methods described in Ref.~\cite{Bazavov:2009bb}
  as applied to the data available in June, 2007~\cite{Bernard:2007ux}. 
  Specifically, to smooth $r_1/a$, $\ln(r_1/a)$ is fit to a polynomial in $\beta$ and $2am'_l + am'_s$.
  Values of $u_0$ can also be found in this reference.
%
\bibitem{Bernard:2007ux}
  C.~Bernard {\it et al.}  [MILC Collaboration],
  PoS {\bf LAT2007}, 137 (2007)
  [arXiv:0711.0021 [hep-lat]].
 


\bibitem{Gray:2005ur}
  A.~Gray, I.~Allison, C.~T.~H.~Davies, E.~Gulez, G.~P.~Lepage, J.~Shigemitsu and M.~Wingate,
  Phys.\ Rev.\  D {\bf 72}, 094507 (2005)
  [arXiv:hep-lat/0507013].
  
  
  
\bibitem{MILCr1_upsilon}
  C.~Bernard {\it et al.}  [MILC Collaboration],
  PoS {\bf LAT2005}, 025 (2006)
  [arXiv:hep-lat/0509137].

  
\bibitem{Bernard:2007ps}
  C.~Bernard {\it et al.},
  PoS {\bf LAT2007}, 090 (2007)
  [arXiv:0710.1118 [hep-lat]].




\bibitem{Davies:2009tsa}
  C.~T.~H.~Davies, E.~Follana, I.~D.~Kendall, G.~P.~Lepage and C.~McNeile,
  [arXiv:0910.1229 [hep-lat]].
  
  
  
  


\bibitem{Menscher:2005kj}
  D.~P.~Menscher,
  ``Charmonium and charmed mesons with improved lattice QCD,'' (University of Illinois Ph.~D.\ thesis, 2005).
%
\bibitem{Richardson:1978bt}
  J.~L.~Richardson,
  Phys.\ Lett.\  B {\bf 82}, 272 (1979).
  
  
  
  
\bibitem{Wingate:2002fh}
  M.~Wingate, J.~Shigemitsu, C.~T.~H.~Davies, G.~P.~Lepage and H.~D.~Trottier,
  Phys.\ Rev.\  D {\bf 67}, 054505 (2003)
  [arXiv:hep-lat/0211014].





\bibitem{Golterman:1986jf}
  M.~F.~L.~Golterman,
  Nucl.\ Phys.\  B {\bf 278}, 417 (1986).
  
  
\bibitem{Gliozzi:1982ib}
  F.~Gliozzi,
  Nucl.\ Phys.\  B {\bf 204}, 419 (1982).

\bibitem{KlubergStern:1983dg}
  H.~Kluberg-Stern, A.~Morel, O.~Napoly and B.~Petersson,
  Nucl.\ Phys.\  B {\bf 220}, 447 (1983).
  
  

 
\bibitem{Lepage:2001ym}
  G.~P.~Lepage, B.~Clark, C.~T.~H.~Davies, K.~Hornbostel, P.~B.~Mackenzie, C.~Morningstar and H.~Trottier,
  Nucl.\ Phys.\ Proc.\ Suppl.\  {\bf 106}, 12 (2002)
  [arXiv:hep-lat/0110175];
  C.~Morningstar,
  Nucl.\ Phys.\ Proc.\ Suppl.\  {\bf 109A}, 185 (2002)
  [arXiv:hep-lat/0112023].
  
  
  
\bibitem{Bayes}
For a pedagogical introduction see D.~S.~Sivia, {\it Data Analysis: A Bayesian Tutorial} (Oxford University Press, USA, 1996); 
a review can be found in K. Nakamura et al. (Particle Data Group), J. Phys. G 37, 075021 (2010) .
  



\bibitem{PDG06}
  W.~M.~Yao {\it et al.}  [Particle Data Group],
  J.\ Phys.\ G {\bf 33}, 1 (2006) and 2007 partial update for edition 2008 (URL: http://pdg.lbl.gov).
  The 2009-10 updates for the $D_s$ hyperfine splitting have not changed its value.  
  For the $B_s$ hyperfine splitting, the ``average value'' of 46.1(1.5) has remained consistent; the ``fit'' value has increased slightly to 49.0(1.5).





\bibitem{Kronfeld:1996uy}
A.~S.~Kronfeld,
Nucl.\ Phys.\ B Proc.\ Suppl.\  {\bf 53}, 401 (1997)
[arXiv:hep-lat/9608139].
%


\bibitem{Jenkins:1992hx}
  E.~E.~Jenkins,
  Nucl.\ Phys.\  B {\bf 412}, 181 (1994)
  [arXiv:hep-ph/9212295].


\bibitem{Arnesen:2005ez}
  C.~M.~Arnesen, B.~Grinstein, I.~Z.~Rothstein and I.~W.~Stewart,
  Phys.\ Rev.\ Lett.\  {\bf 95}, 071802 (2005)
  [arXiv:hep-ph/0504209].

\bibitem{Eichten:1990vp}
 E.~Eichten and B.~R.~Hill,
 Phys.\ Lett.\  B {\bf 243}, 427 (1990).
 
 
\bibitem{MNobes}
  M.~Nobes and H.~Trottier,
  PoS {\bf LAT2005}, 209 (2006)
  [arXiv:hep-lat/0509128];
   %
   M.~Nobes, ``Automated Lattice Perturbation Theory for Improved Quark and Gluon Actions,'' (Simon Fraser University) Ph.~D.\ thesis, 2004.
 
 
\bibitem{twistedBC}
  C.~T.~Sachrajda and G.~Villadoro,
  Phys.\ Lett.\  B {\bf 609}, 73 (2005)
  [arXiv:hep-lat/0411033];
  P.~F.~Bedaque,
  Phys.\ Lett.\  B {\bf 593}, 82 (2004)
  [arXiv:nucl-th/0402051].
  

  
  



\bibitem{Freeland:2006nd}
  E.~D.~Freeland, A.~S.~Kronfeld, J.~N.~Simone and R.~S.~Van de Water
                  [for the Fermilab Lattice and MILC Collaborations],
  PoS {\bf LAT2006}, 083 (2006)
  [arXiv:hep-lat/0610108].
  
  
\bibitem{Aubert:2006bk}
  B.~Aubert {\it et al.}  [BABAR Collaboration],
  Phys.\ Rev.\  D {\bf 74}, 032007 (2006)
  [arXiv:hep-ex/0604030].


\bibitem{diPierro:2003iw}
  M.~Di Pierro {\it et al.},
  Nucl.\ Phys.\ Proc.\ Suppl.\  {\bf 129}, 328 (2004)
  [arXiv:hep-lat/0310045].


\bibitem{Berestetskii:1971vol6}
V.~I. Berestetskii, E.~M. Lifshitz, and L.~P. Pitaevskii,
\emph{Relativistic Quantum Theory} (Pergamon, Oxford, 1971).

%



\bibitem{Aubin:2005aq}
  C.~Aubin and C.~Bernard,
  Phys.\ Rev.\  D {\bf 73}, 014515 (2006)
  [arXiv:hep-lat/0510088].

\bibitem{Aubin:2003mg}
C.\ Aubin and C.\ Bernard
Phys.\ Rev.\  D {\bf 68} (2003) 034014 [arXiv:hep-lat/0304014].

\bibitem{Aubin:2003uc}
C.\ Aubin and C.\ Bernard,
Phys.\ Rev.\ D {\bf 68} (2003) 074011 [arXiv:hep-lat/0306026].

\end{thebibliography}
